\documentclass[twocolumn,tighten]{aastex62}

\received{July 5, 2018}
\revised{October 31, 2018}
\accepted{November 5, 2018}
\submitjournal{ApJ}

\usepackage{url, soul, bm}
\usepackage{amsmath}      

\shorttitle{Kinematics of Young Clusters}
\shortauthors{Kuhn et al.}

\begin{document} 

\title{Kinematics in Young Star Clusters and Associations with {\it Gaia} DR2}

\correspondingauthor{Michael A. Kuhn}
\email{mkuhn@astro.caltech.edu}

\author[0000-0002-0631-7514]{Michael A. Kuhn}
\affil{Department of Astronomy, California Institute of Technology, Pasadena, CA 91125, USA}

\author{Lynne A. Hillenbrand}
\affil{Department of Astronomy, California Institute of Technology, Pasadena, CA 91125, USA}

\author{Alison Sills}
\affiliation{Department of Physics \& Astronomy, McMaster University, 1280 Main Street West, Hamilton, ON L8S 4M1, Canada}

\author{Eric D. Feigelson}
\affiliation{Department of Astronomy \& Astrophysics, 525 Davey Laboratory, Pennsylvania State University, University Park, PA 16802, USA}

\author{Konstantin V. Getman}
\affiliation{Department of Astronomy \& Astrophysics, 525 Davey Laboratory, Pennsylvania State University, University Park, PA 16802, USA}

\begin{abstract}
The {\it Gaia} mission has opened a new window into the internal kinematics of young star clusters at the sub-km s$^{-1}$ level, with implications for our understanding of how star clusters form and evolve. We use a sample of 28 clusters and associations with ages from $\sim$1--5~Myr, where lists of members are available from previous X-ray, optical, and infrared studies. Proper motions from {\it Gaia} DR2 reveals that at least 75\% of these systems are expanding; however, rotation is only detected in one system. Typical expansion velocities are on the order of $\sim$0.5~km~s$^{-1}$, and, in several systems, there is a positive radial gradient in expansion velocity. Systems that are still embedded in molecular clouds are less likely to be expanding than those that are partially or fully revealed. One-dimensional velocity dispersions, which range from $\sigma_{1D}=1$ to 3~km~s$^{-1}$, imply that most of the stellar systems in our sample are supervirial and that some are unbound. In star-forming regions that contain multiple clusters or subclusters, we find no evidence that these groups are coalescing, implying that hierarchical cluster assembly, if it occurs, must happen rapidly during the embedded stage.
\end{abstract}


\section{Introduction}\label{intro.sec}

While star formation in the Galaxy occurs in large star-forming complexes, most of the stars formed in these complexes quickly disperse into the field without remaining bound to one another as members of open clusters \citep{2010ARA&A..48...47A, 1538-3873-130-989-072001}. Star formation takes place in turbulent, clumpy giant molecular clouds, and yet the star clusters that do remain bound tend to have smooth stellar density distributions. 
 The processes of cluster assembly, equilibration, and dissolution have remained poorly constrained
by observation for two reasons: the difficulty of obtaining reliable samples of cluster members in nebulous regions with many field star contaminants and the absence of kinematic information for faint stars. The nearby Orion Nebula Cluster (ONC) has been extensively characterized, but these difficulties have been overcome only more recently for other massive star-forming complexes. In these regions, samples of members emerge from multi-wavelength surveys \citep{2018ASSL..424..119F}, while the {\it Gaia} mission \citep{2016A&A...595A...1G} is expected to revolutionize the study of their internal kinematics \citep{2012MNRAS.421.3338A,clarke2015dynamics,2016EAS....80...73M}.

The kinematics of young stars should reflect the processes of star cluster formation. Two principal paths from theoretical studies are: ``monolithic'' cluster formation, in which a young star cluster is born in a single molecular cloud core,  and ``hierarchical'' cluster formation, in which larger clusters are built via the accumulation of smaller subclusters \citep{2000ApJ...530..277E,2003MNRAS.343..413B,2015MNRAS.447..728B}. Young stellar objects (YSOs) that are embedded or partially embedded in star-forming clouds typically have clumpy distributions, where groups are known as subclusters \citep[e.g.,][]{1972A&A....21..255A,1995AJ....109.1682L,2014ApJ...787..107K,2016AJ....151....5M,2018MNRAS.tmp..655S}.  It has been unclear, however, whether the subclusters will disperse once the molecular gas disperses, or if they will merge into larger, possibly bound clusters that survive gas expulsion. Examination of the motions of these subclusters can help constrain cluster formation scenarios.

The evolution of a young cluster depends on its dynamical state \citep[e.g.,][]{2014MNRAS.438..620P,2018MNRAS.tmp..655S}, as well as changes in the gravitational potential due to the dispersal of the molecular cloud  \citep[e.g.,][]{1978A&A....70...57T,1980ApJ...235..986H} and tidal interactions with clouds and clusters \citep[e.g.,][]{2012MNRAS.426.3008K}.
Much attention has been focused on the role of gas expulsion, which will cause a cluster to expand and can cause bound embedded clusters to become unbound \citep{1983ApJ...267L..97M, 1983MNRAS.203.1011E, 2000ApJ...542..964A, 2001MNRAS.321..699K, 2006MNRAS.373..752G, 2012MNRAS.420.1503P, 2017A&A...600A..49B}. However, recent simulations have suggested that gas expulsion may play a less important role in cluster disruption if there is spatial decoupling between stars and gas \citep{2015MNRAS.451..987D}. 

The empirical relationship between size and density of young star clusters and associations provides indirect evidence for their expansion \citep{2009A&A...498L..37P, 2011A&A...536A..90P,2014ApJ...794..147P}. 
For subclusters, size has been found to be negatively correlated with density, and positively correlated with age, as would be expected if they were expanding \citep{2015ApJ...812..131K, 2018arXiv180405075G}. Furthermore, \citet{2012A&A...543A...8M} argue that the low binary fractions observed in young star clusters imply these clusters were more compact upon formation. 

Direct evidence of cluster expansion via proper-motion measurements has been difficult to obtain. Recent studies of OB associations have yielded either no evidence of expansion \citep{2018MNRAS.475.5659W} or no evidence of expansion from a single compact system \citep{2016MNRAS.460.2593W,2018MNRAS.476..381W}.  For the ONC, there has long been debate about whether the system is expanding or contracting \citep[][]{2008hsf1.book..483M}. In recent work, \citet{2017ApJ...834..139D} find no evidence of expansion or contraction of the ONC using proper motions from radio observations, while \citet{2017ApJ...845..105D} report a correlation between radial velocity (RV) and extinction that can be explained by expansion, and \citet{2018arXiv180504649K} report a preference for expansion among stars around the outer edges of Orion~A. 

In this paper, we use the superb astrometry of {\it Gaia}'s Second Data Release \citep[DR2;][]{GaiaBrown} to elucidate the formation (evidence for or against hierarchical assembly) and dynamical state (expansion, contraction, or equilibrium) of young star clusters. 
We use a sample of young star clusters and associations, ranging from 0.3--3.7~kpc, that were the focus of a series of studies combining NASA's {\it Chandra} X-ray, {\it Spitzer} mid-infrared, and ground-based optical and near-infrared images to provided reliable samples of tens-of-thousands of YSOs  \citep{2013ApJS..209...26F,2017ApJS..229...28G,2017AJ....154...87K}. 

Section~\ref{data.sec} describes the data, and Section~\ref{properties.sec} derives basic cluster properties that are used in the analysis. Section~\ref{bulk.sec} addresses the question of whether clusters are in equilibrium, expanding, or contracting, and whether they are rotating. Section~\ref{sd.sec} examines the velocity dispersions in clusters. Section~\ref{subcluster_motions.sec} addresses the question of whether young clusters form by merging of subclusters. Section~\ref{physical_properties.sec} relates internal cluster kinematics to other physical properties. A discussion of the observational results, along with a comparison to a cluster formation simulation, are provided in Section~\ref{discussion.sec}. Section~\ref{conclusions.sec} is the conclusion.

\section{Data Sets}\label{data.sec}

\subsection{YSOs in Clusters and Associations}

 This {\it Gaia} study is based on samples of YSOs, with typical ages of 1--5~Myr, from the Massive Young Star-Forming Complex Study in Infrared and X-ray survey \citep[MYStIX;][]{2013ApJS..209...32B,2013ApJS..209...26F}, the Star-Formation in Nearby Clouds study \citep[SFiNCs;][]{2017ApJS..229...28G}, and a similar study of NGC~6231 \citep{2017AJ....154...87K}.  In each of these studies, X-ray emission was used to classify probable cluster members based on the higher expected X-ray luminosities of pre--main-sequence stars compared to main-sequence field stars. MYStIX and SFiNCs also include sources selected by infrared excess, which indicates YSOs with disks and envelopes that may or may not be X-ray emitters. The infrared-selected samples come from \citet{2013ApJS..209...31P} and \citet{2017ApJS..229...28G}, with careful attention to reduce contamination by post--main-sequence dusty red giants. 
Reducing field star contamination is particularly important for these massive star forming regions that lie in the Galactic Plane.  

In this study, we include only the regions that contain rich clusters visible in the optical.  We omit the regions with few stars, and those for which {\it Gaia} is limited by high optical extinction (e.g., Serpens South or DR~21). For inclusion in our study, a region must contain at least 20 cluster members with reliable {\it Gaia} proper motions. These criteria yield a sample of 28 star clusters and associations which reside in 21 star-forming regions (Table~\ref{clusters.tab}).

In this paper, we use the term ``stellar system'' to indicate a major group of spatially associated stars, which includes embedded clusters, bound open clusters, and compact associations of unbound stars \citep{2014ApJ...787..107K}.\footnote{Historically, it has been difficult to distinguish between bound and unbound young stellar systems from observations. Studies of cluster density have shown that there is no density threshold that divides ``clustered'' and ``distributed'' star formation \citep[e.g.,][]{2010MNRAS.409L..54B,2012MNRAS.426L..11G,2015ApJ...802...60K}, and uncertainties in measurements of system masses and kinematics limit dynamical modeling. The {\it Gaia} mission will undoubtedly improve the situation for the last two issues.} Some of the regions contain multiple stellar systems that are analyzed individually. Notable examples include NGC~2264  (containing a loose association around S~Mon and embedded clusters around IRS~1 and IRS~2 adjacent to the Cone Nebula), NGC~6357 (containing Pismis~24, G353.1+0.6, and
G353.2+0.7), the Carina Nebula (including Tr~14, Tr~15, and Tr~16), and the Cep~OB3b association (containing a group to the east adjacent to the Cep~B cloud and a group to the west around V454~Cep). For the Rosette Nebula region, which includes stars both in the cluster NGC~2244 and in the Rosette Molecular Clouds, we use only NGC~2244 for expansion analysis. In the Orion star-forming region, we focus only on the ONC. The sample of 28 stellar systems is given in Table~\ref{clusters.tab}.

\subsection{Cross-matching to  {\it Gaia} DR2}

Our study is primarily based on astrometric measurements from the {\it Gaia} DR2 catalog \citep{GaiaBrown,2018arXiv180409366L}. We use the astrometric notation $\alpha$ and $\delta$ for right ascension and declination, $\varpi$ for parallax in units of milliarcseconds (mas) and $\mu_{\alpha^\star}$ and $\mu_\delta$ for proper motions in units of mas~yr$^{-1}$, where $\mu_{\alpha^\star}\equiv\mu_\alpha\cos\delta$. 

We cross-matched 30,839 objects from the YSO catalogs to sources in the {\it Gaia} catalog. Significant effort has already been devoted to identifying the best match between X-ray and optical/infrared sources in the MYStIX, SFiNCs, and NGC~6231 catalogs \citep[e.g.,][]{2013ApJS..209...30N,2017ApJS..229...28G}. The optical or infrared source coordinates are often more precise than the X-ray positions, and in such cases we use those coordinates for cross-matching with {\it Gaia}. The match radius for matching to {\it Gaia} sources was set to 1.2~arcsec, and we select the nearest {\it Gaia} source within that match radius. 

The cross matching lead to 20,716 matches with the {\it Gaia} DR2 catalog, 17,509 of which have the 5-parameter ``astrometric global iterative solution'' (AGIS) involving position, parallax, and proper motion \citep{2018arXiv180409366L}.
The median magnitude of these sources is $G=18.1$~mag (inter-quartile range: 16.6--19.1~mag) and the median proper-motion precision is $\sigma_\mu=0.4$~mas~yr$^{-1}$ (inter-quartile range: 0.2--0.8~mas~yr$^{-1}$). The {\it Gaia} catalog includes statistical uncertainties on astrometric properties calculated from the astrometric model. For AGIS models that do not converge, solutions are provided with relaxed criteria, with up to 20~mas of astrometric excess noise as defined by \citet{2012A&A...538A..78L}. We only accept sources with $\mathtt{astrometric\_excess\_noise} < 1$~mas as providing reliable kinematics.  We also omit likely non-member contaminants (Section~\ref{refined.sec}) and sources with  statistical uncertainties $>$3~km~s$^{-1}$ on tangential velocity (Section~\ref{kinematics.sec}). 
The final sample contains 6507 objects.

\begin{figure*}[t]
\centering
\includegraphics[width=0.45\textwidth]{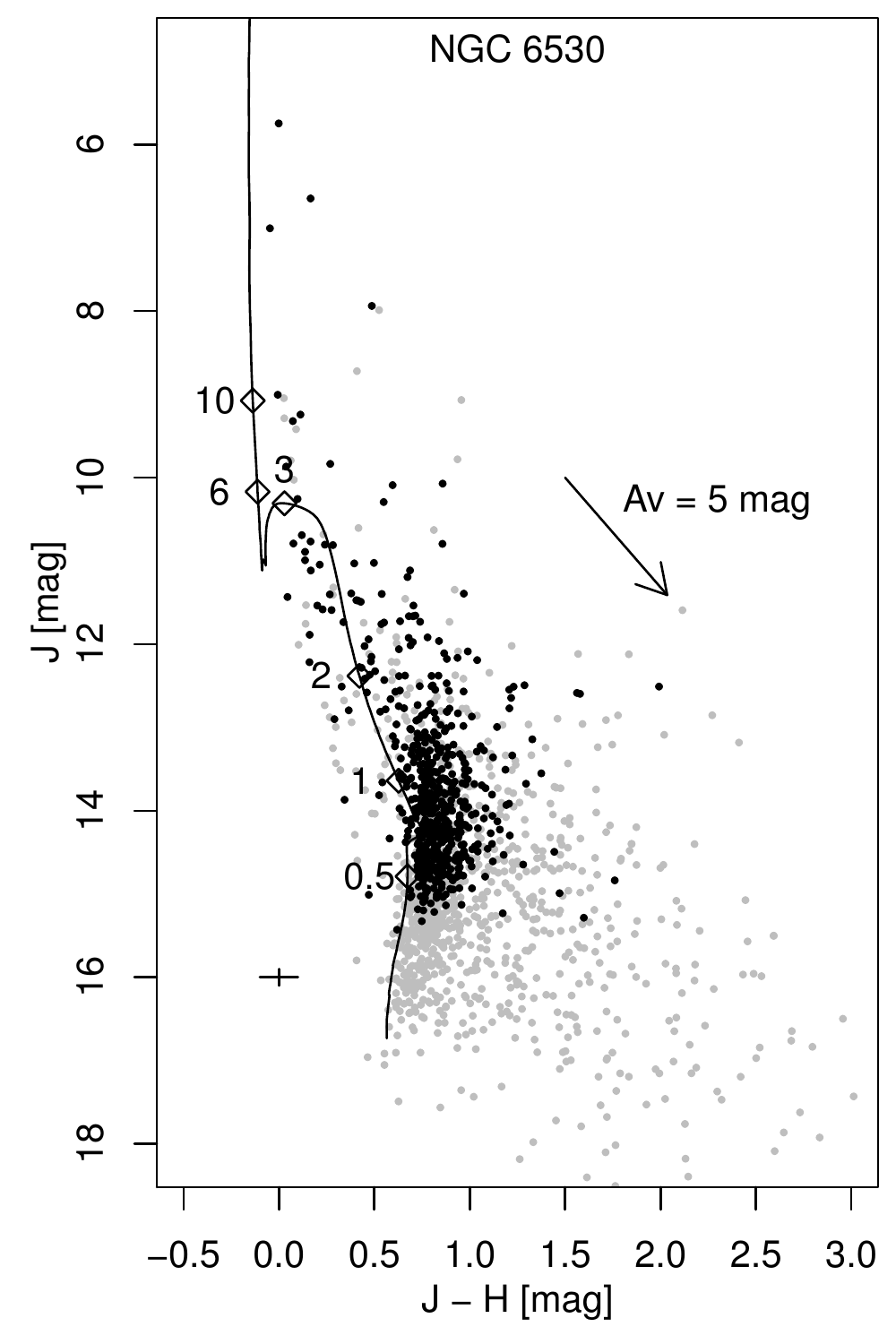} 
\includegraphics[width=0.45\textwidth]{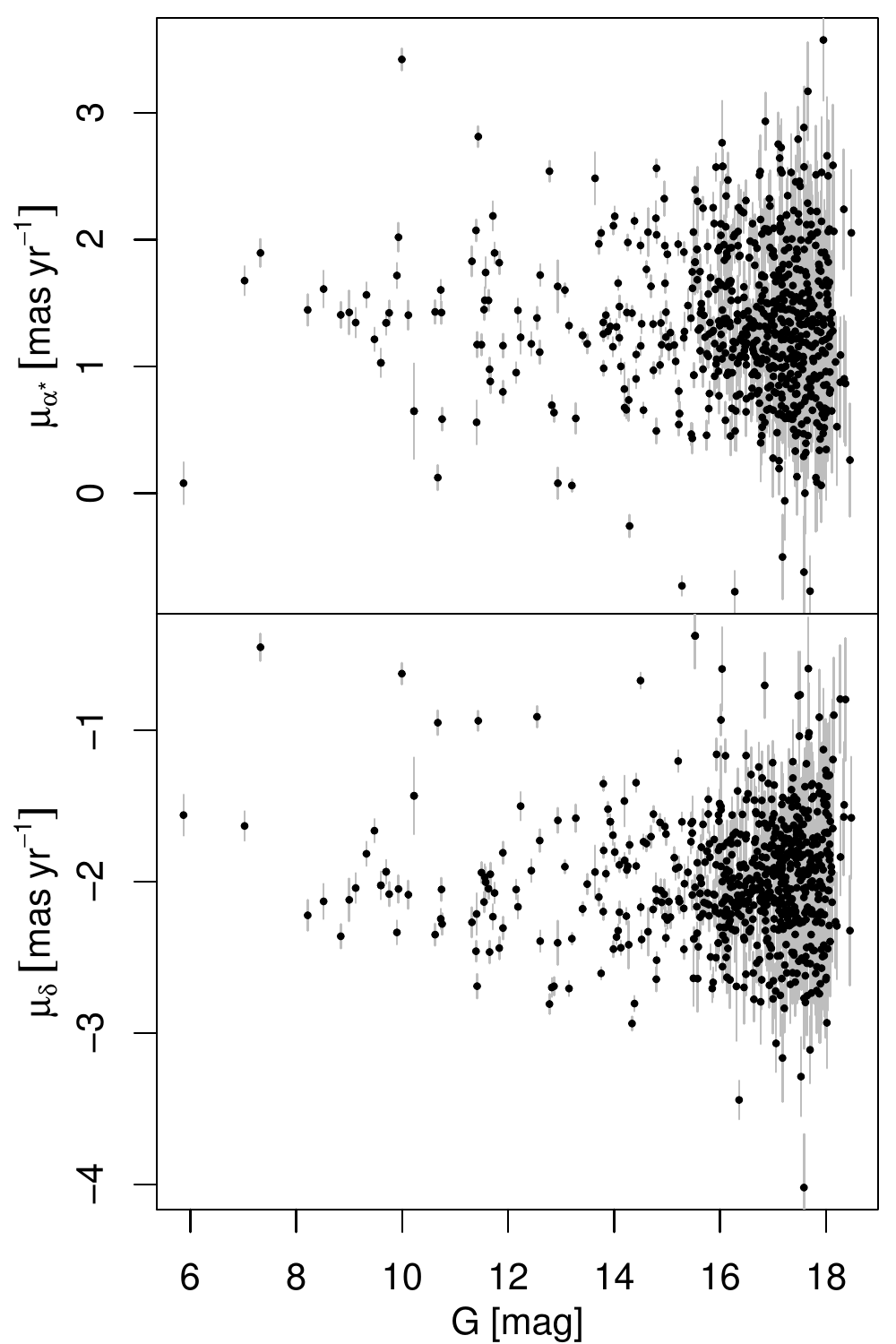} 
\caption{
Left: J vs.\ J-H color--magnitude diagram for sources in NGC~6530. {\it Gaia} sources used in the analysis are shown in black, while other MYStIX probable members are shown in gray. Only sources with uncertainties in $J$ and $J-H$ less than 0.1~mag are plotted. 
A reddening vector \citep{1985ApJ...288..618R}, a 2 Myr isochrone \citep{2012MNRAS.427..127B} with several masses indicated, and a cross showing maximum allowed 1$\sigma$ uncertainties are shown for guidance. Right: Proper motion measurements and uncertainties vs. $G$-band magnitude for YSOs in NGC~6530. Only sources included in the study are plotted. Plots for other stellar systems are included in an online figure set.
 \label{error.fig}}
\end{figure*}

In Figure~\ref{error.fig} the left panel shows a near-infrared color-magnitude diagram for both the full catalog of YSOs in NGC~6530 and the subsample used for the {\it Gaia} analysis. In this region, at a distance of $d=1.34$~kpc, the {\it Gaia} sources include stars down to $\sim$0.5~$M_\odot$, and tend to be the least absorbed sources. The right panel shows proper motions, and their uncertainties, as a function of {\it Gaia}'s $G$ magnitude for these sources. The mass range of the {\it Gaia} sample is different in different regions, depending on distance and extinction.
However, in nearly all the regions, stars down to 0.5--1~$M_\odot$ are included. Several nearby regions have {\it Gaia} data that probes to even lower mass stars, notably NGC~1333, IC~348, the ONC, and NGC~2264. Conversely, our sample for the distant region NGC~1893 contains only stars with masses $>$2~$M_\odot$. 

The presence of a visual binary or the acceleration of a source can cause the {\it Gaia} astrometric solution to be rejected \citep{2018arXiv180409366L}. Nevertheless, astrometric binaries remain a possible contributor to scatter in proper motion distributions. The velocity dispersion induced by binaries depends on the individual stellar masses, binary separations, eccentricity of orbits, and inclinations of the systems. 
Binary orbital motions are unlikely to have a preferential orientation, so they should not bias observed bulk shifts in velocity, but they can contribute a high-velocity tail to velocity distributions. 

The effect of binaries on velocity dispersions can be partially mitigated by filtering out sources with high astrometric excess noise. We use the ONC as a testbed to examine the link between binarity and astrometric excess noise. \citet{2018MNRAS.tmp.1126D} provides a list of ONC stars with and without companions at separations of 10--60~AU, based on HST imaging. For the visual binaries, 50\% have $\mathtt{astrometric\_excess\_noise}>1$~mas, while only 7\% of the non-visual binary stars in their sample exceed this threshold. This result supports our decision to only use sources below a $1$~mas threshold for measuring median properties of stellar kinematics.  We also note that the well-known O-star system, $\theta^1$~Ori~C, has $\mathtt{astrometric\_excess\_noise} > 1$~mas, and thus is not included in our sample. For applications that require  accurate estimates of measurement uncertainty, such as measuring velocity dispersions, we limit the sample to data with $\mathtt{astrometric\_excess\_noise} = 0$~mas.

\subsection{Subclusters}

Some of the systems we analyze as a unit in Sections~\ref{bulk.sec}--\ref{sd.sec} can be further decomposed into multiple subclusters. These include the clumpy distributions of stars that make up systems like NGC~6530 or M17 or smaller groups of stars surrounding a main cluster in systems like NGC~6611 and M20. In general, these subclusters contain too few stars to investigate their internal kinematics with DR2 data. Instead, we examine the relative velocities of different subclusters to test for signs of merger or dispersal (Section~\ref{subcluster_motions.sec}). 

Subclusters in the star-forming regions studied here were cataloged by \citet{2014ApJ...787..107K} for MYStIX, \citet{2018arXiv180405075G} for SFiNCs, and \citet{2017AJ....154..214K} for NGC~6231.  Subcluster identification in these papers was based on mixture models, a statistical cluster analysis method that is well adapted to cases where size and density of clusters can vary and the number of clusters is uncertain \citep{Mclachlan00,2017arXiv171111101K}. In some star-forming regions subclusters can be found outside the main cluster, while, in others, subclusters are clumps of stars that make up the main cluster. 

\startlongtable
\onecolumngrid
\begin{deluxetable*}{lccccccc}
\tablecaption{Cluster/Association Sample \label{clusters.tab}}
\tabletypesize{\small}\tablewidth{0pt}%\rotate
\tablehead{
\colhead{Region}  & \colhead{$\alpha_0$} &\colhead{$\delta_0$} & \colhead{$n_\mathrm{samp}$} & \colhead{$ \mu_{\alpha^\star,0}$} & \colhead{$ \mu_{\delta,0} $} & \colhead{$\varpi_0$} & \colhead{distance}\\
\colhead{} &  \colhead{J2000} &\colhead{J2000} & \colhead{stars} & \colhead{mas~yr$^{-1}$} & \colhead{mas~yr$^{-1}$} & \colhead{mas} & \colhead{pc}\\
\colhead{(1)} & \colhead{(2)} & \colhead{(3)} &\colhead{(4)} & \colhead{(5)} & \colhead{(6)} & \colhead{(7)} & \colhead{(8)}
}
\startdata
Berkeley 59  &   0  02 14.91 &+67 25  07.6  &  225  &  $-$1.61$\pm$0.10  &  $-$1.92$\pm$0.09  &   0.91$\pm$0.04  &  1100$^{+  50}_{-  50}$  \\ 
NGC 1333  &   3 29  07.96 &+31 20 39.3  &  47  &   7.24$\pm$0.28  &  $-$9.68$\pm$0.13  &   3.36$\pm$0.06  &   296$^{+   5}_{-   5}$  \\ 
IC 348  &   3 44 33.88 &+32 09 31.8  &  180  &   4.63$\pm$0.14  &  $-$6.45$\pm$0.13  &   3.09$\pm$0.05  &   324$^{+   5}_{-   5}$  \\ 
LkH$\alpha$~101  &   4 30 10.17 &+35 16  04.9  &  65  &   2.16$\pm$0.20  &  $-$5.18$\pm$0.42  &   1.77$\pm$0.05  &   564$^{+  15}_{-  14}$  \\ 
NGC 1893  &   5 22 53.74 &+33 26 54.5  &  88  &  $-$0.24$\pm$0.08  &  $-$1.40$\pm$0.08  &   0.26$\pm$0.04  &  3790$^{+ 700}_{- 510}$  \\ 
ONC  &   5 35 15.68 &$-$05 23 40.1  &  378  &   1.51$\pm$0.11  &   0.50$\pm$0.12  &   2.48$\pm$0.04  &   403$^{+   7}_{-   6}$  \\ 
Mon R2  &   6  07 47.58 &$-$06 22 42.6  &  97  &  $-$2.91$\pm$0.11  &   1.05$\pm$0.18  &   1.06$\pm$0.04  &   948$^{+  42}_{-  38}$  \\ 
Rosette  &   6 32 26.76 &+04 47 37.1  &  468  &  $-$1.63$\pm$0.07  &   0.15$\pm$0.07  &   0.64$\pm$0.04  &  1560$^{+ 110}_{-  90}$  \\ 
--- NGC 2244  &   6 31 55.77 &+04 55  7.8  &  272  &  $-$1.70$\pm$0.07  &   0.20$\pm$0.07  &   0.65$\pm$0.04  &  1550$^{+ 100}_{-  90}$  \\ 
NGC 2264  &   6 40 57.93 &+09 40 49.0  &  519  &  $-$1.76$\pm$0.08  &  $-$3.72$\pm$0.07  &   1.35$\pm$0.04  &   738$^{+  23}_{-  21}$  \\ 
--- S Mon  &   6 40 49.83 &+09 51  03.3  &  242  &  $-$1.62$\pm$0.08  &  $-$3.71$\pm$0.07  &   1.36$\pm$0.04  &   738$^{+  23}_{-  21}$  \\ 
--- NGC 2264 IRS 2  &   6 41  00.64 &+09 35 55.7  &  151  &  $-$2.29$\pm$0.14  &  $-$3.61$\pm$0.08  &   1.34$\pm$0.04  &   748$^{+  24}_{-  23}$  \\ 
--- NGC 2264 IRS 1  &   6 41  07.03 &+09 28  09.3  &  126  &  $-$2.05$\pm$0.18  &  $-$3.90$\pm$0.09  &   1.36$\pm$0.04  &   736$^{+  23}_{-  22}$  \\ 
NGC 2362  &   7 18 42.90 &$-$24 57 44.0  &  246  &  $-$2.83$\pm$0.07  &   2.95$\pm$0.08  &   0.75$\pm$0.04  &  1332$^{+  75}_{-  68}$  \\ 
Carina  &  10 45  02.23 &$-$59 41 59.8  &  1285  &  $-$6.55$\pm$0.07  &   2.17$\pm$0.07  &   0.38$\pm$0.04  &  2620$^{+ 310}_{- 250}$  \\ 
--- Tr 14  &  10 44  01.68 &$-$59 32 48.1  &  401  &  $-$6.54$\pm$0.07  &   2.06$\pm$0.07  &   0.38$\pm$0.04  &  2640$^{+ 310}_{- 250}$  \\ 
--- Tr 15  &  10 44 42.08 &$-$59 22 40.6  &  194  &  $-$6.25$\pm$0.08  &   2.06$\pm$0.08  &   0.38$\pm$0.04  &  2630$^{+ 310}_{- 250}$  \\ 
--- Tr 16  &  10 44 53.49 &$-$59 43 10.1  &  268  &  $-$6.90$\pm$0.07  &   2.63$\pm$0.08  &   0.38$\pm$0.04  &  2610$^{+ 310}_{- 250}$  \\ 
NGC 6231  &  16 54 15.90 &$-$41 49 59.0  &  615  &  $-$0.55$\pm$0.07  &  $-$2.17$\pm$0.07  &   0.59$\pm$0.04  &  1710$^{+ 13}_{- 110}$  \\ 
RCW 120  &  17 12 23.88 &$-$38 29 15.7  &  29  &  $-$0.82$\pm$0.12  &  $-$2.38$\pm$0.19  &   0.59$\pm$0.04  &  1680$^{+ 130}_{- 110}$  \\ 
NGC 6357  &  17 25 18.73 &$-$34 17 24.8  &  178  &  $-$0.90$\pm$0.08  &  $-$2.29$\pm$0.10  &   0.56$\pm$0.04  &  1770$^{+ 140}_{- 120}$  \\ 
--- Pismis 24  &  17 24 44.06 &$-$34 13 20.3  &  75  &  $-$0.83$\pm$0.08  &  $-$2.08$\pm$0.10  &   0.56$\pm$0.04  &  1790$^{+ 150}_{- 130}$  \\ 
--- G353.1+0.6  &  17 25 34.09 &$-$34 24 49.5  &  53  &  $-$0.98$\pm$0.09  &  $-$2.34$\pm$0.11  &   0.56$\pm$0.04  &  1780$^{+ 150}_{- 130}$  \\ 
--- G353.2+0.7  &  17 25 59.77 &$-$34 16  04.4  &  47  &  $-$1.09$\pm$0.09  &  $-$2.66$\pm$0.12  &   0.56$\pm$0.04  &  1780$^{+ 150}_{- 130}$  \\ 
M20  &  18  02 23.10 &$-$23 01 50.0  &  116  &   0.41$\pm$0.12  &  $-$1.69$\pm$0.09  &   0.79$\pm$0.05  &  1264$^{+  76}_{-  68}$  \\ 
NGC 6530  &  18  04 14.84 &$-$24 21 45.9  &  669  &   1.32$\pm$0.08  &  $-$2.07$\pm$0.08  &   0.75$\pm$0.04  &  1336$^{+  76}_{-  68}$  \\ 
NGC 6611  &  18 18 42.20 &$-$13 47  03.0  &  356  &   0.21$\pm$0.08  &  $-$1.56$\pm$0.08  &   0.57$\pm$0.04  &  1740$^{+ 130}_{- 120}$  \\ 
M17  &  18 20 28.50 &$-$16 10 58.0  &  82  &  $-$0.04$\pm$0.17  &  $-$1.40$\pm$0.13  &   0.60$\pm$0.04  &  1680$^{+ 130}_{- 110}$  \\ 
IC 5146  &  21 53 31.47 &+47 15 54.5  &  115  &  $-$2.87$\pm$0.12  &  $-$2.52$\pm$0.18  &   1.28$\pm$0.04  &   783$^{+  26}_{-  25}$  \\ 
NGC 7160  &  21 53 46.36 &+62 35  07.6  &  71  &  $-$3.53$\pm$0.10  &  $-$1.43$\pm$0.13  &   1.04$\pm$0.04  &   961$^{+  41}_{-  38}$  \\ 
Cep OB3b  &  22 55 31.19 &+62 38 54.1  &  678  &  $-$0.75$\pm$0.09  &  $-$2.31$\pm$0.08  &   1.16$\pm$0.04  &   863$^{+  31}_{-  29}$  \\ 
--- Cep B  &  22 56 40.29 &+62 42  06.2  &  482  &  $-$0.61$\pm$0.10  &  $-$2.28$\pm$0.09  &   1.15$\pm$0.04  &   868$^{+  32}_{-  30}$  \\ 
--- V454 Cep  &  22 53 47.06 &+62 35 46.6  &  196  &  $-$1.26$\pm$0.11  &  $-$2.61$\pm$0.11  &   1.18$\pm$0.04  &   847$^{+  31}_{-  29}$  \\  
\enddata
\tablecomments{Column~1: Region name. Systems that are part of larger star-forming complexes are indented. 
Columns~2--3: Coordinates of the system center.
Column~4: The sizes of the sample of stars used in the analysis. 
Column~5--6: Proper motion of the system center.
Column~7: Parallax of the system center. }
\end{deluxetable*}

%\clearpage\clearpage

\section{Basic Properties of Stellar Systems}\label{properties.sec}

\subsection{Parallaxes}

Distances to stellar systems can be estimated with the assumption that their members span a small range of distances \citep[cf.][their Section~3.5]{2018arXiv180410121B}. 
This assumption is approximately true for our sample --- the nearest region NGC~1333 has a diameter only $\sim$0.4\% of the distance to the system. {\it Gaia} parallax measurements can then be considered to be random variables drawn from a distribution where the mean is $\varpi_0$, the parallax of the system center, and the standard deviation is the measurement uncertainties given in the {\it Gaia} catalog. Using multiple stars to estimate the parallax of the center of a system, $\varpi_0$, will yield a measurement with smaller uncertainties than for the individual stars. However, gains in precision from pooling stars are limited due to correlated uncertainties of up to $\sim$0.04~mas that are noticeable on spatial scales smaller than 1~degree \citep{2016arXiv160904303L}.

We estimate system parallax using the weighted median of individual stellar parallax measurements. This method is robust against contaminants while taking into account the measurement uncertainties. For this analysis, we use the conventional $1/\mathrm{error}^2$ weights and the {\it weighted.median} function from the 
CRAN package {\it spatstat} \citep{baddeley2015spatial} within the {\it R} statistical software environment \citep{RCoreTeam2018}. Uncertainties on the weighted median parallax are calculated using bootstrap analysis, with random sampling with replacement from the set of measurements (with added random errors) and weights. Finally, we add in quadrature the systematic uncertainty of 0.04~mas described by \citet{2016arXiv160904303L}, which provides a noise floor.\footnote{Several papers have proposed correction factors for systematic errors in {\it Gaia} astrometry \citep[e.g.,][]{2018arXiv180503526S,2018arXiv180508742M,2018arXiv180504649K}, all of which are consistent with the systematic uncertainties reported by \citet{2016arXiv160904303L}. The parallax and proper motion values reported in Table~\ref{clusters.tab} are based on {\it Gaia} astronomy with no correction applied.} 

Table~\ref{clusters.tab} provides the new parallax estimates that we use in this study. Parallaxes are calculated independently for each stellar system in a star-forming complex. In all cases, the uncertainties on median parallaxes are dominated by the systematic uncertainties in {\it Gaia} astrometry and not by statistical dispersion. 

Several highlights from revised distances in Table~\ref{clusters.tab} are mentioned here. 
\begin{itemize}
\item Our analysis of the {\it Gaia} data places the ONC at a distance of 403$^{+7}_{-6}$~pc, which is slightly closer than the estimate of 414$\pm$7~pc from \citet{2007A&A...474..515M}, but farther than 388$\pm$5~pc found by \citet{2017ApJ...834..142K}, both based on Very Long Baseline Array (VLBA) data. In Appendix~\ref{orion.sec} we compare our distance estimate to 386$\pm$8~pc from \citet{2018arXiv180504649K} and discuss effects of the three-dimensional structure of Orion~A. 

\item Our distance estimate for NGC~1333 of 296$\pm$5~pc is $\sim$25\% larger than the distance from \citet{2008PASJ...60...37H} of 235$\pm$18~pc. A larger distance to NGC~1333 will move stellar age estimates downwards, which would resolve some anomalies in age estimates noted by \citet{2016ApJ...827...52L} and \citet{2018MNRAS.477.5191R}. 

\item The {\it Gaia} distance estimate to M20 is 1260$\pm$70~pc -- a factor of $\sim$2 nearer than the distance estimate of 2700~pc from \citet{2011A&A...527A.141C}. The revised distance places M20 at a similar distance as NGC~6530, from which it is separated by only 1.4$^{\circ}$ on the sky. 

\item In star-forming regions made up of multiple stellar systems, the nearly identical parallax measurements for the different components provides confidence in the accuracy of the measurements and reassurance that these systems are physically associated. For example, we have confirmed that Tr~14, 15, and 16 are all at a distance of $\sim$2600~pc, putting to rest claims in older literature that Tr~15 is in the background \citep[cf.][]{1995RMxAC...2...51W}.
\end{itemize}

\subsection{Proper Motions}

The proper motions of the system centers, $\mu_{\alpha^\star,0}$ and $\mu_{\delta,0}$, are estimated using the same weighted-median strategy as above. For systematic uncertainties due to correlated errors, we use $\pm$0.07~mas~yr$^{-1}$ \citep{2016arXiv160904303L}. These proper motions are reported in Table~\ref{clusters.tab}. Given that the systems each subtend $<$1~degree on the sky, we ignore spherical geometry effects in determination of the median -- the small corrections calculated below for individual stars do not affect the proper motion of the system center. 

The effects of correlated errors in {\it Gaia} DR2 on relative velocities of stars within systems was investigated in simulations by \citet{2018arXiv180602580B}. They found the effects were $<$0.5~$\mu$as~yr$^{-1}$ on length scales of 10~arcmin, which suggests that these effects will be negligible in our study of internal kinematics. 

\subsection{Refined Sample}\label{refined.sec}

Parallax and proper motion measurements allow us to identify likely field-star contaminants in the MYStIX/SFiNCS/NGC6231 catalogs. Stars are removed from the sample if their parallax measurements are inconsistent with the median parallax by more than 3 standard deviations, taking into account uncertainty on the measurement and on the median. Outliers in proper motion by more than $>$5 standard deviations (using maximum absolute deviation as a robust estimator of standard deviation) are also filtered out as likely contaminants.

Estimation of median parallax and proper motion and refinement of membership is performed iteratively until convergence. For example, in NGC~6231 with 1760 {\it Gaia} counterparts, 121 (7\% of the total) are removed as likely contaminants, leaving 1639 {\it bona fide} system members in {\it Gaia}. (Only 615 of these have sufficient astrometric precision for inclusion in the analysis.) Overall, contamination rates were about 13\%, with contamination rates for individual systems mostly falling into the range 7--15\%. Several systems with much higher contamination among the {\it Gaia} sources include M17 (23\%), NGC~2362 (25\%), NGC~7160 (28\%), and RCW~120 (38\%). Table~\ref{clusters.tab} gives the number of stars in the final, refined sample. 

\subsection{Stellar Kinematics}\label{kinematics.sec}

We are interested in obtaining two components of stellar velocities (from a three-dimensional Cartesian coordinate system) based on the proper motions $\mu_{\alpha^\star}$ and $\mu_\delta$. Since the systems in our sample are sufficiently distant, compact, and slow-moving, the relative proper motions of their stars are dominated by their physical velocities, so only small correction factors, calculated below, are necessary. 

Observed proper motions of stars in stellar systems can be affected by perspective and the motion of the center of the system \citep{2009A&A...497..209V}. In particular, radial motion of a system can produce an effect known as ``perspective expansion'' which is seen in {\it Gaia} measurements of globular clusters \citep{2018arXiv180409381G}, many of which have RVs of hundreds of kilometers per second.  A first-order approximation to the perspective expansion can be obtained from Equation~13 in \citet{2009A&A...497..209V}. Here, $\alpha_0$, $\delta_0$, $\varpi_0$, $\mu_{\alpha^\star,0}$ and $\mu_{\delta_0}$ are the astrometric parameters of the system center, and $\Delta\alpha_i$ and $\Delta\delta_i$ are the difference in right ascension and declination between an individual star and the system center. The equations for the additional shift in proper motion of a star are
\begin{align}\label{perspective_expansion.eqn}
\Delta\mu_{\alpha^\star,\mathrm{per}} &\approx 
\Delta\alpha_i\left(
\mu_{\delta,0}\sin\delta_0 - \frac{v_r \varpi_0}{\kappa}\cos \delta_0
\right)\\
\label{perspective_expansion2.eqn}
\Delta\mu_{\delta,\mathrm{per}} &\approx 
-\Delta\alpha_i\mu_{\alpha^\star,0}\sin\delta_0 - \Delta\delta_i \frac{v_r\varpi_0}{\kappa},
\end{align}
where $\kappa = 4.74$ is the conversion factor from mas~yr$^{-1}$ to km~s$^{-1}$ at a distance of 1~kpc. 
The first term in each equation relates to motion in a spherical coordinate system while the second term relates to the apparent expansion/contraction of an object as it gets farther/nearer. 

These equations, calculated using RVs in Appendix~\ref{radial_velocity.sec}, contribute small shifts to apparent stellar proper motions.  Following the strategy recommended by \citet{1997MNRAS.285..479B}, these contributions can be subtracted from the observed proper motions (relative to the system center) to isolate the effects of internal kinematics. 
The region with the largest corrections is the Orion~A cloud with corrections on the order of $\sim$0.07~mas~yr$^{-1}$. Most other regions have corrections with magnitudes $<$0.02~mas~yr$^{-1}$.
 Similarly, for regions spanning $<$1~degree, projection effects of spherical geometry are small ($<$0.02\%), and are not expected to affect science results. 

The correction factors computed above allow us to approximate a two-dimensional velocity of each star $\mathbf{v}=(v_x,v_y)$, relative to the rest frame of the center of the system. First, we calculate the components of the corrected velocities parallel to lines of constant right ascension and declination,
\begin{align}
v_\alpha &\approx -\kappa\left(
\frac{
\Delta\mu_{\alpha^\star,\mathrm{obs}} - \Delta\mu_{\alpha^\star,\mathrm{per}}
}{\varpi_0}
\right)\\
v_\delta &\approx \kappa\left(
\frac{
\Delta\mu_{\delta,\mathrm{obs}} - \Delta\mu_{\delta,\mathrm{per}}
}{\varpi_0}
\right).
\end{align}  
Then, these velocities are transformed to velocities in a Cartesian coordinate system, $v_x$ and $v_y$, using the orthographic projection \citep{2018arXiv180409381G}. Uncertainties on $\mathbf{v}$ can be described by a covariance matrix $\Sigma_{\mathrm{err},i}$. This is obtained from the covariance matrix for the astrometric solution, reconstructed from the DR2 catalog using Equation~B.3 in \citet{2018arXiv180409366L}, multiplied by a correction factor of $1.1^2$  \citep[Section~5.2 in][]{2018arXiv180409366L}, and scaled by $(\kappa/\varpi)^2$ to convert angular motions into velocities. This covariance matrix does not include the systematic effects of uncertainties in system parallax or RV.   

Velocities can also be expressed in different coordinate systems. For example, the outward and azimuthal components of the projected velocity with respect to the center of the system are
\begin{align}
v_\mathrm{out}&=\mathbf{v}\cdot\mathbf{\hat{r}}\\
v_\mathrm{az}&=\mathbf{v}\cdot\bm{\hat{\varphi}},
\end{align}
where $\mathbf{\hat{r}}$ is a unit vector pointing away from the system center and $\bm{\hat{\varphi}}$ is a unit vector pointing in the azimuthal direction relative to the system center. Uncertainty on a velocity component in the $\mathbf{\hat{u}}$ direction can be obtained from the covariance matrix 
\begin{equation}\label{projection.eqn}
\sigma_{\mathrm{err},i} =\left( \mathbf{\hat{u}}^\intercal\,\Sigma_{\mathrm{err},i}\,\mathbf{\hat{u}}\right)^{1/2}.
\end{equation}

\subsection{Visualizations of Kinematics}

Several approaches to visualization of the observed 4-dimensional kinematic structure of stellar systems are shown in Figures~\ref{arrow.fig}, \ref{arrowhead.fig}, and \ref{v_vs_pos.fig}. The ONC and NGC~6530 are shown as examples, but plots for the other 26 stellar systems are available online as figure sets. In each case, we depict velocities in the rest frames of their system centers.

Kinematics of stellar systems can be displayed using arrows to show velocity vectors for each star (Figure~\ref{arrow.fig}). In regions with large numbers of stars, these produce crowded plots that are difficult to interpret. The figure highlights sources with the highest-quality proper-motion measurements. 

Figure~\ref{arrowhead.fig} shows the direction of motion (no amplitude) for individual stars. The color of the arrowhead is also determined by the direction of motion (as indicated by the compass wheel) for clearer visualization. In an area of the diagram where arrowheads are mostly one color, the stars are mostly moving in one direction. The color saturation of the symbol is related to the statistical weighting of the data point. Only points with uncertainties $<$3~km~s$^{-1}$ are shown, with the darkest points representing points with uncertainties $<$0.5~km~s$^{-1}$. The X's mark the system centers used for measuring outward velocities $v_\mathrm{out}$. 

Figure~\ref{v_vs_pos.fig} shows the relationship between star positions and velocities. For each system, the four scatter plots show velocities perpendicular and parallel to each coordinate axis.
A simple radial contraction or expansion will produce a gradient in the $v_x - R.A.$ and $v_y - Dec$ diagrams (upper left and lower right) but will not affect the other diagrams (upper right and lower left).
The different patterns of stellar motion within  different systems are analyzed in the next section. 

\section{Bulk Stellar Motions}\label{bulk.sec}

\subsection{Contraction vs Expansion Velocities}

The expansion or contraction of a stellar system would produce a bulk outward or inward motion of its stars. In studies of stellar systems in star-forming regions, positive correlations between cluster (or subcluster) size and age suggest that these systems expand with velocities of $\sim$0.25--2~km~s$^{-1}$ \citep{2009A&A...498L..37P,2015ApJ...812..131K,2018MNRAS.477..298G}, where the slope of the relation is interpreted as an expansion velocity. Expansion velocities in this range would be detectable with {\it Gaia} in the star-forming regions in our sample.

\begin{figure*}[t]
\centering
\includegraphics[width=0.45\textwidth]{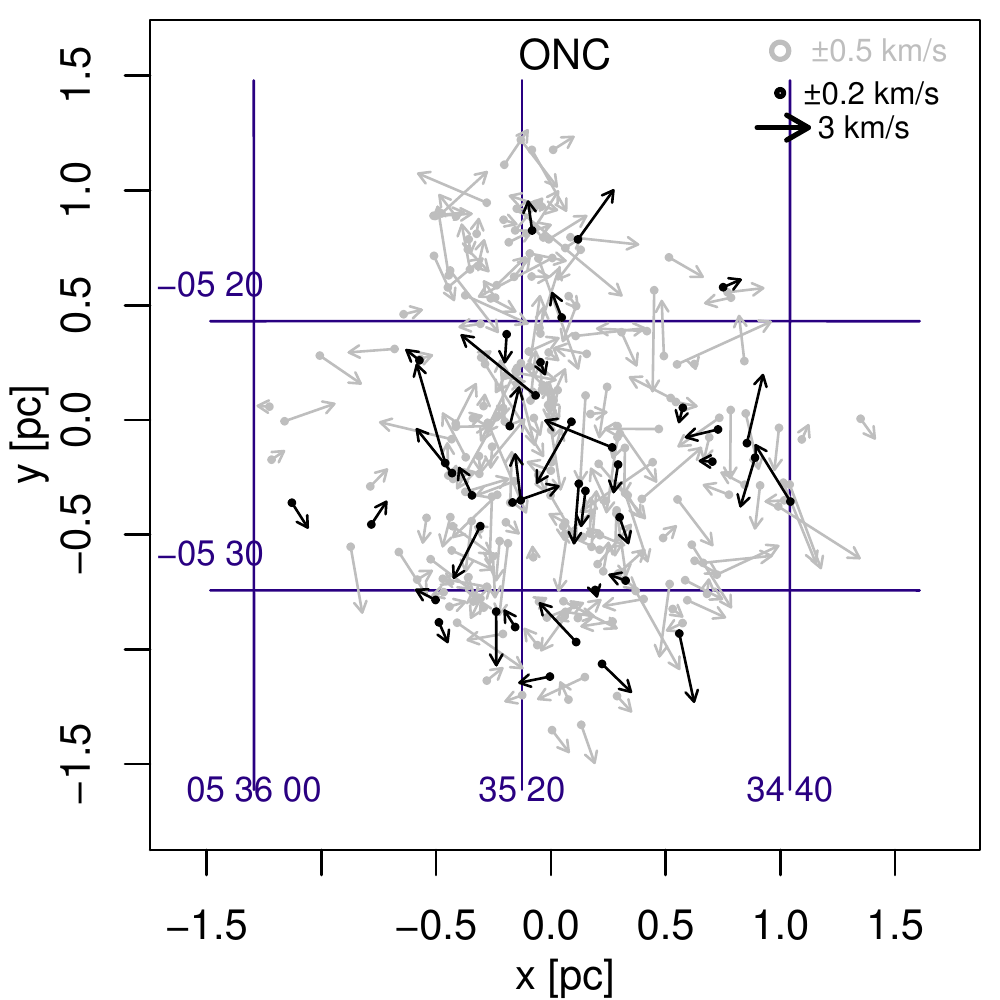} 
\includegraphics[width=0.45\textwidth]{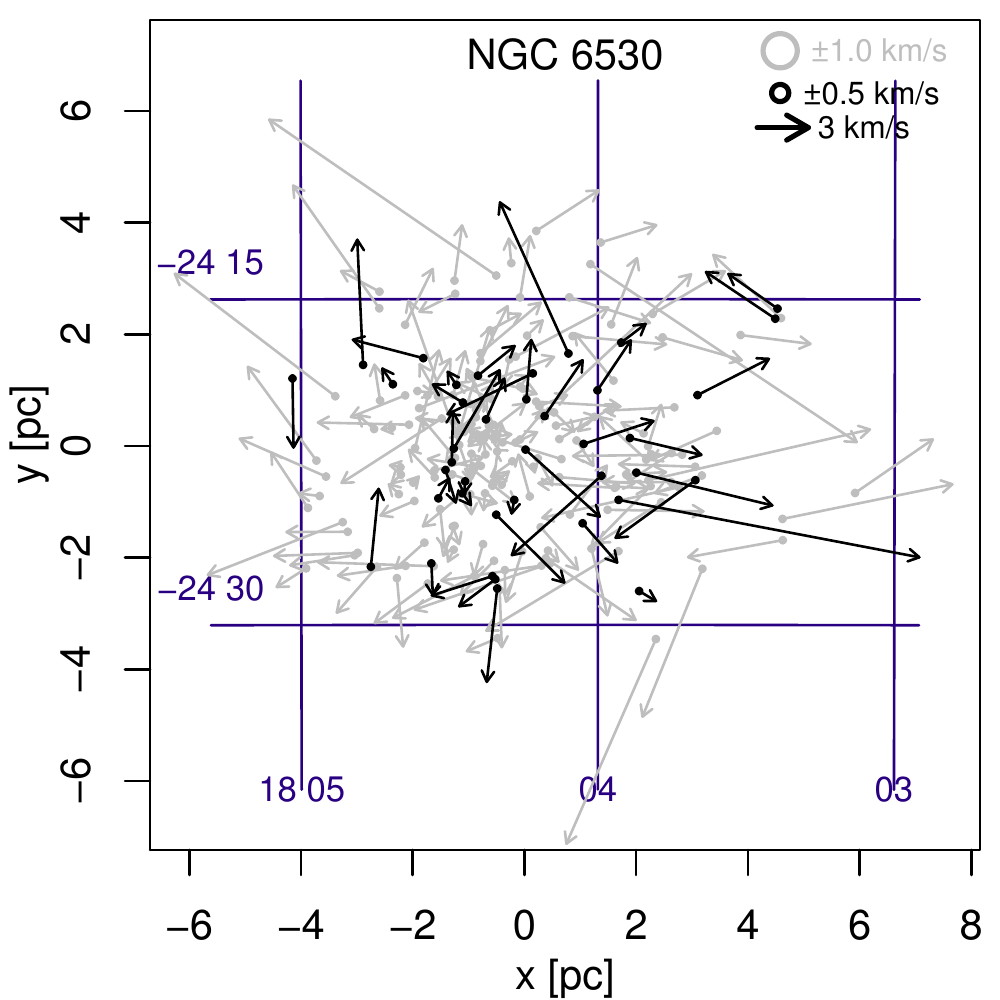} 
\caption{Maps of velocity vectors (magnitude and direction) for the best quality {\it Gaia} data in the ONC (left) and NGC~6530 (right). The quality of the velocity measurements are indicated by the arrow color (black for most precise and gray for less precise), with maximum allowed uncertainties for each group shown by the circles in the legend. Plots for other stellar systems are included in an online figure set. 
 \label{arrow.fig}}
\end{figure*}

\begin{figure*}[t]
\centering
\includegraphics[width=0.45\textwidth]{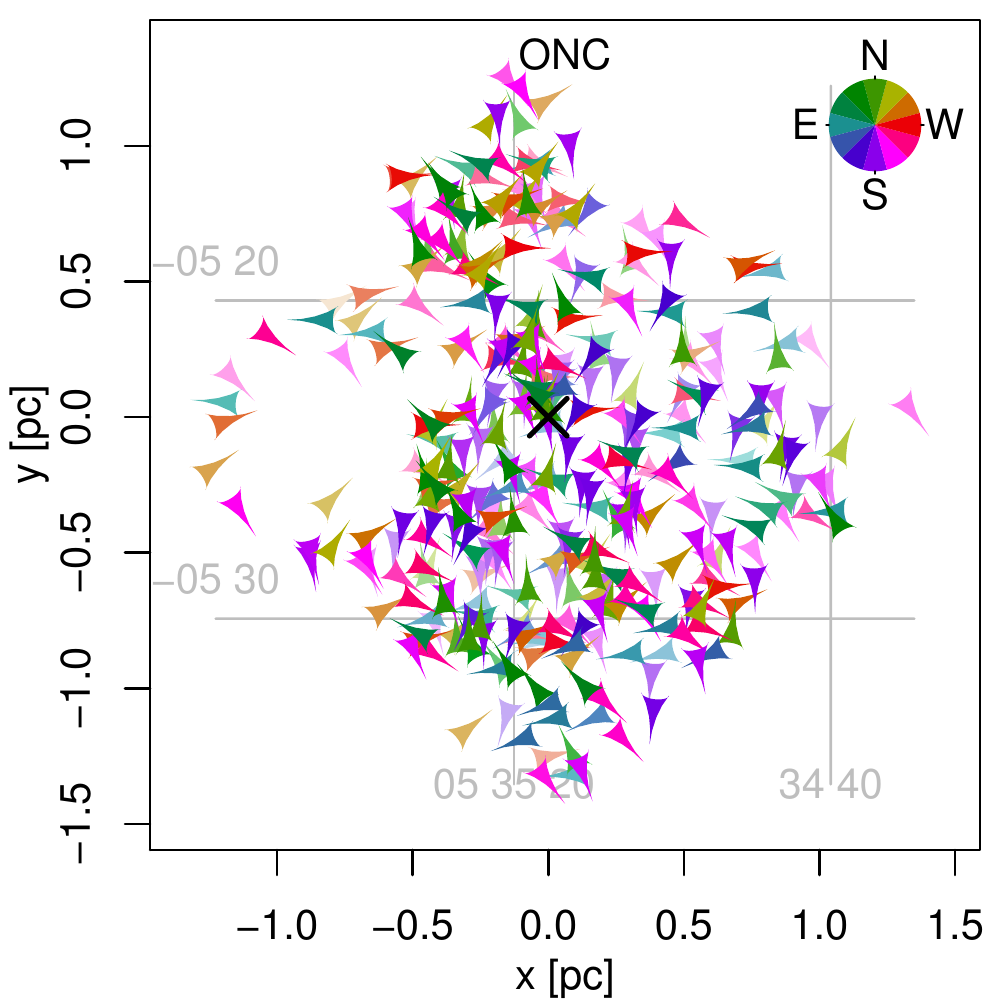} 
\includegraphics[width=0.45\textwidth]{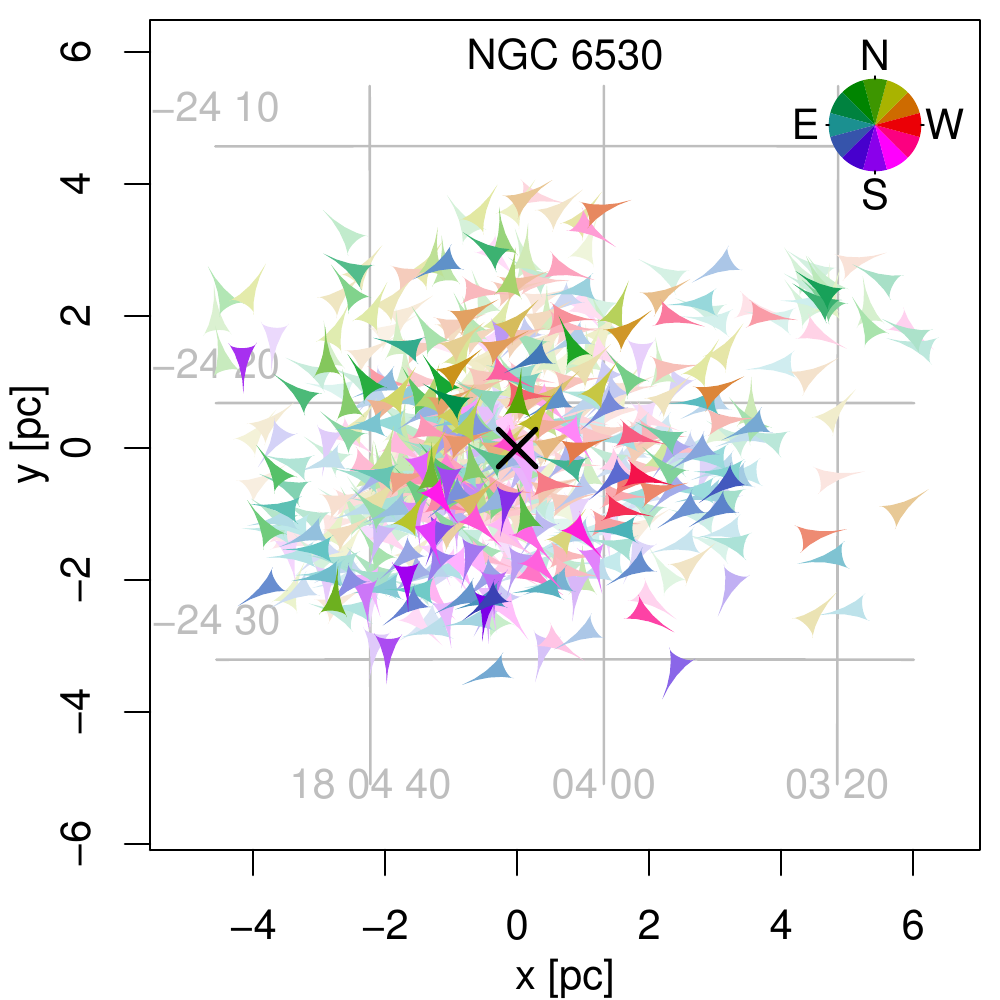} 
\caption{Direction of motion of individual stars in the rest frame of the system. Diagrams are shown for the ONC (left) and NGC~6530 (right). The orientations of the arrows and their hues indicate their direction, while saturation indicates weighting based on statistical uncertainty. In the ONC, stars with different velocities are mixed together, while, in NGC~6530, many stars have directions of motions away from the system center (as indicated by the outward pointing arrows and color segregation by azimuth). Plots for other stellar systems are included in an online figure set. 
\label{arrowhead.fig}}
\end{figure*}

\begin{figure*}[t]
\centering
\includegraphics[width=0.45\textwidth]{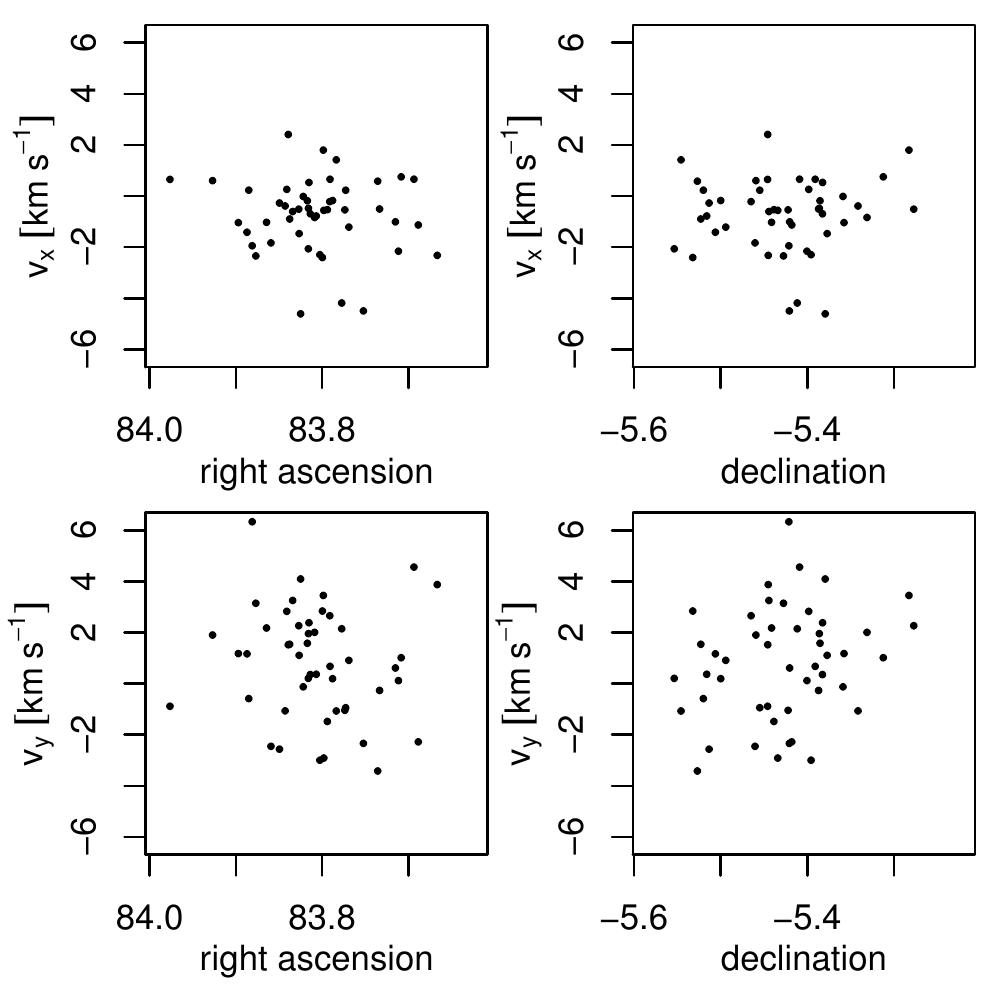} 
\includegraphics[width=0.45\textwidth]{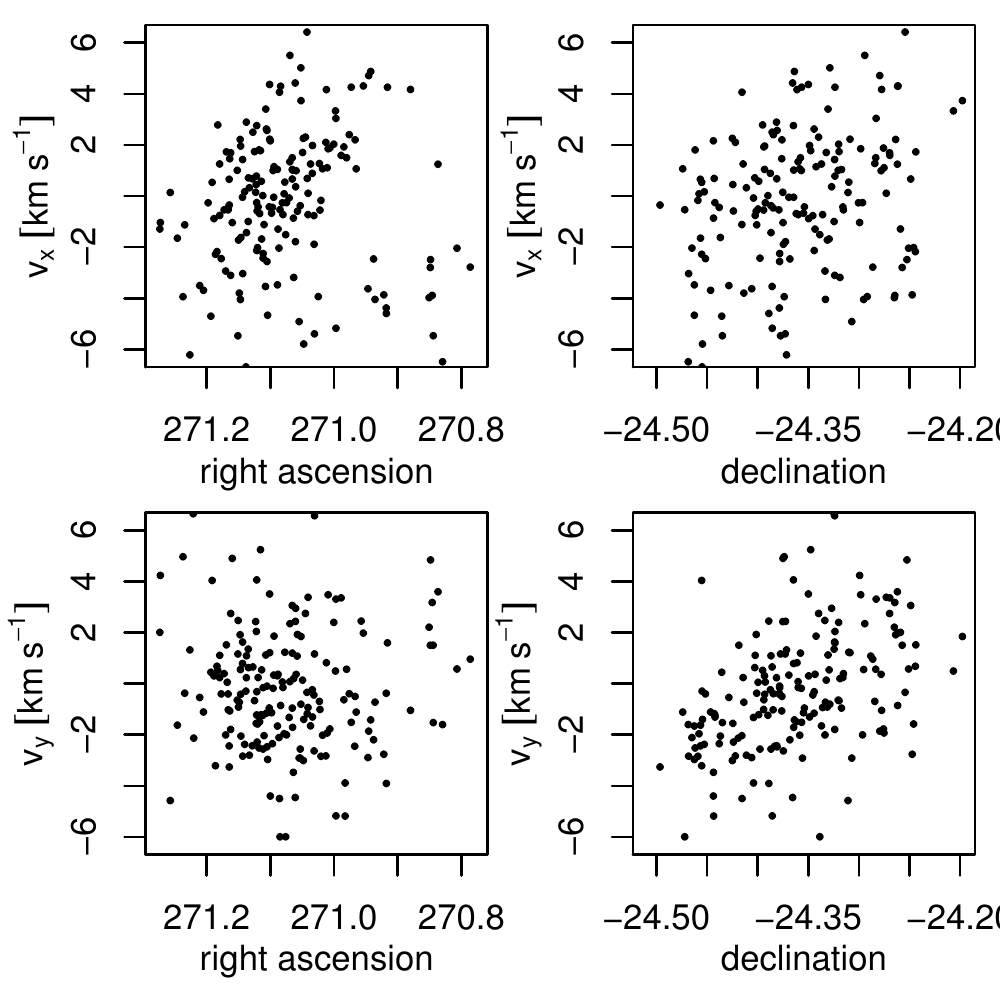} 
\caption{Dispersions of stars in position and velocity for the ONC (left) and NGC~6530 (right). For each system, the four panels show all combinations of right ascension, declination, $v_x$, and $v_y$. (Recall that $v_x$ is motion east to west and $v_y$ is motion south to north). All panels show the same velocity range to facilitate comparison of velocity dispersions. Expansion may show up as a correlation between $v_x$ and right ascension or between $v_y$ and declination. Only stars with no astrometric excess noise are included to reduce the presence of velocity outliers. Plots for other stellar systems are included in an online figure set.
 \label{v_vs_pos.fig}}
\end{figure*}

Velocities that are preferentially oriented outward can be seen in some stellar systems, and not in others. Expansion would show up as arrows pointing radially outward in Figure~\ref{arrow.fig}, coherent patterns in arrow direction and hue in Figure~\ref{arrowhead.fig}, and correlations in positions and velocity in Figure~\ref{v_vs_pos.fig}. 

NGC~6530 exhibits these characteristics. Stellar velocity vectors are primarily oriented away from the center, making the arrows in Figures~\ref{arrow.fig} point preferentially outward and showing up as a gradient in the color of arrowheads in Figure~\ref{arrowhead.fig}. There are also correlations between $v_x$ and $\alpha$ and $v_y$ and $\delta$ (Figure~\ref{v_vs_pos.fig}) that are consistent with the expectations for an expanding system. 
In contrast, for the ONC, these figures show that stars with differently oriented velocity vectors are more mixed up and there is no visually obvious correlation between position and velocity of stars.

We quantify system expansion (or contraction) using the weighted median value of $v_\mathrm{out}$ for each system using all members of the ``refined sample'' of stars. Uncertainties on the median are calculated by bootstrap resampling, as earlier. The expansion velocities are provided in Table~\ref{expansion.tab}. 

\begin{figure*}[t]
\centering
\includegraphics[width=0.4\textwidth]{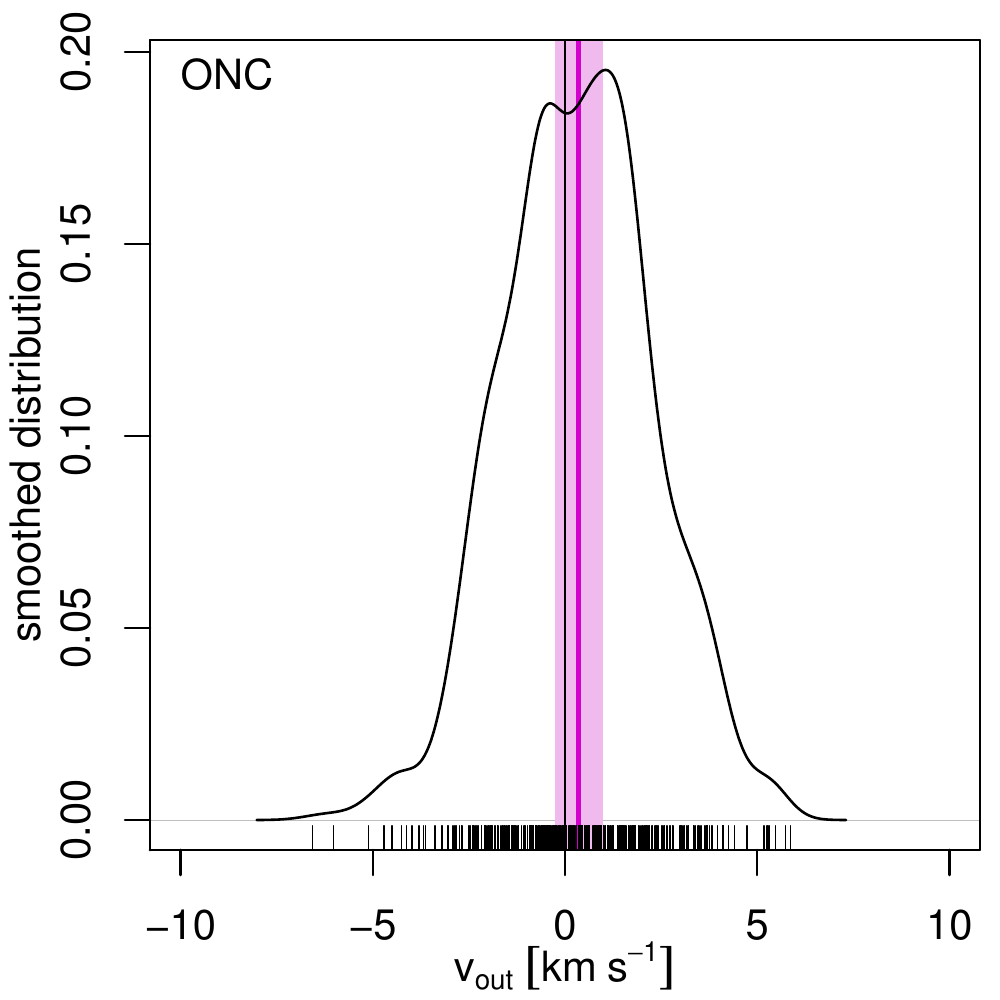} 
\includegraphics[width=0.4\textwidth]{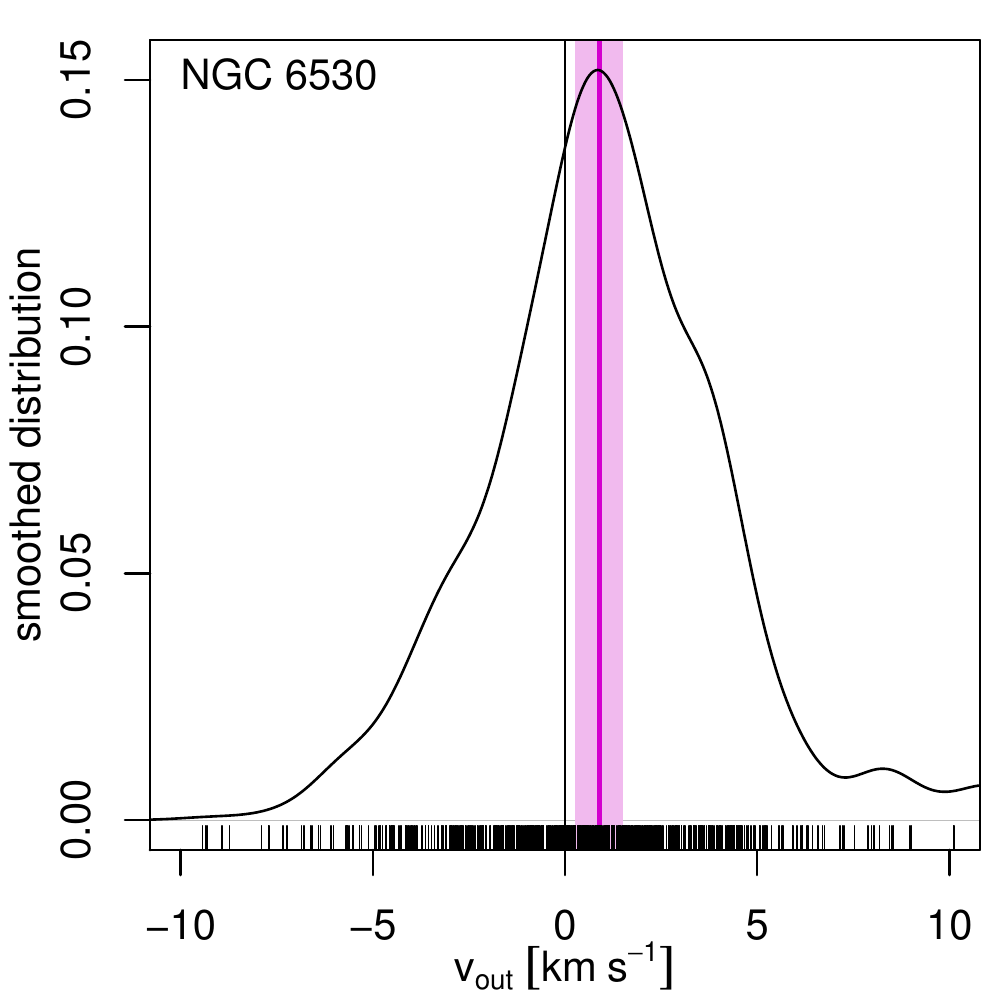} 
\caption{Distributions of $v_\mathrm{out}$ shown as KDE plots for stars in the ONC (left) and  NGC~6530 (right). The black lines indicates $v_\mathrm{out}=0$~km~s$^{-1}$, the solid magenta line indicates the median of the distribution, and the shaded magenta region indicates the 3$\sigma$ uncertainty on the median. Medians greater than 0 indicate bulk expansion while medians less than zero indicate bulk contraction. Plots for other stellar systems are included in an online figure set. 
 \label{kde.fig}}
\end{figure*}

The distribution of $v_\mathrm{out}$ values that go into calculating the median are illustrated in Figure~\ref{kde.fig}, with weighted kernel-density estimates (KDE) of $v_\mathrm{out}$ obtained with the function {\it density} in {\it R}. Shifts of the distribution toward positive values indicate expansion while shifts to negative values indicate contraction. 

\begin{figure*}[t]
\centering
\includegraphics[width=0.45\textwidth]{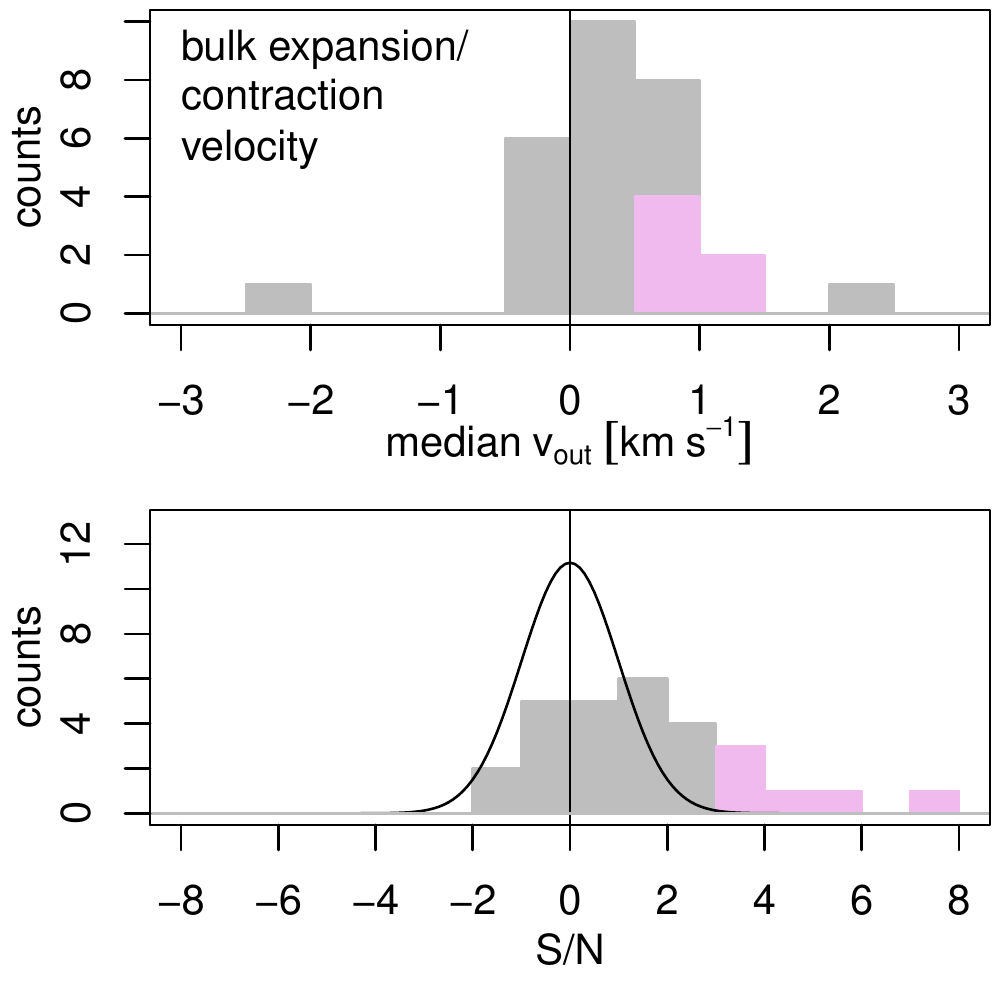} 
\caption{Histograms of expansion velocities (top) and signal-to-noise of expansion velocities (bottom) for the stellar systems in our sample. Expansion significant at the $>$3$\sigma$ level is marked in magenta. The Gaussian curve in the bottom plot shows the distribution that would be expected under the null-hypothesis that all measurements of expansion and contraction were products of measurement uncertainty. \label{vout_hist.fig}}
\end{figure*}

In our sample of systems, median $v_\mathrm{out}$ values are more likely to be directed outward than inward. Figure~\ref{vout_hist.fig} shows a histogram of these expansion/contraction velocities and a histogram of the signal-to-noise, calculated as the ratio of the median $v_\mathrm{out}$ to the estimated uncertainty on the median $v_\mathrm{out}$. Although most systems show effect sizes $<$3$\sigma$, the distribution is clearly shifted to the right from what would be expected if systems had zero expansion and all non-zero measurements were due to measurement uncertainty.
There are 6 systems with statistically significant outward velocities at the $>$3$\sigma$ level, while no individual systems show statistically significant inward velocities at this level. The velocity shifts range from $-2.0$ to $+2.0$~km~s$^{-1}$, but are predominantly positive with a mean value of $0.5$~km~s$^{-1}$. 

Systems where expansion is detected at the $>$3$\sigma$ level include NGC 1893, NGC~2244, Tr~16, NGC~6530, NGC~6611, and Cep~B.  In these regions, the expansion pattern is often visually apparent on the plots in Figures~\ref{arrow.fig}--\ref{v_vs_pos.fig}. These cases happen to be the ones with high expansion velocities greater than 0.9~km~s$^{-1}$. However, a number of other systems have measured expansion velocities on the order of $\sim$0.4~km~s$^{-1}$, but the magnitude of the expansion velocity is not sufficient to reach the $>$3$\sigma$ level given typical uncertainties on expansion velocity of $\sim$0.2~km~s$^{-1}$. This group includes the ONC, S~Mon, Tr~14, Tr~15, Pismis~24, G353.1+0.6, IC~5146, and V454 Cep. Overall, it is likely that most of the systems in this second group have mild expansion (Section~\ref{fraction.sec}), but the evidence for expansion of any individual system is not definitive.

Not all systems are likely to be in a state of expansion. The systems NGC~1333, IC~348, NGC~2264 IRS~2, NGC~2362, and NGC~7160 show no evidence for either expansion or contraction, and M17 and NGC~6231 show signs of contraction, albeit not at a high significance level. M17 has a fairly fast contraction rate ($-2$~km~s$^{-1}$), but the {\it Gaia} data are limited by high extinction in this region, so the result is significant only at the 2$\sigma$ level. NGC~6231 has a fairly slow contraction velocity of $-0.2$~km~s$^{-1}$. Contraction of this system would be interesting because NGC~6231 is physically larger than most systems, giving it the appearance of having expanded in the past. Finally, the systems Berkeley~59, LkH$\alpha$~101, Mon~R2,      NGC~2264 IRS~1,        RCW 120,     G353.2+0.7, and M20 have large enough uncertainties relative to the expansion/contraction velocity that results are ambiguous.

\begin{deluxetable}{lrrr}
\tablecaption{Bulk Expansion and Rotation Velocities\label{expansion.tab}}
\tabletypesize{\small}\tablewidth{0pt}%\rotate
\tablehead{
 \colhead{Region} & 
\colhead{$n_\mathrm{samp}$} &
\colhead{median~$v_\mathrm{out}$} & \colhead{median~$v_\mathrm{az}$} \\
 \colhead{} & \colhead{stars} &\colhead{km~s$^{-1}$} &  \colhead{km~s$^{-1}$} \\
\colhead{(1)} & \colhead{(2)} & \colhead{(3)} &\colhead{(4)} 
}
\startdata
Berkeley 59  &  225  &  0.36$\pm$0.24  &  $-$0.39$\pm$0.31  \\ 
NGC 1333  &  47  &  0.23$\pm$0.28  &  $-$0.45$\pm$0.21  \\ 
IC 348  &  180  &  0.16$\pm$0.18  &  $-$0.27$\pm$0.23  \\ 
LkH$\alpha$ 101  &  65  &  0.97$\pm$0.68  &  0.72$\pm$0.33  \\ 
NGC 1893  &  88  &  1.34$\pm$0.32  &  0.19$\pm$0.53  \\ 
ONC  &  378  &  0.43$\pm$0.20  &  $-$0.30$\pm$0.14  \\ 
Mon R2  &  97  & $-$0.12$\pm$0.53  &  0.58$\pm$0.30  \\ 
NGC 2244  &  272  &  1.23$\pm$0.17  &  $-$0.10$\pm$0.22  \\ 
S Mon  &  242  &  0.39$\pm$0.15  &  $-$0.03$\pm$0.07  \\ 
NGC 2264 IRS 2  &  151  &  $-$0.27$\pm$0.28  &  $-$0.13$\pm$0.21  \\ 
NGC 2264 IRS 1  &  126  &  0.36$\pm$0.40  &  $-$0.25$\pm$0.32  \\ 
NGC 2362  &  246  &   0.02$\pm$0.28  & $-$0.10$\pm$0.15  \\ 
Tr 14  &  401  &  0.39$\pm$0.34  &  0.52$\pm$0.19  \\ 
Tr 15  &  194  &  0.64$\pm$0.38  &  1.72$\pm$0.47  \\ 
Tr 16  &  268  &  0.84$\pm$0.22  &  $-$0.06$\pm$0.43  \\ 
NGC 6231  &  615  &  $-$0.23$\pm$0.14  &  $-$0.09$\pm$0.17  \\ 
RCW 120  &  29  &  $-$0.28$\pm$1.45  &  $-$0.12$\pm$0.50  \\ 
Pismis 24  &  75  &   0.91$\pm$0.44  &   0.54$\pm$0.44  \\ 
G353.1+0.6  &  53  &  2.07$\pm$1.10  &  0.07$\pm$0.52  \\ 
G353.2+0.7  &  47  &  $-$0.24$\pm$0.65  &  $-$0.95$\pm$0.48  \\ 
M20  &  116  &  0.33$\pm$0.37  &  $-$0.21$\pm$0.35  \\ 
NGC~6530  &  669  &  0.99$\pm$0.19  &  $-$0.29$\pm$0.15  \\ 
NGC~6611  &  356  &  0.90$\pm$0.23  &  0.29$\pm$0.18  \\ 
M17  &  82  &  $-$2.06$\pm$1.00  &   $-$0.06$\pm$1.09  \\ 
IC 5146  &  115  &  0.48$\pm$0.25  &  0.05$\pm$0.49  \\ 
NGC 7160  &  71  &  $-$0.20$\pm$0.30  &  0.21$\pm$0.18  \\ 
Cep B  &  482  &  0.95$\pm$0.29  &  0.39$\pm$0.14  \\ 
V454 Cep  &  196  &  0.55$\pm$0.34  &  $-$0.14$\pm$0.35  \\ 
\enddata
\tablecomments{Column~1: System name. Column~2: Number of stars used to calculate median velocities. Column~3: Median $v_\mathrm{out}$ -- measure of expansion or contraction. Column~4: Median $v_\mathrm{az}$ -- measure of rotation.}
\end{deluxetable}

\begin{deluxetable*}{lrccccrccr}
\tablecaption{Velocity Dispersions in Select Clusters\label{velocities.tab}}
\tabletypesize{\small}\tablewidth{0pt}%\rotate
\tablehead{
 \colhead{Region} & \colhead{$n_\mathrm{samp}$} &   \colhead{$\sigma_\mathrm{1D}$} &
 \colhead{$\sigma_\mathrm{pc1}$} & \colhead{$\sigma_\mathrm{pc2}$} & \colhead{$\sigma_{pc1}/\sigma_{pc2}$} & 
\colhead{PA} & \colhead{$\eta_\mathrm{outliers}$} & \colhead{$\sigma_\mathrm{outliers}$} &
\colhead{$\Delta$BIC} \\
 \colhead{} & \colhead{stars} &
 \colhead{km~s$^{-1}$} & \colhead{km~s$^{-1}$} & \colhead{km~s$^{-1}$} & \colhead{} & \colhead{deg} & \colhead{\%} & \colhead{km~s$^{-1}$} & \colhead{}\\
\colhead{(1)} & \colhead{(2)} & \colhead{(3)} &\colhead{(4)} & \colhead{(5)} & \colhead{(6)}& \colhead{(7)}& \colhead{(8)} & \colhead{(9)} & \colhead{(10)}
}
\startdata
Berkeley 59  &  18  		&  1.2$\pm$0.2  &  1.5$\pm$0.3  &  0.8$\pm$0.2  &  1.9$\pm$0.6  &  178  \\ 
NGC 1893  &  54  		&  2.6$\pm$0.5  &  3.3$\pm$0.7  &  1.7$\pm$0.7  &  1.9$\pm$0.9  &  43  &  15$\pm$10  &  13  &  $-$40  \\ 
ONC  &  48  			&  1.8$\pm$0.1  &  2.2$\pm$0.2  &  1.4$\pm$0.1  &  1.5$\pm$0.2  &  5  \\ 
Mon R2  &  13  			&  1.6$\pm$0.4  &  2.0$\pm$0.5  &  1.1$\pm$0.5  &  1.7$\pm$0.9  &  172  \\ 
NGC 2244  &  89  		&  2.0$\pm$0.1  &  2.2$\pm$0.2  &  1.8$\pm$0.2  &  1.2$\pm$0.2  &  31  \\ 
S Mon  &  67 			&  1.1$\pm$0.1  &  1.3$\pm$0.1  &  0.9$\pm$0.1  &  1.4$\pm$0.2  &  89  \\ 
NGC 2264 IRS 2  &  29  	&  1.8$\pm$0.2  &  2.3$\pm$0.4  &  1.2$\pm$0.2  &  1.9$\pm$0.5  &  97  \\ 
NGC 2264 IRS 1  &  30  	&  1.5$\pm$0.2  &  1.9$\pm$0.3  &  1.0$\pm$0.2  &  1.9$\pm$0.4  &  76  \\ 
NGC 2362  &  98  		&  0.8$\pm$0.1  &  0.9$\pm$0.2  &  0.7$\pm$0.1  &  1.2$\pm$0.3  &  153  &  14$\pm$17  &  2.0  &  $-$6.9  \\ 
Tr 14  &  145  			&  2.5$\pm$0.2  &  2.8$\pm$0.2  &  2.1$\pm$0.3  &  1.3$\pm$0.2  &  46  &  19$\pm$7  &  7.3  &  $-$42  \\ 
Tr 15  &  105  			&  2.4$\pm$0.2  &  2.5$\pm$0.2  &  2.2$\pm$0.2  &  1.1$\pm$0.2  &  114  &  11$\pm$5  &  9.3  &  $-$34  \\ 
Tr 16  &  121  			&  2.8$\pm$0.2  &  3.4$\pm$0.4  &  1.9$\pm$0.2  &  1.8$\pm$0.3  &  126  &  9:  &  5.5  &  5.4  \\ 
NGC 6231  &  278  		&  1.6$\pm$0.1  &  1.7$\pm$0.1  &  1.5$\pm$0.1  &  1.2$\pm$0.1  &  132  &  10$\pm$4  &  5.4  &  $-$106  \\ 
M20  &  36  			&  1.6$\pm$0.2  &  1.9$\pm$0.3  &  1.2$\pm$0.3  &  1.6$\pm$0.4  &  73  &  11$\pm$10  &  6.1  &  $-$18  \\ 
NGC 6530  &  185  		&  2.3$\pm$0.1  &  2.7$\pm$0.2  &  1.8$\pm$0.2  &  1.5$\pm$0.2  &  106  &  13$\pm$5  &  7.3  &  $-$73  \\ 
NGC 6611  &  94 		&  1.8$\pm$0.2  &  2.2$\pm$0.3  &  1.4$\pm$0.2  &  1.6$\pm$0.4  &  22  &  23$\pm$9  &  5.8  &  $-$37  \\ 
IC 5146  &  11  			&  0.9$\pm$0.2  &  1.1$\pm$0.3  &  0.7$\pm$0.2  &  1.5$\pm$0.5  &  179  \\ 
NGC 7160  &  25  		&  1.5$\pm$0.2  &  1.8$\pm$0.2  &  1.1$\pm$0.2  &  1.7$\pm$0.4  &  51  \\ 
Cep B  &  27  			&  1.9$\pm$0.2  &  2.1$\pm$0.3  &  1.6$\pm$0.3  &  1.3$\pm$0.3  &  177  \\ 
\enddata
\tablecomments{Column~1: System name. Column~2: Number of stars used to calculate velocity dispersions. Column~3: Characteristic one-dimensional velocity dispersion. Column~4--5: Velocity dispersion in the first and second principal components for the two-dimensional velocity model. 
Column~6: Ratio of velocity dispersions in the first and second velocity principal components -- a measure of overall velocity anisotropy. 
Column~7: Position angle (east from north) of the semi-major axis of the velocity dispersion. Column~8: Fraction of stars belonging to the second component of the mixture model. Column~9: Velocity dispersion for the second component of the mixture model. Column~10: Change in BIC when the second component was added. The last three columns are only used when a multiple component velocity model was required.   }
\end{deluxetable*}
\vspace*{0.3in}

\subsection{Radial Dependence of Expansion Velocity}\label{radial_gradient.sec}

Models for the expansion of OB associations suggest linear relationships between expansion velocity and distance of stars from a system center \citep{1964ARA&A...2..213B,1997MNRAS.285..479B}. This occurs because, in an unbound system, faster stars will travel farther from their point of origin, causing the stars to spatially sort themselves by velocity.

\begin{figure*}[t]
\centering
\includegraphics[width=0.45\textwidth]{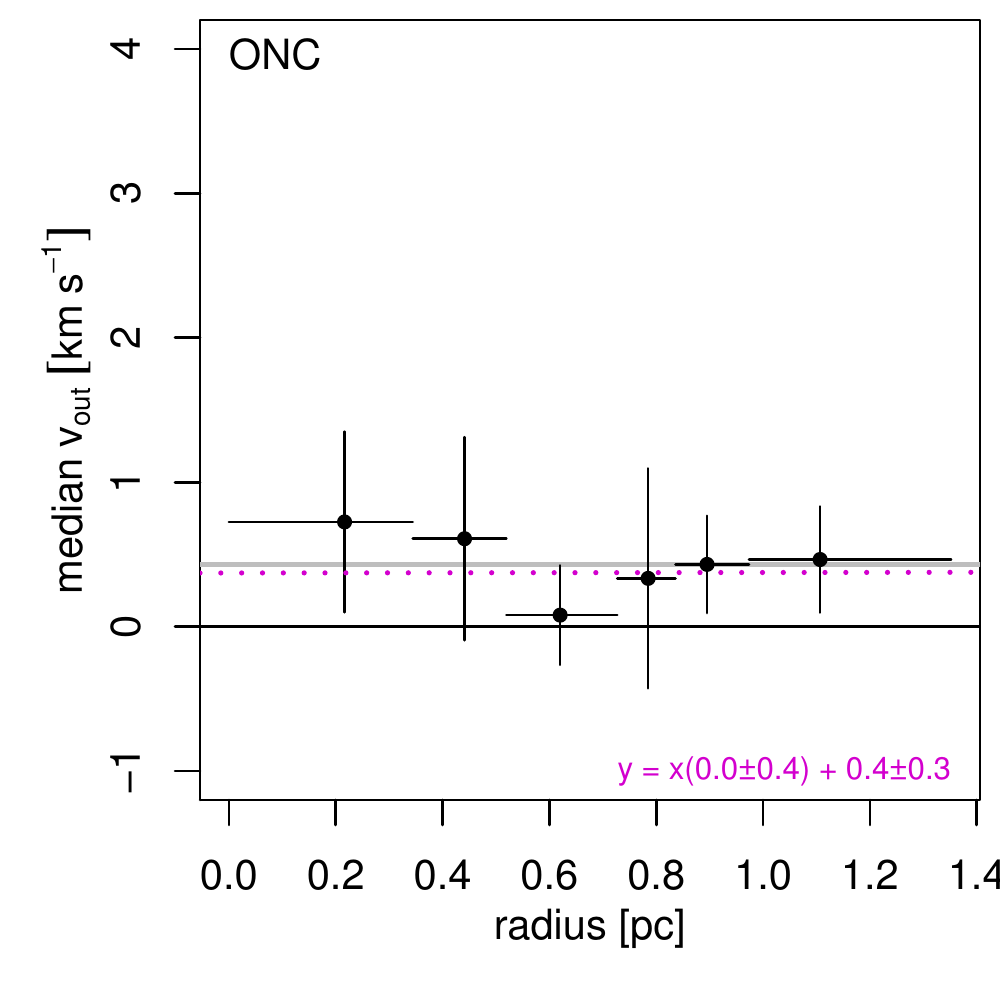} 
\includegraphics[width=0.45\textwidth]{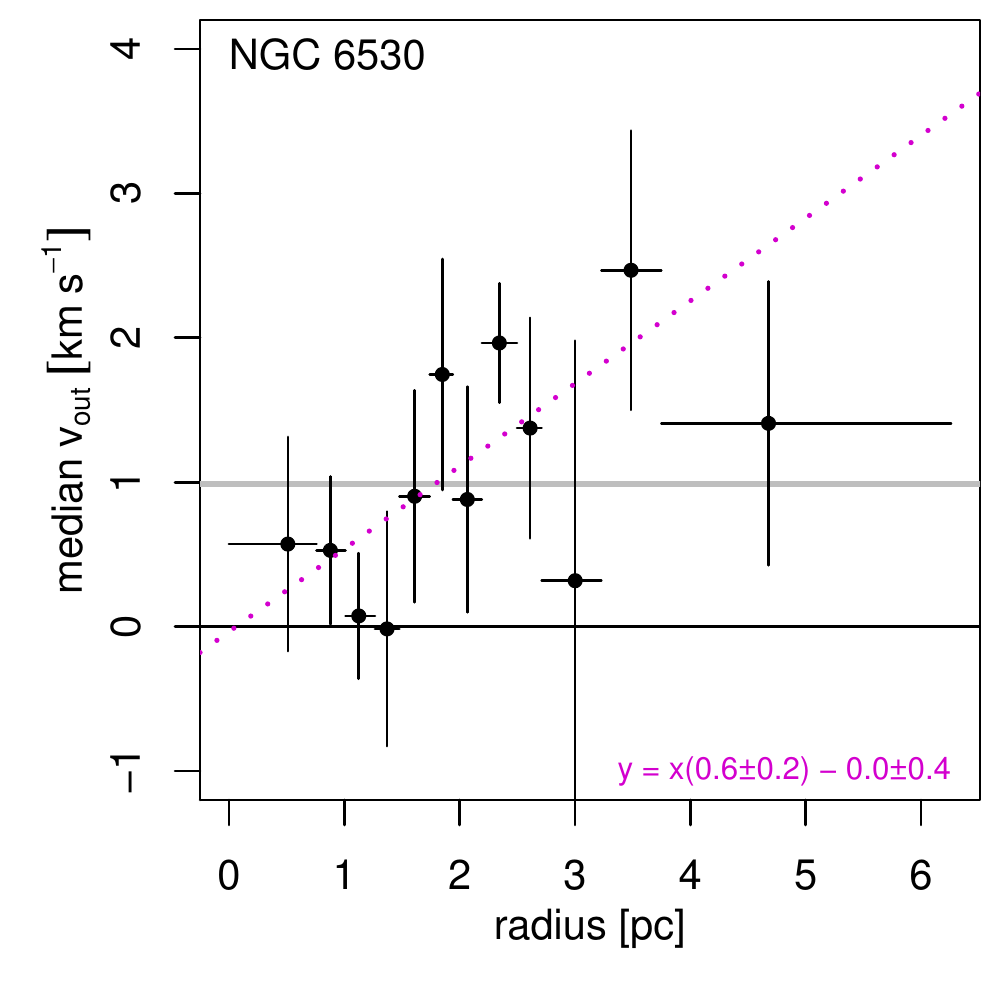} 
\caption{Expansion velocity as a function of distance from the system center. Stars are binned by radial distance in groups of 60. In each panel, the black line indicates the division between contraction and expansion, the gray line is the median expansion velocity for all stars, the points are the binned data, and the magenta line is a linear regression to the data. The equation of the regression line is included on the plot.  Plots for other stellar systems are included in an online figure set. \label{expand_rad.fig}}
\end{figure*}

Figure~\ref{expand_rad.fig} shows expansion velocity as a function of radius. For this plot, stars are binned by radial distance using bin sizes of $\sim$60 stars, and expansion velocities are estimated using the same method as above. We fit the points with a line of the form $y=ax+b$ using weighted least squares regression, where weights are proportional to the reciprocal of the uncertainty squared. The slopes of regression lines have units of km~s$^{-1}$~pc$^{-1}$ ($\approx \mathrm{Myr}^{-1}$) and intercepts have units of km~s$^{-1}$.

For the ONC, which shows evidence for only mild expansion, all points are consistent with the mean value. Both slope (0.0$\pm$0.4) and intercept (0.4$\pm$0.3) are consistent with there being either no or mild expansion. For NGC~6530, most of the points follow a positive linear relationship between radius and expansion velocity, and the two points that deviate from this relationship have large uncertainties. The slope of 0.6$\pm$0.2 is statistically significant, while the intercept $0.0\pm0.4$ is not statistically significant. The regression analysis, using the {\it lm} function in {\it R}, gives a $p$-value of 0.002, providing strong evidence that NGC~6530 has a radially dependent expansion velocity. 

Several other systems fall into each class. Examples that illustrate possible linear relationships between radius and expansion rate include Cep~B ($p=0.02$), NGC~2244 ($p=0.003$), and Tr~16 ($p=0.08$).  For these systems, the slopes of the relationships range from $\sim$0.5 to 1~km~s$^{-1}$~pc$^{-1}$.  NGC~6611 also shows fastest expansion in the outer regions, consistent with this pattern, but the slope of the relation is not statistically significant ($p=0.06$). Others systems like Berkeley~59, NGC 2362, NGC~6231, S~Mon, Tr~14, and Tr~15 show flat relationships -- the latter group includes both systems with and without evidence for net expansion. 

\subsection{Fraction of Systems that are Expanding}\label{fraction.sec}

Overall, 21 out of 28 systems (i.e.\ 75\%) have positive values of median~$v_\mathrm{out}$. However, calculating the fraction of systems that are expanding is complicated by measurement errors, which can change the sign of median~$v_\mathrm{out}$ for systems with small or zero expansion velocity. To evaluate this effect, we include uncertainties on median~$v_\mathrm{out}$ in a statistical model akin to the Extreme Deconvolution method of \citet{2011AnApS...5.1657B}. The construction of such a model is described more thoroughly in Section~\ref{sd.sec}, where it is used for a different application.

The observed distribution of expansion velocities (Figure~\ref{vout_hist.fig}), can be modeled of as an intrinsic distribution convolved with measurement uncertainties. We examine several models for the intrinsic distribution, which are fit to the data using the Monte Carlo Markov Chain (MCMC) method. We first try modeling the distribution of median~$v_\mathrm{out}$ as a single Gaussian, which yields $f=0.86$ of systems in expansion (0.65--0.98 95\% credible interval). We next use a mixture of two Gaussians, noting that Gaussian mixture models can be used as flexible models for unknown probability density functions \citep[e.g.,][]{2008ApJ...682..874K}. This model yields a fraction of $f=0.88$ (0.79--0.94 95\% credible interval). For MCMC analysis we use ``non-informative'' priors for mixture models: for the standard deviation of the Gaussians we use a uniform distribution between $0$ and $6$, for the mixing parameters we use a Dirichlet distribution with $\alpha=1$, and for the mean we use a uniform distribution between $-3$ and $3$. We use the Metropolis-Hastings algorithm to sample from the posterior. Through experimentation we find that variations in the functional form of the priors has relatively minor effects on results.

Our conclusion from this analysis is that our assumptions on the distribution of expansion velocities for systems have little effect on the fraction that are expanding ($\sim$85\%) and that the effect of measurement errors is to slightly decrease the fraction of systems observed undergoing expansion. Thus, we can reliably claim that $\gtrsim$75\% of systems in our sample are expanding.

\subsection{Cluster Rotation}\label{rotation.sec}

The stellar velocity measurements can also be used to look for evidence of bulk rotation. The angular momentum of a star with mass $m$ and velocity $\mathbf{v}$ at a position $\mathbf{r}$ relative to the center of the system is
\begin{equation}
\mathbf{L} = \mathbf{r}\times m\,\mathbf{v},
\end{equation}
so the component of the angular momentum along the line of sight is
\begin{equation}
L \cos i 
= - m\,R\,v_\mathrm{az},
\end{equation}
where $i$ is the inclination of the angular momentum vector and $R$ is the projected distance of the star from the center of the system. Thus, a non-zero median value of $v_\mathrm{az}$ can indicate bulk rotation of a system. For $v_\mathrm{az}$, we use the same method to calculate the median and error on the median that we used for $v_\mathrm{out}$. These values are provided in Table~\ref{expansion.tab}.

For the 28 systems in the sample, the values of median $v_\mathrm{az}$ are distributed around zero with an average of $0.06\pm0.10$~km~s$^{-1}$. For most systems, the value of median~$v_\mathrm{az}$ is within 2$\sigma$ of $v_\mathrm{az}=0$~km~s$^{-1}$. A few cases have more statistically significant values, including  LkH$\alpha$~101, the ONC, Tr~14, and G353.2+0.7 at 2$\sigma$ significance, and Tr~15 (median~$v_\mathrm{az}=1.7$~km~s$^{-1}$) at $>$3$\sigma$ significance. Typical uncertainties on median azimuthal velocities are $<$0.4~km~s$^{-1}$, so rotational velocities less than this value may not be detectable.

\begin{figure*}[t]
\centering
\includegraphics[width=0.45\textwidth]{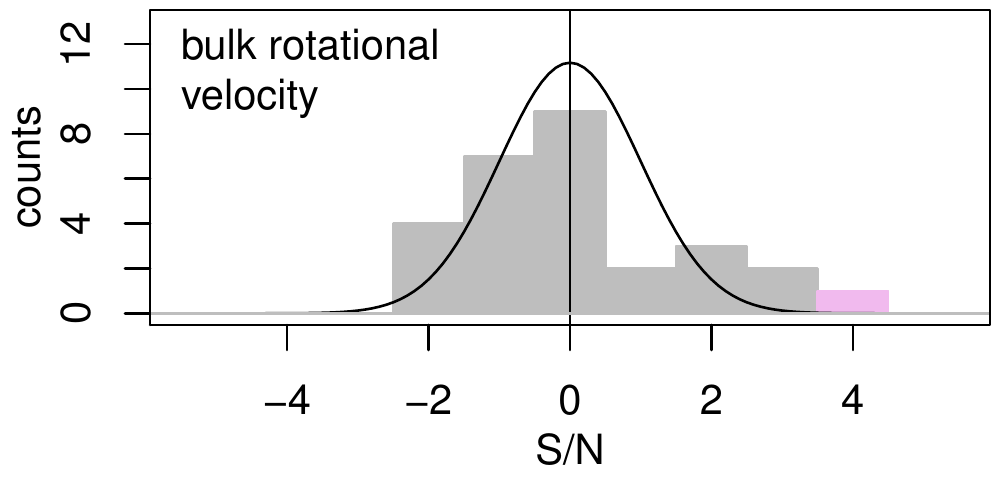} 
\caption{Histogram of signal-to-noise in rotational velocity (median~$v_\mathrm{az}$ divided its uncertainty). The observed distribution is compared to the expected Gaussian distribution (black curve) for the null hypothesis that there is no rotation and non-zero values are effects of measurement uncertainty. Tr~15, significant at $>$3$\sigma$, is highlighted in magenta.
 \label{rotation_qq.fig}}
\end{figure*}

Under the assumption that the median values of $v_\mathrm{az}$ are all results of measurement uncertainty, the ratios of these values to their uncertainties should be drawn from a normal distribution with a mean 0 and standard deviation 1.
Figure~\ref{rotation_qq.fig} shows a histogram of the ratio of observed rotation to measurement uncertainty, which is compared to the normal distribution expected given the null hypothesis. The distributions are remarkably similar, suggesting that no real rotation is detected in most of the stellar systems. However, Tr~15 is difficult to explain using the null hypothesis, so rotation may be real in this individual case. 

Although bulk rotation has been reliably measured in globular clusters \citep[e.g.][]{kamann2018,2018arXiv180602580B}, very few attempted measurements exist in the literature for open clusters.  A recent study of {\it Gaia} DR1 data by \cite{reino2018} concluded that there was no evidence for bulk rotation in the $\sim$600 Myr old Hyades cluster, based on measurement of azimuthal motion at only the 2$\sigma$ level. Our results are consistent with cluster rotation being rare, but we are less sensitive to rotational velocities below several tenths of a km~s$^{-1}$. 

\section{Velocity Dispersions}\label{sd.sec}

Calculation of velocity dispersion is particularly sensitive to measurement errors, which can broaden the observed distribution. The astrometric measurement uncertainties reported in the {\it Gaia} catalog are heteroscedastic and comparable to the velocity dispersion, so their effect must be carefully modeled. For {\it Gaia} sources with $\mathtt{astrometric\_excess\_noise} > 0$, the statistical uncertainties derived from the {\it Gaia} DR2 AGIS model may not represent the real errors in relative proper motions \citep{2012A&A...538A..78L}. 
Thus, we use only sources with $\mathtt{astrometric\_excess\_noise} = 0$ when modeling velocity distributions.  

We use maximum likelihood to estimate the intrinsic velocity dispersion in the presence of measurement error \citep[cf.][]{2006AJ....131.2114W}. We model the observed velocity of a star $i$ as the sum of its intrinsic velocity $\mathbf{v}_i$ and an error term $\bm{\epsilon}_i$, 
\begin{equation}
\mathbf{v}_{\mathrm{obs},i} = \mathbf{v}_i + \bm{\epsilon}_i.
\end{equation}
Both $\bf{v}_i$ and $\bm{\epsilon}_i$ are assumed to be multivariate normally distributed, with
\begin{align}
\mathbf{v}_i & \sim N\left(\bm{\mu}_v,\Sigma_v\right) \\
\bm{\epsilon}_i & \sim N\left(\bm{0},\Sigma_{\mathrm{err},i}\right),
\end{align}
where $\bm{\mu}_v$ and $\Sigma_v$ are the mean and covariance matrix of the intrinsic velocity distribution, and $\Sigma_{\mathrm{err},i}$ is the covariance matrix for measurement uncertainty on the $i$th star. Then the log-likelihood is
\begin{equation}\label{likelihood1.eqn}
\mathcal{L}\left(\bm{\mu}_v,\Sigma_v\middle|\mathbf{v}_\mathrm{obs,i},\Sigma_{\mathrm{err},i} \right) = 
\sum_{i=1}^N \log\phi\left(\bm{\mu}_v, \Sigma_v + \Sigma_{\mathrm{err},i} \right),
\end{equation}
where $\phi$ is the probability density of the normal distribution. The maximum likelihood $\bm{\mu}_v$ and $\Sigma_v$ parameters can be found by numerical optimization. We used the BFGS algorithm \citep{fletcher1964function} implemented in the {\it R} function {\it optim}. Examination of the likelihood function shows it to be approximately normal, so we use {\it optim} to numerically calculate the Hessian matrix at the maximum (also called the Fisher Information Matrix) and invert it to estimate the covariance matrix. 

\subsection{Velocity Phase Space}

\begin{figure*}[t]
\centering
\includegraphics[width=0.45\textwidth]{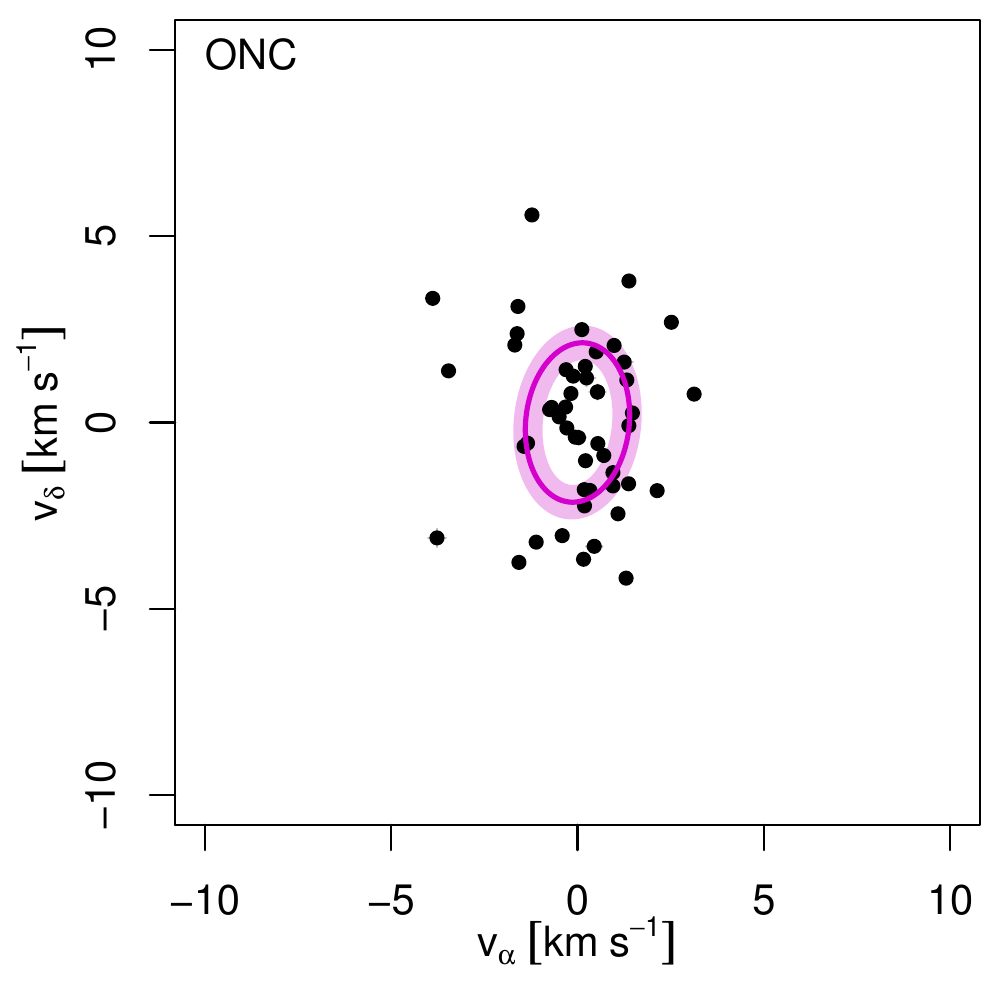} 
\includegraphics[width=0.45\textwidth]{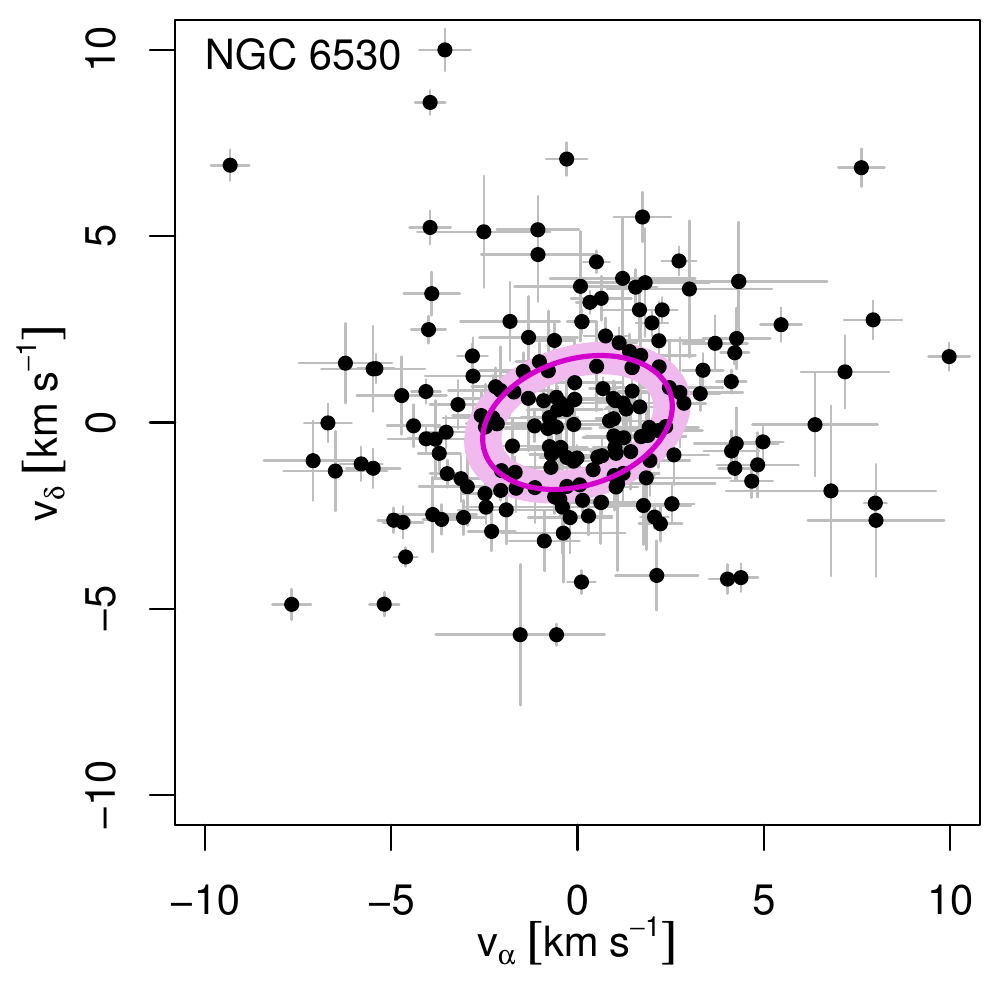} 
\caption{Velocity distributions for the ONC (left) and NGC~6530 (right). The magenta ellipses shows velocity dispersion from the best-fit bivariate normal distribution -- the ellipse is the iso-density contour for this normal distribution at the 1$\sigma$ level. The shaded regions shows uncertainty resulting from 2 times the standard error on the velocity dispersions. Only sources with no astrometric excess noise are used. Plots for other stellar systems are included in an online figure set. 
 \label{phase.fig}}
\end{figure*}

The total velocity dispersions of the individual stellar systems will incorporate both the bulk expansion velocity (if non-zero) and a velocity spread due to the orbital motions of individual stars. Figure~\ref{phase.fig} shows stars plotted in velocity phase space with coordinates $(v_x, v_y)$, where the ONC and NGC~6530 are used as examples. 

The observed velocity distributions are not entirely isotropic. Stars in the ONC are preferentially moving north or south, rather than east or west, while stars in  NGC~6530 are preferentially moving east or west, rather than north or south.  However, the referencing of stellar motions to a frame that is based on the equatorial coordinate system is arbitrary. 

The Gaussian model of the velocity distribution provides two velocity dispersion components.  These are the two principal components of the velocity dispersion, where the first component is defined to be the one with the largest variance. Thus $\sigma_{pc1}$ is the semi-major axis of the ellipses in Figure~\ref{phase.fig}, while $\sigma_{pc2}$, is the semi-minor axes. The use of Equation~\ref{likelihood1.eqn} allows for the heteroscedastic uncertainties to be taken into account in the principal component analysis.  Table~\ref{velocities.tab} provides $\sigma_{pc1}$ and $\sigma_{pc2}$, as well as their ratio, the position angle of the first principal component, and statistical uncertainties.

The ratios of $\sigma_{pc1}$ to $\sigma_{pc2}$ show that most systems do not have statistically significant velocity anisotropy (i.e.\ values that are significantly greater than 1). However, the ONC has $\sigma_{pc1}/\sigma_{pc2}=1.5$ and NGC~6530 has $\sigma_{pc1}/\sigma_{pc2}=1.5$, both significant at $\sim$2.5$\sigma$ significance. In both cases, the orientation of the velocity anisotropy is approximately aligned with the spatial elongation of the system. 

\subsection{Shape of the Velocity Distribution}

We can examine the shape of velocity distributions by comparison to bivariate normal distributions. \citet{henze1990} provide a test of multivariate normality, implemented in the {\it R} package {\it MVN} \citep{korkmaz2014mvn}. The systems where velocity distributions are consistent with multivariate normal distributions include Berkeley~59, the ONC, IRS~1, and IC~5146 ($p>0.05$). The systems Mon~R2, NGC~2244, NGC 7160, and Cep~B show moderate statistically significant deviation from normality ($0.05<p<0.001$), while NGC~1893, S~Mon, IRS~2, NGC~2362, Tr~14/15/16, NGC~6231, M20, NGC~6530, and NGC~6611 show large statistically significant deviations from normality ($p<0.001$). We note that this hypothesis test is more sensitive to deviations when sample sizes are larger and it does not indicate how the distribution deviates from normality.  

Deviations from normality can be visualized using plots of observed data quantiles versus theoretical quantiles (the Q--Q plot). The data quantiles are the difference between each measurement and the mean value, normalized by the standard deviation,
\begin{equation}
Q_{\mathrm{data},i} = \frac{v_{\mathrm{obs},i}-\mu}{\sqrt{\sigma_v^2 + \sigma^2_{\mathrm{err},i}}}\\
\end{equation}
while the theoretical quantiles are
\begin{equation}
Q_{\mathrm{theo},i} = \sqrt{2}\,\mathrm{erf}^{-1}\left(2(r_i-0.5)/n - 1\right),\label{qnorm.eqn}
\end{equation}
the quantile function of the Gaussian distribution, where $n$ is the number of velocity measurements, $r_i$ is the rank of the $i$th measurement, and $\mathrm{erf}^{-1}$ is the inverse of the error function. We calculate a test envelope (95\%) for the null hypothesis that the data are correctly described by our model using Monte Carlo simulations. 

\begin{figure}[t]
\centering
\includegraphics[width=0.45\textwidth]{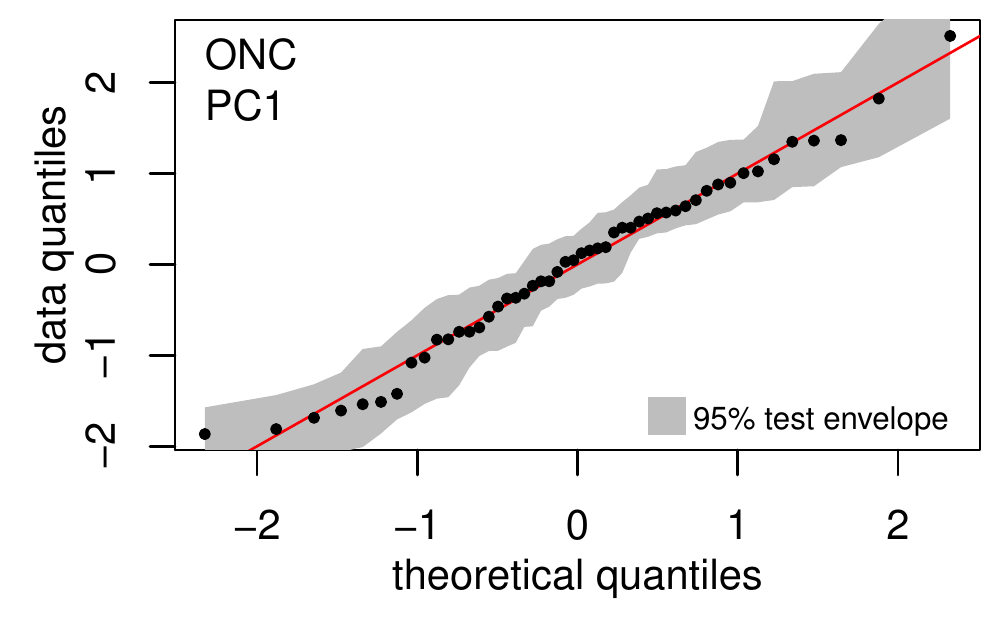} 
\includegraphics[width=0.45\textwidth]{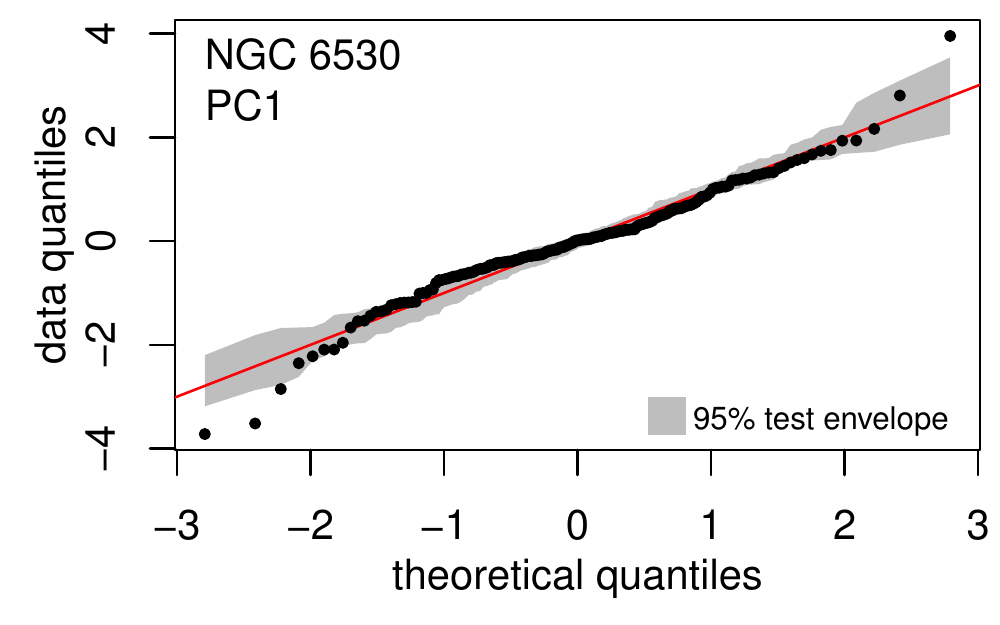} 
\caption{Q--Q plot to evaluate the Gaussianity of stellar velocity distributions. The plots show theoretical quantiles versus data quantiles, assuming that the data are fit by a normal distribution. Lower values on the left and higher values on the right show an excess number of outliers with higher than expected velocities. The red line shows the expected value for a normal distribution, and the gray shaded region shows a 95\% test envelope. Only sources with no astrometric excess noise are used. Plots of $v_{pc1}$ and $v_{pc2}$ for other stellar systems are included in an online figure set.  
 \label{qq.fig}}
\end{figure}

Q--Q plots are produced for each velocity component (Figure~\ref{qq.fig}). The ONC velocity distribution is remarkably well fit by a normal distribution. This result is astrophysically interesting in itself (see Section~\ref{discussion.sec}), but also implies that standard deviations of the distribution calculated using Equation~\ref{likelihood1.eqn} will be reliable. 

For NGC~6530, the distribution closely follows a normal distribution out to $\pm$2$\sigma$, but beyond this threshold there is a significant excess of stars with higher velocities than would be expected from a normal distribution. Even a small number of outliers can have a large effect on estimates of standard deviation, so the observed deviations from a normal distribution suggest that a standard deviation estimated from Equation~\ref{likelihood1.eqn} will overestimate the width of the distribution.

For nearly half the systems, the shape of the observed velocity distribution indicates the presence of outliers. The nature of the outliers is uncertain. While these could represent a population of higher velocity stars that are escaping the system at a faster rate, they could also represent points with large errors not represented well by the AGIS uncertainties or field stars contaminating the ``refined sample.'' In Table~\ref{velocities.tab}, the more distant systems tend to be the ones where outliers are detected. While the astrometric excess noise parameter is effective at flagging known binary stars in the ONC, it is not likely to be as effective for more distant regions. Overall, the outliers do not seem to have a preferential spatial location within the systems.

A possible strategy to cope with outliers is to use a mixture model to represent the main population and the outliers as two distinct Gaussian components. In our model, these components will have the same means but different covariance matrices \citep[cf.][]{2018arXiv180301908B}. The complete data likelihood for this mixture model is
\begin{multline}\label{mix.eqn}
\mathcal{L}\left(\bm{\mu}_v,\Sigma_{v,1},\Sigma_{v,2},\eta\middle|\mathbf{v}_\mathrm{obs,i},\Sigma_{\mathrm{err},i} \right) =\\
\sum_{i=1}^N \log
(
(1-\eta)\,\phi\left(\bm{\mu}_v, \Sigma_{v,1} + \Sigma_{\mathrm{err},i} \right) + \\
\eta\,\phi\left(\bm{\mu}_v, \Sigma_{v,2} + \Sigma_{\mathrm{err},i} \right)
),
\end{multline}
where $\Sigma_{v,1}$ and $\Sigma_{v,2}$ are the covariance matrices describing the velocity distribution and   
$0<\eta<1$ is the mixing parameter. 
For the second component, we use a radially symmetric normal distribution (i.e.\ $\Sigma_{v,2} = a\,I_n$) and require the dispersion to be larger than for the first component. 
If the second component is significantly wider than the first but has a much smaller fraction of the stars, than it can be considered a model for the ``outliers.''

For stellar systems where outliers can be seen on the Q--Q plot, we use Equation~\ref{mix.eqn} to estimate the velocity dispersion of the main component, the fraction of sources that are outliers (the mixing parameter $\eta$ in the model), and a velocity dispersion for the outliers. For both the single component and the mixture model, we calculate the change in the Bayesian Information Criterion (BIC), which is a penalized likelihood used for selecting between models with different numbers of parameters where the model with the lowest BIC is preferred \citep{schwarz1978estimating}. For cases where the model appears to correctly identify outliers (i.e.\ $\eta$ is small and the velocity dispersion is large for the second component) we report values from the mixture model method in Table~\ref{velocities.tab}. In 8 out of 9 cases where this model was applied, the two-component model produced a significant improvement in the BIC ($\Delta BIC < -6$). We note that in all cases where the outlier model improves the fit,   Henze-Zirkler's test showed strong deviation from normality.  

\subsection{One-Dimensional Velocity Dispersions}

Formulas for stellar dynamics are often given in terms of a one-dimensional velocity dispersion, $\sigma_{1D}$, because this quantity can be obtained from radial velocity measurements alone. A characteristic one-dimensional velocity dispersion can also be obtained from multi-dimensional velocity dispersions by taking the mean variance 
\begin{equation}
\sigma^2_{1D} = \frac{\sigma^2_{\mathrm{pc1}}+\sigma^2_{\mathrm{pc2}}}{2}.
\end{equation}
These values are also recorded in Table~\ref{velocities.tab}.

The one-dimensional velocity dispersions found for clusters in our sample range from 0.8 to 2.8~km~s$^{-1}$, with a mean of 1.8~km~s$^{-1}$. 
For the ONC the {\it Gaia}-based velocity dispersion is $\sigma_{1D}=1.8\pm0.1$~km~s$^{-1}$. Estimates from earlier studies include $\sim$2.3~km~s$^{-1}$ from a proper-motion study by \citet{1988AJ.....95.1755J}, 3.1~km~s$^{-1}$ from an RV study by \citet{2008ApJ...676.1109F}, $\sim$2.3~km~s$^{-1}$ from an RV study by \citet{2009ApJ...697.1103T}, $\sim$2.5~km~s$^{-1}$ from a RV study by \citet{2016ApJ...821....8K}, $\sim$2.3~km~s$^{-1}$ from a radio proper-motions study by \citet{2017ApJ...834..139D}, 
and 1.7~km~s$^{-1}$ from an RV study by \citet{2017ApJ...845..105D} after corrections to take into account spatial variations in the mean velocity dispersion. Our value is smaller than most of these estimates, but approximately equal to the estimate by \citet{2017ApJ...845..105D}. We note that our estimate is based only on the central cluster, rather than on larger areas that were the focuses of studies by \citet{2008ApJ...676.1109F} and \citet{2016ApJ...821....8K}, for which the total velocity dispersions will include broadening effects from the velocity gradients identified by \citet{2017ApJ...845..105D}.

\begin{figure*}[t]
\centering
\includegraphics[width=0.45\textwidth]{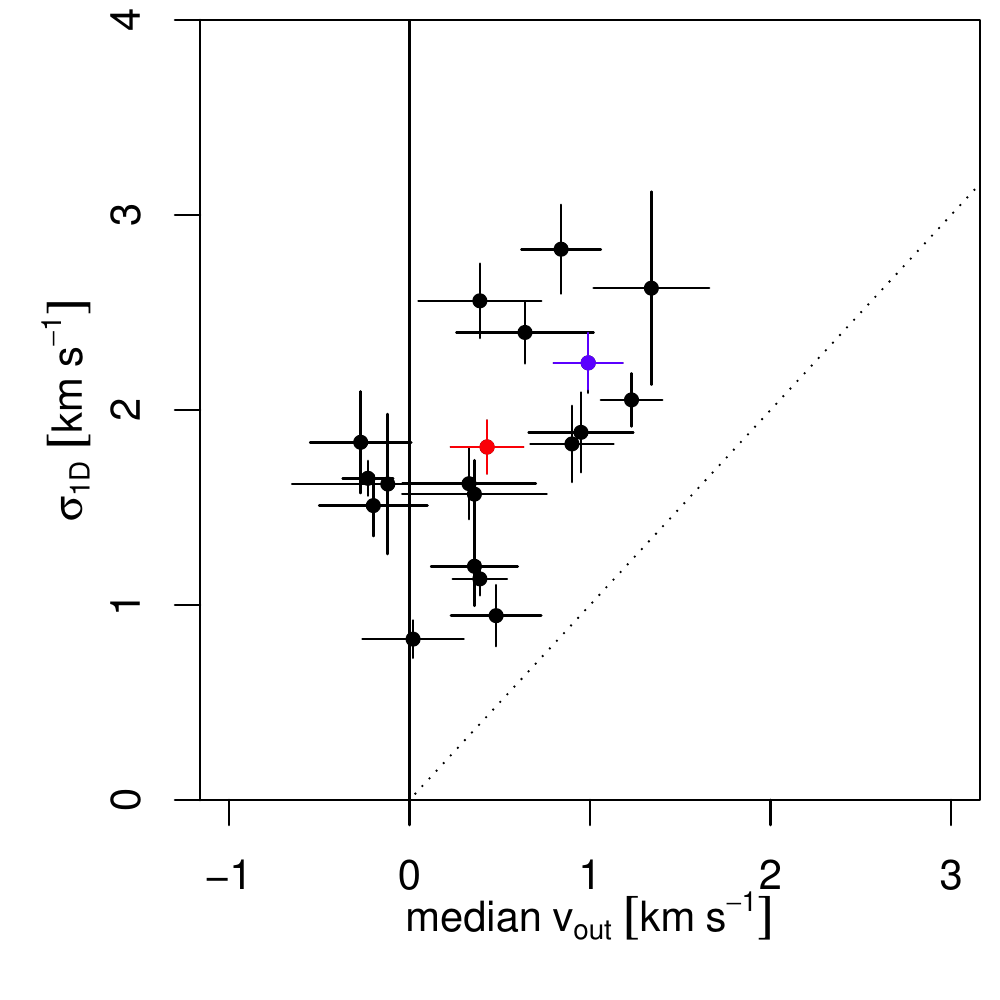} 
\caption{One-dimensional velocity dispersion versus expansion velocity. The solid line indicates the division between contracting and expanding systems, and the dashed line indicates where both quantities are equal. The data points for the ONC (red) and NGC~6530 (blue) are marked.   
 \label{sigma_1d.fig}}
\end{figure*}

There is a positive correlation between $\sigma_{1D}$ and expansion velocity  (Figure~\ref{sigma_1d.fig}); Kendall's rank correlation test shows the dependence to be marginally statistically significant ($p=0.03$). The one-dimensional velocity dispersions are significantly higher than the expansion velocities, mostly exceeding the bulk expansion velocities by a factor of 2--3. 
For cases where expansion rate varies with radius (e.g., Figure~\ref{expand_rad.fig}) this effect will contribute to velocity dispersions. For example, in a toy model with pure expansion from a central point, we would expect $\sigma_{1D}\approx 1.5\, \bar{v}_\mathrm{out}$. However, velocity dispersions that are much larger than this imply that not all stars are moving outward. 

In this initial {\it Gaia} study, the comparison of velocity dispersions in different regions comes with the caveat that the mass ranges of stars in the samples for different regions are not identical (e.g., Figure~\ref{error.fig}). If there is a relationship between a star's mass and its velocity, then the selection of stars with good {\it Gaia} astrometry, which tend to be the brightest stars in a region, can affect the observed velocity dispersion. The systems for which velocity dispersions are derived in Table~\ref{velocities.tab} typically have samples that include stars with masses down to 0.5~$M_\odot$. 

\begin{figure*}
\centering
\includegraphics[width=0.45\textwidth]{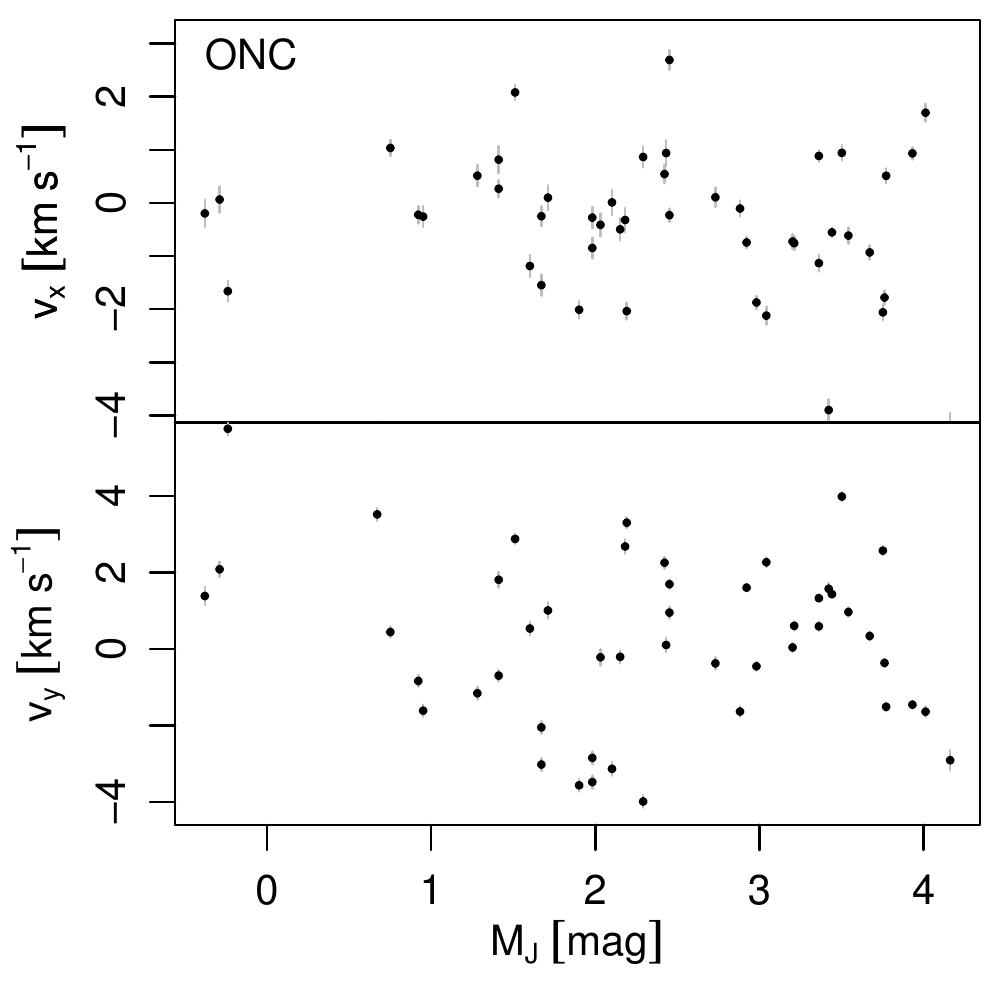} 
\includegraphics[width=0.45\textwidth]{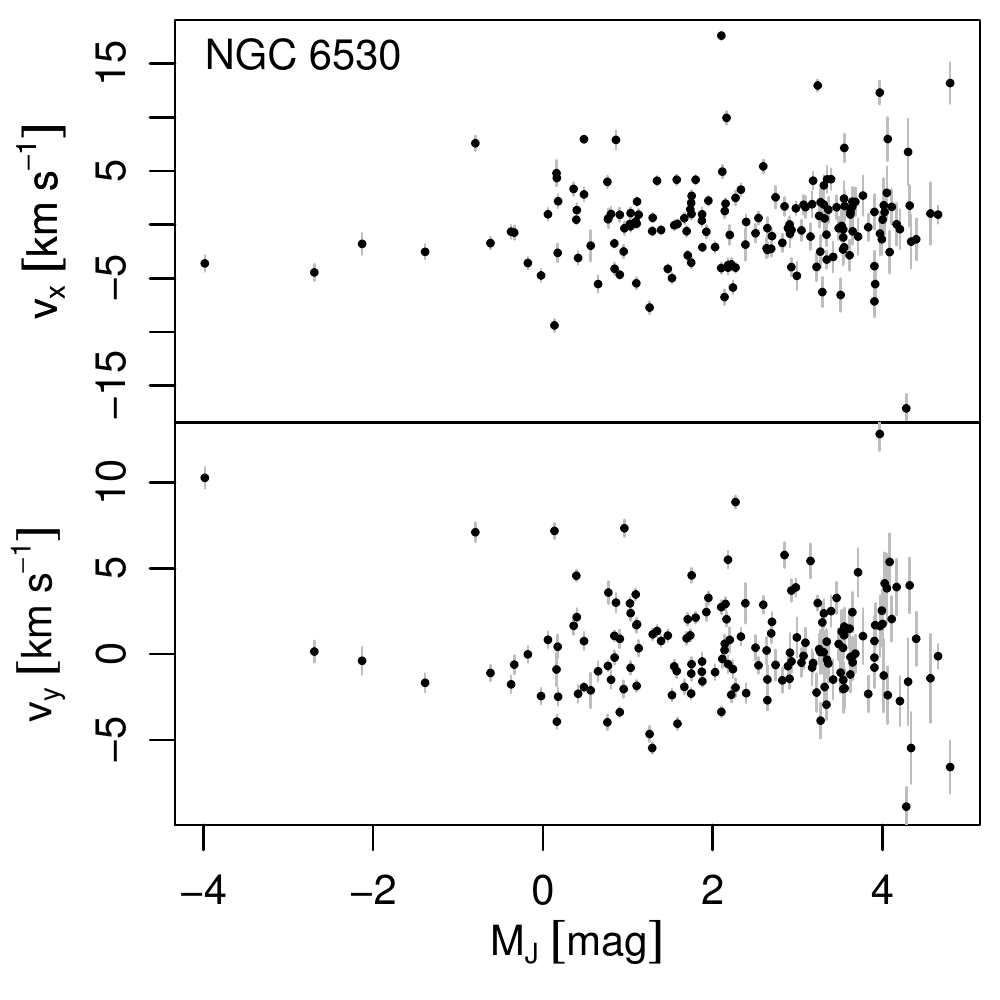} 
\caption{Velocity components $v_x$ and $v_y$ versus $J$-band absolute magnitude. Statistical 1$\sigma$ uncertainties on velocities are shown by the gray error bars. Only sources with no astrometric excess noise are used.
 \label{J_v.fig}}
\end{figure*}

In order to determine how sample selection may affect estimated velocity dispersions, we show plots of velocity versus absolute magnitude in the $J$ band, where absolute magnitude is
\begin{equation}
M_J= J + 5\log\varpi_0[mas] - 10.
\end{equation}
For the ONC and NGC~6530 we show the plots of both $v_x$ and $v_y$ versus $M_J$ in Figure~\ref{J_v.fig}. In both cases, for stars with $1<M_J<4$~mag (approximately a mass range of 0.5--2.5~$M_\odot$; constituting the bulk of the sample) velocity dispersion stays relatively constant with magnitude. 

There has been much interest, both observational and theoretical, of the effect of stellar mass on velocity dispersions in open clusters and globular clusters \citep[e.g.,][]{2016MNRAS.458.3644B,2016MNRAS.460..317S,2016MNRAS.459L.119P,2017MNRAS.464.1977W,2018arXiv180301908B}. Overall, studies suggest that open clusters do not achieve energy equipartition \citep[cf.][]{1969ApJ...158L.139S}.  
We leave the full investigation of the kinematic implications of stellar mass and mass segregation in the {\it Gaia} DR2 data to a future study. 

\subsection{Velocity Dispersion as a Function of Radius} \label{expansion_radius.sec}

Radial variation in velocity dispersion of a stellar system will reflect its dynamical state. Figure~\ref{sigma_rad.fig} shows velocity dispersion as a function of radius for several stellar systems for which there are a sufficient number of stars to measure velocity dispersions in radial bins. For this analysis we only use stars with $\mathtt{astrometric\_excess\_noise} = 0$ that are not classified as ``outliers.''

The top row in Figure~\ref{sigma_rad.fig} shows velocity dispersion as a function of radius in two systems, NGC 6530 and Cep B, that are clearly expanding (Section~\ref{bulk.sec}). In these two systems, velocity dispersion increases with distance from the center. Given that both of them were found to have increasing expansion velocity as a function of radius (Section~\ref{radial_gradient.sec}), the trend in velocity dispersion supports our earlier result. 

The bottom row in Figure~\ref{sigma_rad.fig} shows three systems, ONC, NGC 6231 and NGC 2362, with mild or no expansion. In the ONC, velocity is approximately constant with radius, and for NGC 6231 and NGC 2362 velocity decreases with radius. In the plots for these three systems, we use the cluster core radius $r_c$ as a length scale, in order to better compare with theoretical models for gravitationally bound clusters. The core radii were measured by \citet{1998ApJ...492..540H}, \citet{2017AJ....154..214K}, and \citet{2014ApJ...787..107K} for these three clusters, updated with the new distance estimates (Section~\ref{bound.sec}). For the ONC and NGC 2362, velocity is nearly constant out to a radius of 8 times $r_c$, while for NGC 6231 it decreases steeply with radius -- by a factor of $\sim$2 at a distance of 4 $r_c$ from the center (Figure~\ref{sigma_rad.fig}).

Given the lack of strong expansion of the ONC, NGC~6231, and NGC~2362, these may be some of the best candidates in our sample for gravitationally bound clusters, and it may be useful to compare them to commonly used cluster distribution functions. For the isothermal sphere model, velocity dispersion is independent of position throughout the cluster. On the other hand, for the Plummer sphere and the lowered isothermal sphere models (also known as King models), velocity dispersions decrease monotonically with radius \citep{2008gady.book.....B}. In Plummer spheres the velocity dispersion at a point at radius $r$ is proportional to $(1+r/r_c)^{-1/4}$. The family of King models is characterized by a parameter $W_0$ describing the central potential \citep{1966AJ.....71...64K}, and curves of one-dimensional velocity dispersion as a function of projected radius are shown by \citet[][their Figure 4.11]{2008gady.book.....B} for several values of $W_0$.  For the ONC and NGC 2362, their velocity profiles would be consistent with isothermal spheres or King models with $W_0 \geq 9$. However, the Plummer model is rejected. For NGC 6231, the velocity dispersion is consistent with the Plummer model or King models with $3<W_0<6$. 

\section{Subcluster Motions}\label{subcluster_motions.sec}

Many of the stellar complexes from the MYStIX and SFiNCs studies contain subclusters that have been delineated by \citet{2014ApJ...787..107K} and \citet{2017ApJS..229...28G}. The clearest examples that have good {\it Gaia} data include NGC~2264, Cep~OB3b, NGC~6530, the Rosette Nebula, NGC~6357, NGC~6611, and the Carina OB1 association. For this subset, we now examine the kinematics of the substructures relative to each other. Subclusters with insufficient {\it Gaia} data ($<$10 stars) are omitted.

Table~\ref{subclusters.tab} gives the properties of subclusters, including subcluster centers from earlier studies\footnote{In some case, the subcluster centers may not be perfectly centered among the stars seen by {\it Gaia}. This is an effect of differential absorption across a subcluster, limiting {\it Gaia}'s sensitivity in the high-extinction part.} and the kinematics properties derived from {\it Gaia}. We find the bulk motions of subclusters by calculating the weighted median velocities of their stellar members. In a few cases, subclusters have been combined when overlapping subclusters represent core--halo structures \citep{2017AJ....154..214K} -- these are indicated in Table~\ref{subclusters.tab}. The projected subcluster velocities, relative to the association rest frame, range up to $\sim$8~km~s$^{-1}$, with a median value of $\sim$2~km~s$^{-1}$ and an interquartile range of 0.9--3.5~km~s$^{-1}$. The velocity dispersions of subclusters can be quite different in different regions. The clusters in the Carina OB1 association have relative velocities of 5--8~km~s$^{-1}$, while the various subclusters in NGC~2264 have relative velocities of 0--2~km~s$^{-1}$.

\begin{figure*}[t]
\centering
\includegraphics[width=0.45\textwidth]{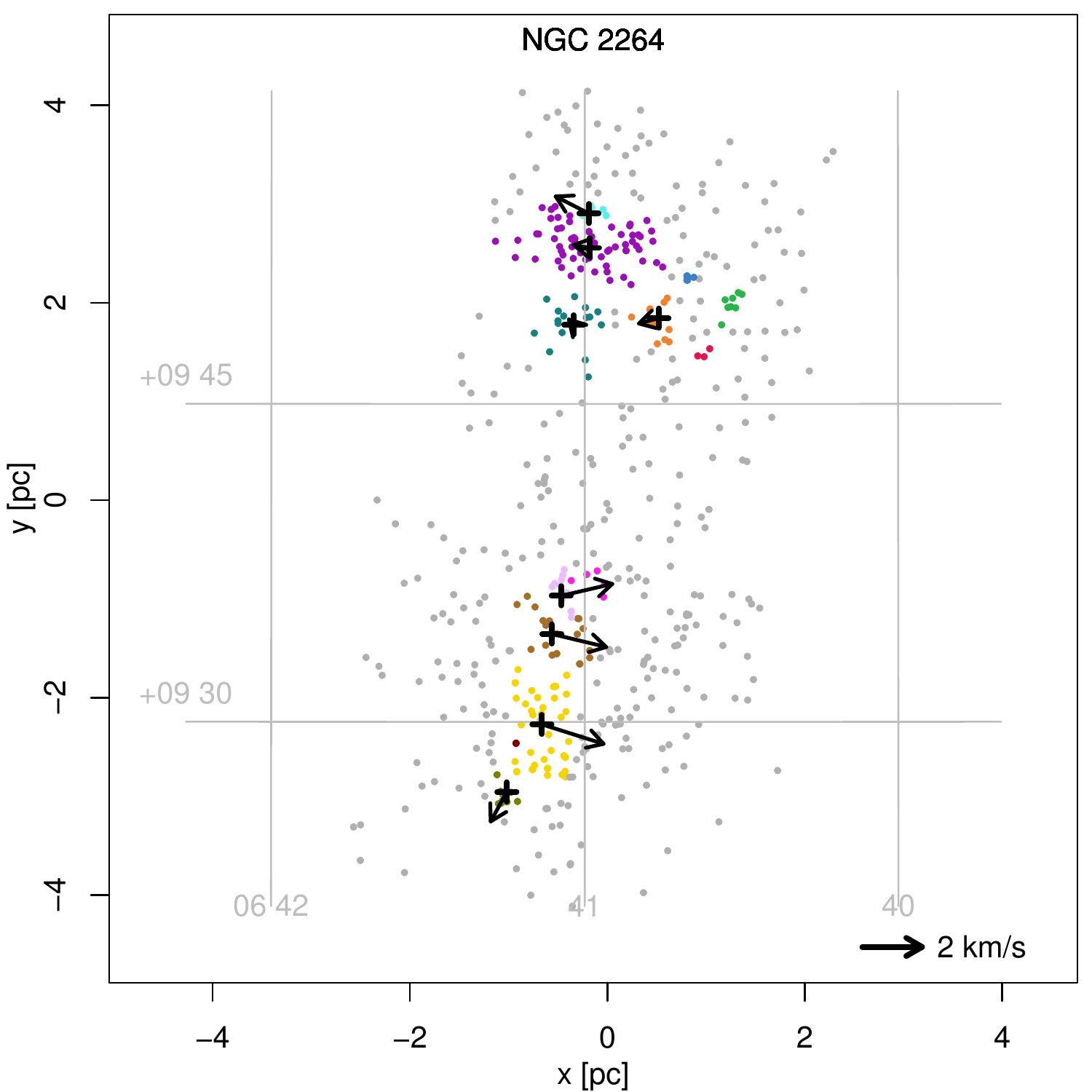} 
\includegraphics[width=0.45\textwidth]{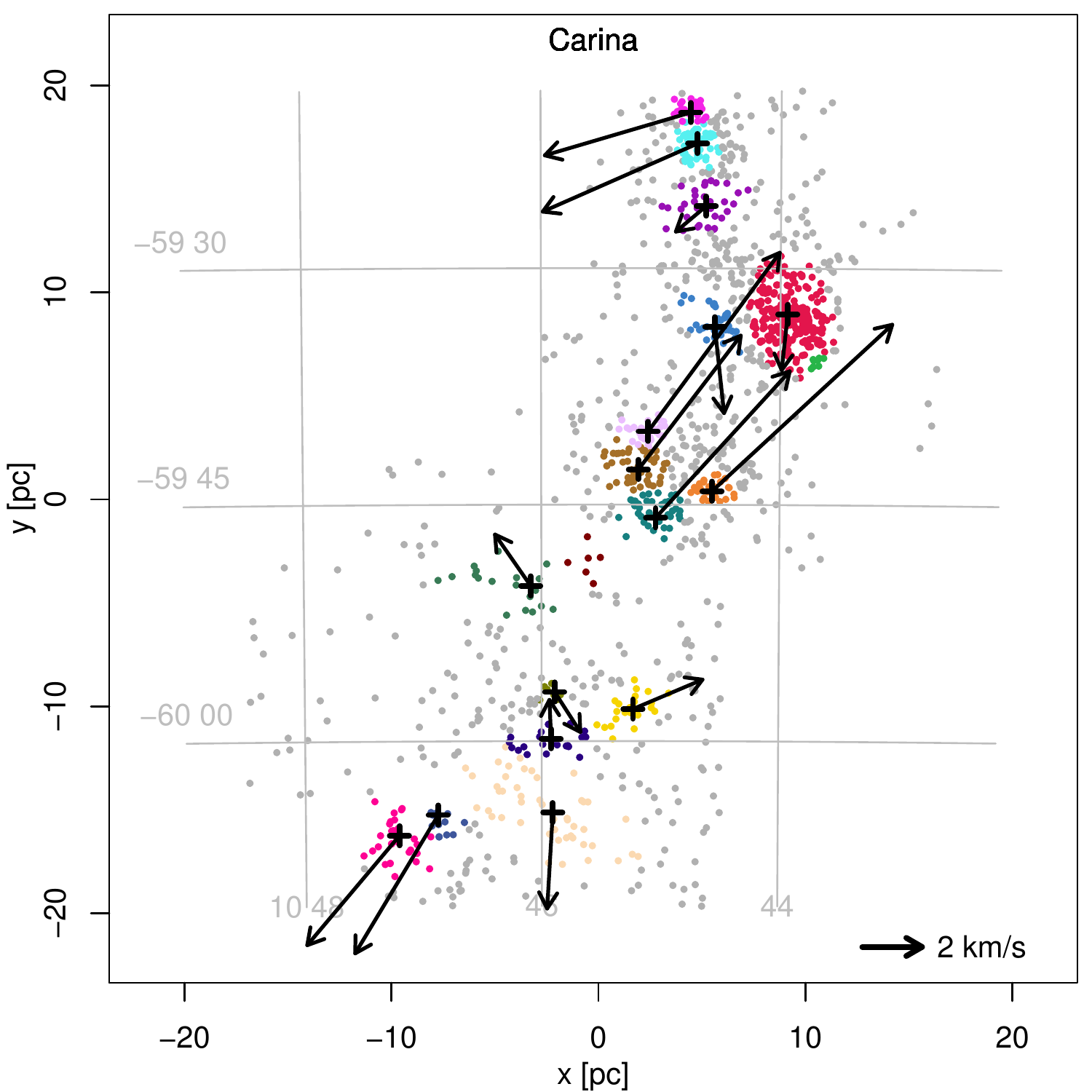} 
\caption{Kinematics of subclusters in NGC~2264 (left) and the Carina OB1 association (right). Stars included in the study are marked with a color symbol indicating the subcluster to which they were assigned in \citet{2014ApJ...787..107K}. The crosses mark subcluster centers and the vectors indicate velocities of the subclusters, as indicated by the velocity scale. Subcluster velocities in Carina tend to be much larger than in the smaller, nearby NGC~2264 region. In both NGC~2264 and Carina, nearby groups of stars tend to move in similar directions, but there is no overall sign of subcluster mergers.  Plots for other regions are included in an online figure set.
 \label{subcluster.fig}}
\end{figure*}

\begin{figure*}[t]
\centering
\includegraphics[width=0.3\textwidth]{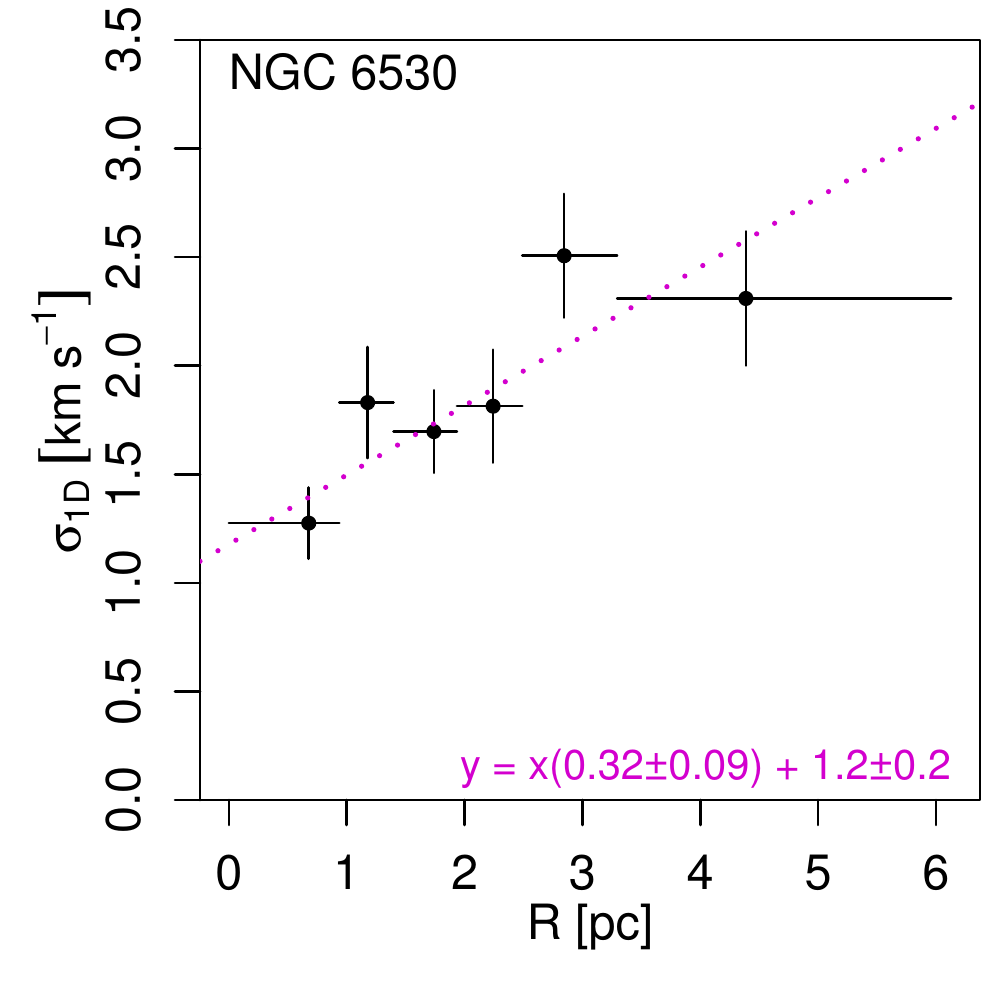} 
\includegraphics[width=0.3\textwidth]{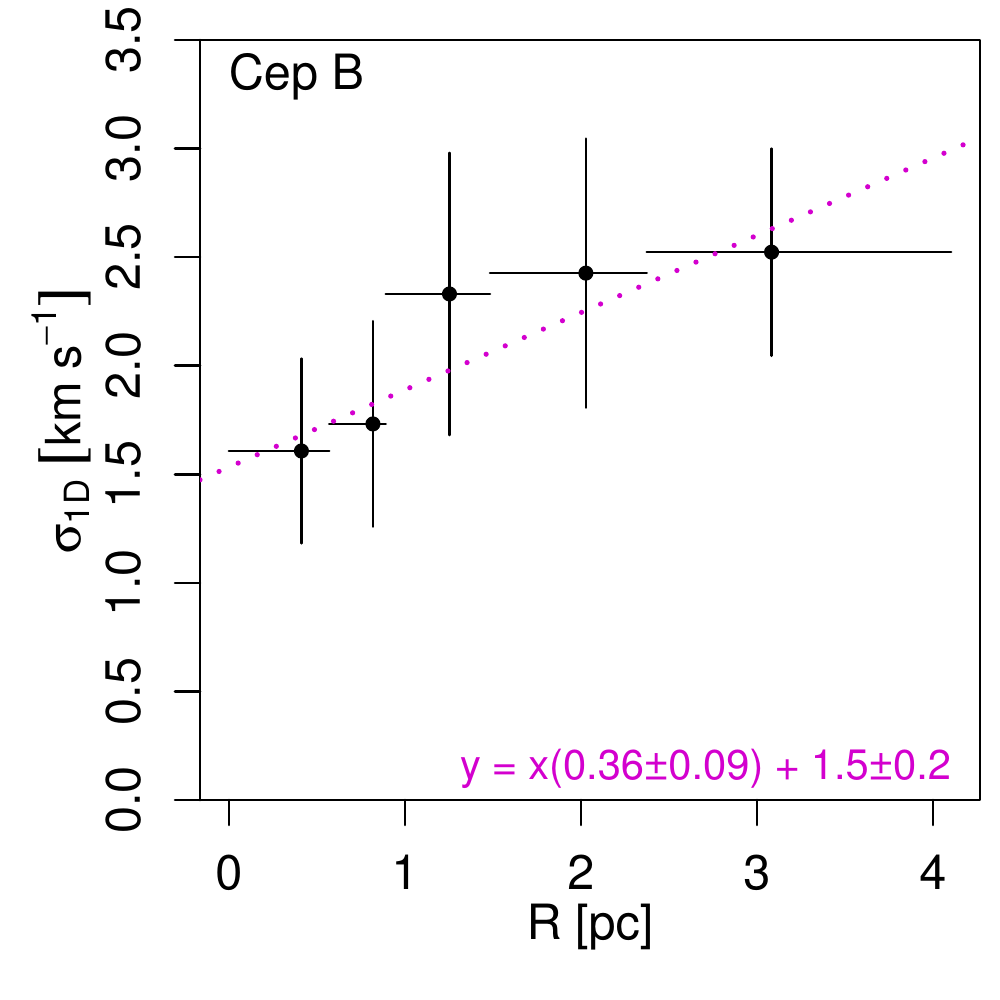} \\
\includegraphics[width=0.3\textwidth]{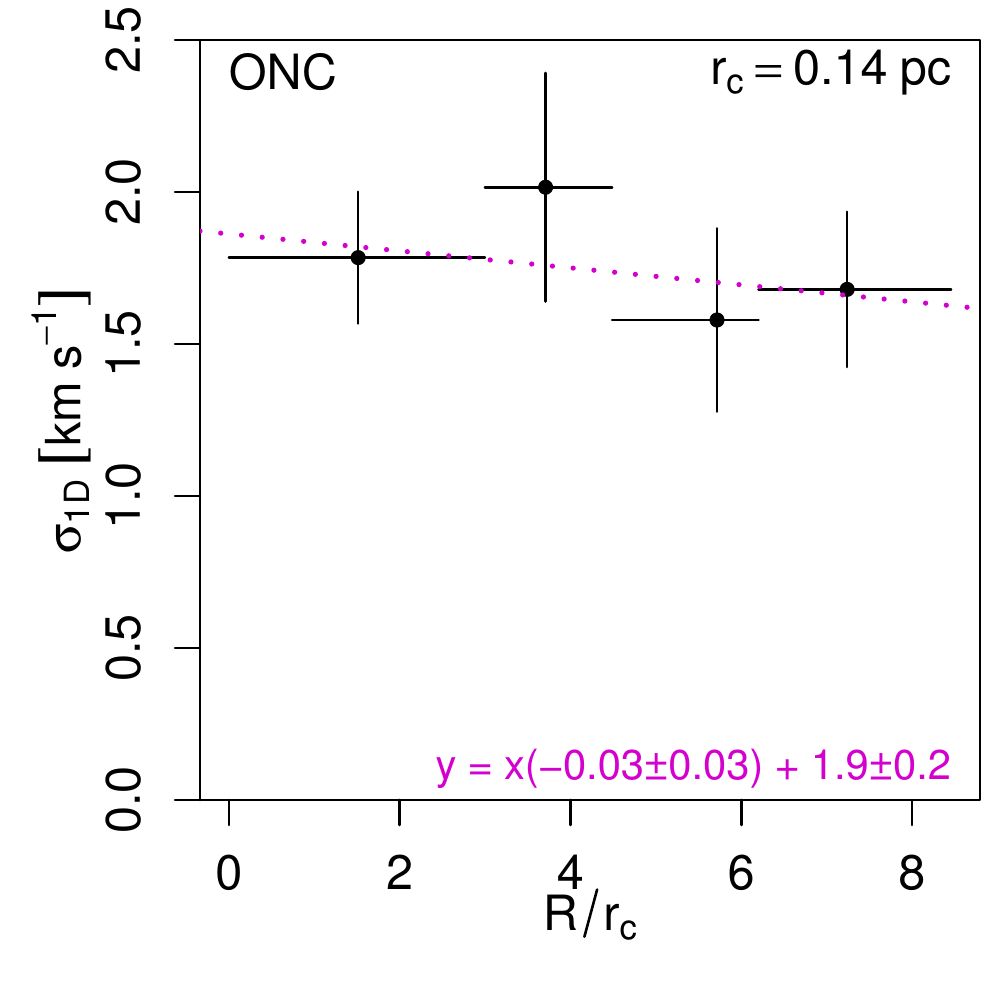} 
\includegraphics[width=0.3\textwidth]{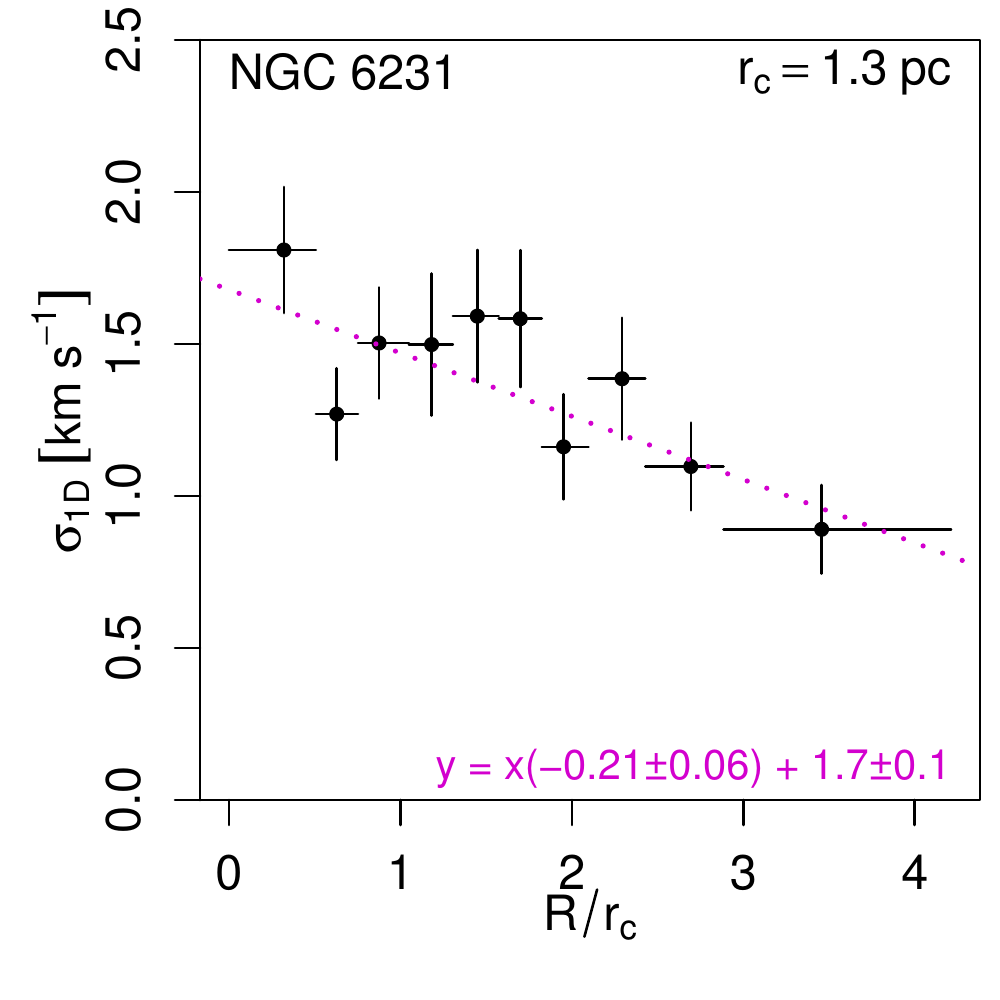} 
\includegraphics[width=0.3\textwidth]{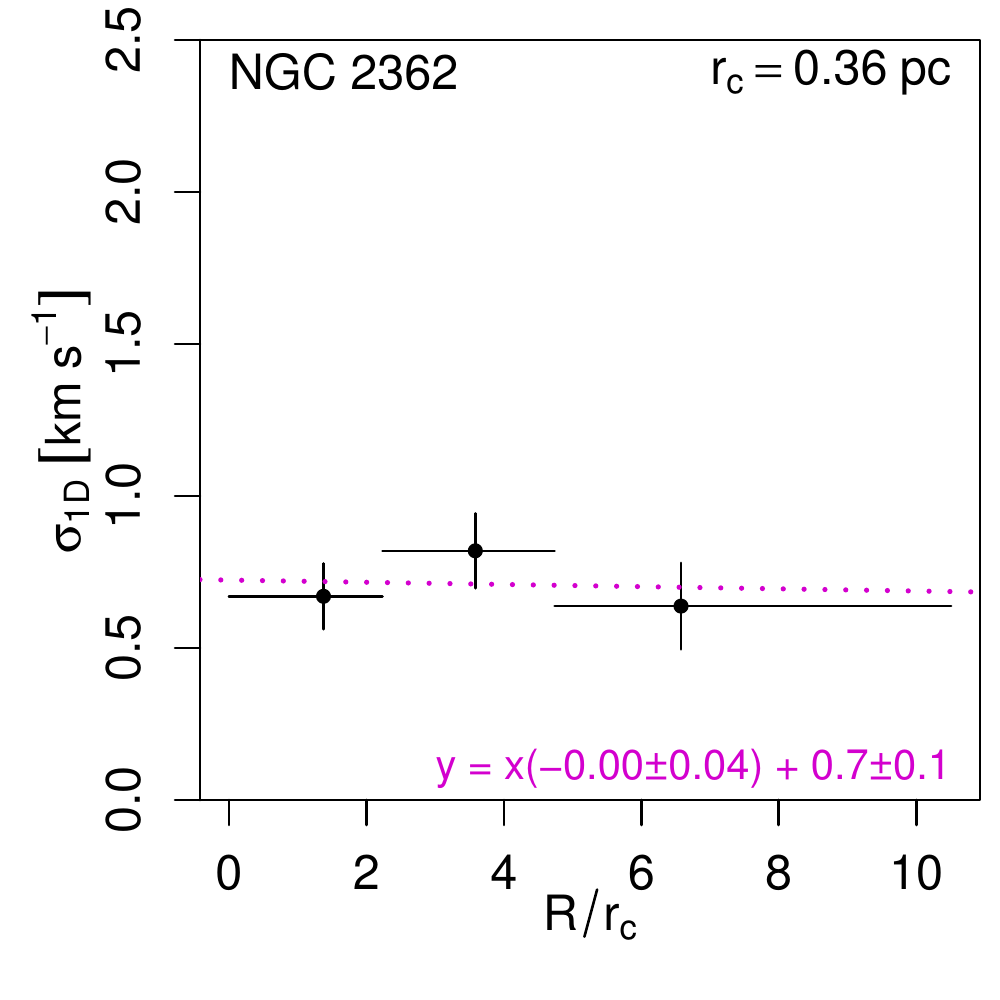} 
\caption{Velocity dispersion as a function of distance from the cluster center. In each panel, the points show the velocity dispersion in each radial bin, the error bars show 1$\sigma$ uncertainties. The weighted least-squares regression line for $y$ as a function of $x$ is shown for each plot (dotted magenta line) and the equation is given in the lower right corner. Top row: Plots for the expanding systems NGC~6530 and Cep~B. Bottom row: Plots for systems with mild (ONC) or no (NGC~6231 and NGC~2362) expansion. For the bottom row, the $x$-axis is normalized by the cluster core radius $r_c$ (Section~\ref{physical_properties.sec}) to facilitate comparison with cluster models. 
 \label{sigma_rad.fig}}
\end{figure*}

Figure~\ref{subcluster.fig} shows spatial maps of the stars assigned to each subcluster along with the velocity vectors of each group. In general, the subcluster motions are not convergent, but appear either randomly oriented or divergent. This pattern is seen in almost every star-forming region investigated, ranging from well-delineated clusters in regions like NGC~6357 to clumpy distributions of stars in regions like NGC~2264. 
The main exception is in Carina where the motions of Tr~14 and Tr~16 are directed towards each other, but the apparent convergence of these clusters could be a chance alignment.

\begin{figure*}[t]
\centering
\includegraphics[width=0.45\textwidth]{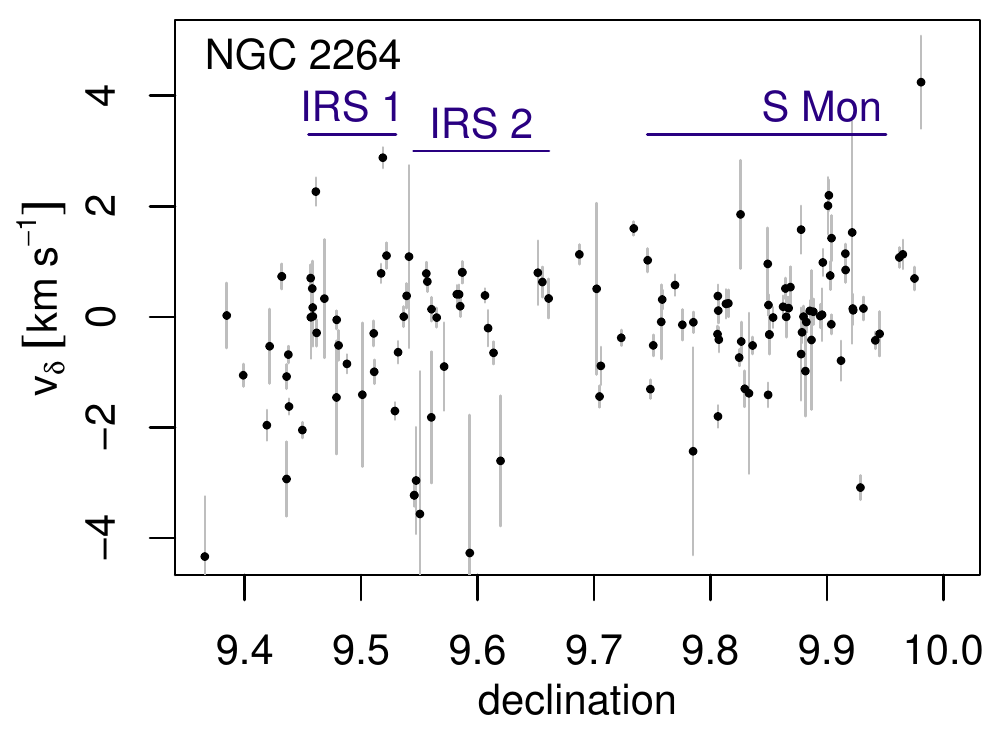} 
\includegraphics[width=0.45\textwidth]{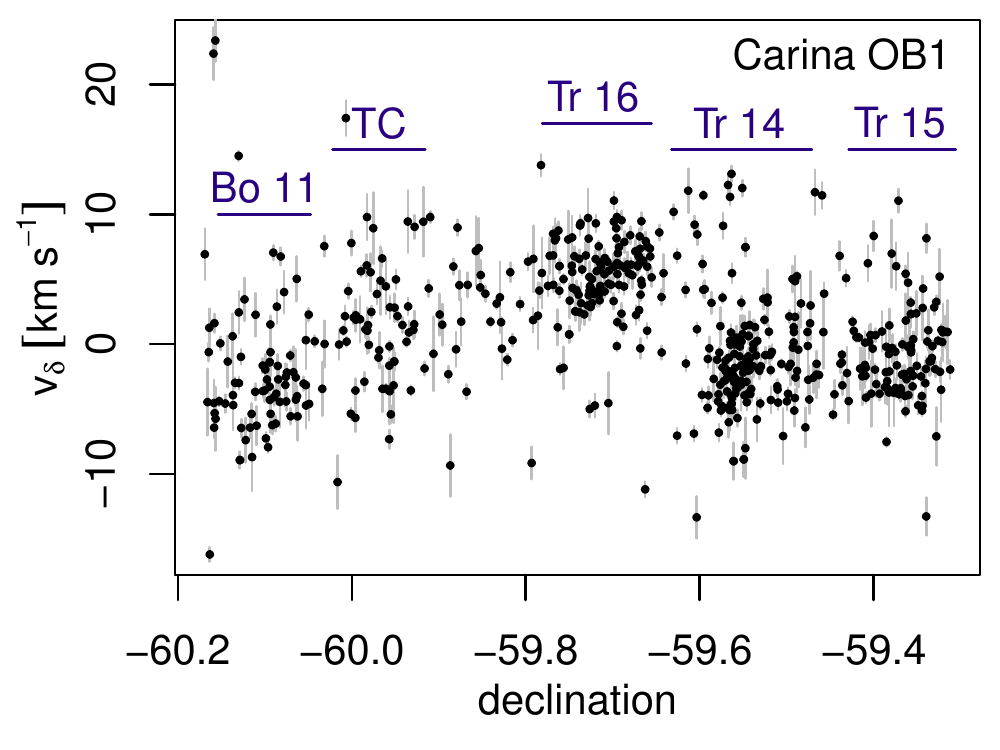} 
\caption{Velocity $v_\delta$ vs.\ declination for stars in NGC~2264 (left) and the Carina OB1 association (right). The declination range of several constituent clusters are marked, including S~Mon, IRS~1 and IRS~2 in NGC~2264, and Tr~14, 15, 16, the Treasure Chest (TC) and Bochum~11 (Bo~11) in Carina. The plot reveals different mean velocities of different clusters as well as cluster expansion within some individual clusters. Note that the velocity range for NGC~2264 is much smaller than for Carina.    Only sources with no astrometric excess noise are used.
 \label{carina_ob1.fig}}
\end{figure*}

The contrast between internal cluster velocities and global kinematics of a complex can be clearly seen in the Carina~OB1 association, which contains several clusters, including Tr~14/15/16 (included in this study) as well as Bochum~11 and the Treasure Chest to the south. 
Figure~\ref{carina_ob1.fig} (right) shows star positions in declination plotted against their $v_\delta$ velocity component. In this complex, the individual clusters are shifted with respect to one another in velocity, but internal velocity dispersions within the individual clusters can also be seen. 
In the south of Carina~OB1, there is a velocity gradient stretching from Bochum~11 to the Treasure Chest 
to Tr~16, while Tr~14 and Tr~15, to the north, have significantly different motions than the clusters to the south. Within Tr~14 and Tr~16, the positive correlation between declination and $v_\delta$ characteristic of an expanding cluster can be seen. The total velocity difference between different clusters is significantly larger than the velocity dispersions within the clusters. 
In contrast, in the lower-mass NGC~2264 region (left panel in Figure~\ref{carina_ob1.fig}), small differences can be seen in the velocities of the different subgroups, but the magnitudes of these differences and the total velocity dispersions are much smaller. 

We find no evidence for hierarchical assembly of rich clusters from subclusters in our sample.  Evidence for this process would be converging motions of subclusters. This failure is expected, as {\it Gaia} does not provide access to the youngest embedded subclusters, but restricts our analysis to older clusters where the molecular cloud is at least partially dispersed. Thus, hierarchical cluster assembly, if it occurs, must occur promptly when a cluster is still embedded and must involve subclusters separated by $<$5~pc and ages $<$1 Myr. 

Subcluster motions are linked to the large-scale kinematics in molecular clouds, which may include effects of supersonic turbulence \citep{1981MNRAS.194..809L, 2004RvMP...76..125M, 2002ApJ...576..870P, 2005ApJ...630..250K}, free-fall velocities of collapsing clouds \citep{2017MNRAS.467.1313V}, and/or accretion of material onto molecular clouds \citep{2014ApJ...780...36F,2017ApJ...850...62I}. The systems observed today at ages 1-5 Myr were formed in different dense molecular cores at widely separated portions in giant molecular clouds.  It is therefore natural that they inherit the motion of their natal cloud cores, and exhibit spatially correlated but essential random motions with respect to each other. 

%\clearpage\clearpage

%\startlongtable
%\onecolumngrid
\begin{deluxetable}{lrrrrrr}
\tablecaption{Relative Subcluster Kinematics \label{subclusters.tab}}
\tabletypesize{\tiny}\tablewidth{0pt}%\rotate
\tablehead{
\colhead{Region} & \colhead{Subcluster}& \colhead{$\alpha_0$} &\colhead{$\delta_0$}  & \colhead{$n_{samp}$}  & \colhead{$\langle v_x \rangle$} & \colhead{$\langle v_y \rangle$}\\
\colhead{} & \colhead{}  & \colhead{(J2000)} &\colhead{(J2000)} & \colhead{stars} & \colhead{km~s$^{-1}$} & \colhead{km~s$^{-1}$}\\
\colhead{(1)} & \colhead{(2)} & \colhead{(3)} &\colhead{(4)} & \colhead{(5)} & \colhead{(6)} & \colhead{(7)}
}
\startdata
Rosette  &  A  &   6 30 57.1& +04 57 57  &  26  &  0.5$\pm$0.5  &  $-$1.2$\pm$0.7  \\ 
 ---  &  C  &   6 31 32.0 &+04 50 58  &  12  &  1.3$\pm$1.2  &    0.0$\pm$1.0  \\ 
 ---  &  D+E  &   6 31 59.3& +04 54 50  &  241  &  0.5$\pm$0.3  &  0.5$\pm$0.3  \\ 
 ---  &  H  &   6 33  07.2& +04 46 57  &  17  &  0.3$\pm$2.0  &  $-$0.8$\pm$0.5  \\ 
 ---  &  L  &   6 34 10.7 &+04 25  06  &  50  &  $-$0.5$\pm$0.4  &  $-$0.6$\pm$1.0  \\ 
NGC 2264  &  D  &   6 40 45.8& +09 49  03  &  11  &  $-$0.7$\pm$0.2  &  $-$0.2$\pm$0.6  \\ 
 ---  &  E  &   6 40 59.1& +09 52 22  &  67  &  $-$0.5$\pm$0.4  &  0.09$\pm$0.1  \\ 
 ---  &  F  &   6 40 59.2& +09 53 59  &  10  &  $-$1.1$\pm$0.5  &  0.6$\pm$0.3  \\ 
 ---  &  H  &   6 41  02.1& +09 48 44  &  16  &  $-$0.2$\pm$1.0  &  0.2$\pm$0.6  \\ 
 ---  &  I  &   6 41  04.5& +09 35 57  &  13  &  1.7$\pm$0.7  &  0.4$\pm$0.2  \\ 
 ---  &  J  &   6 41  06.3 &+09 34  09  &  20  &  1.8$\pm$0.2  &  $-$0.4$\pm$1.0  \\ 
 ---  &  K  &   6 41  08.2 &+09 29 53  &  43  &  2.1$\pm$0.6  &  $-$0.7$\pm$0.4  \\ 
 ---  &  M  &   6 41 14.9& +09 26 42  &  11  &  $-$0.5$\pm$0.8  &  $-$1.0$\pm$0.4  \\ 
Carina  &  B+C  &  10 43 56.4& $-$59 32 54  &  262  &  $-$0.2$\pm$0.4  &  $-$1.9$\pm$0.2  \\ 
 ---  &  D  &  10 44 32.9& $-$59 33 42  &  36  &  0.3$\pm$0.9  &  $-$2.9$\pm$0.9  \\ 
 ---  &  E  &  10 44 34.0& $-$59 44  08  &  31  &  6.0$\pm$0.7  &    5.6$\pm$1.0  \\ 
 ---  &  F  &  10 44 37.4 & $-$59 26  03  &  34  &  $-$1.0$\pm$0.9  &  $-$0.9$\pm$0.8  \\ 
 ---  &  H  &  10 44 41.8& $-$59 22  05  &  59  &  $-$5.2$\pm$0.7  &  $-$2.3$\pm$0.6  \\ 
 ---  &  I  &  10 44 45.3 & $-$59 20  07  &  24  &  $-$4.9$\pm$0.4  &  $-$1.4$\pm$1.4  \\ 
 ---  &  J  &  10 45  02.4 & $-$59 45 50  &  41  &  4.5$\pm$1.1  &    4.9$\pm$0.7  \\ 
 ---  &  K  &  10 45  06.2& $-$59 40 21  &  30  &  4.4$\pm$0.5  &  6.0$\pm$0.5  \\ 
 ---  &  L  &  10 45 11.1 & $-$59 42 46  &  51  &  3.4$\pm$0.3  &  4.5$\pm$0.6  \\ 
 ---  &  M  &  10 45 13.7& $-$59 57 58  &  23  &  2.3$\pm$1.3  &    1.0$\pm$1.0  \\ 
 ---  &  O  &  10 45 53.4& $-$59 56 53  &  11  &  0.9$\pm$1  &  $-$1.3$\pm$0.7  \\ 
 ---  &  P  &  10 45 54.4& $-$60 04 32  &  47  &  $-$0.2$\pm$1.6  &  $-$3.2$\pm$1.3   \\ 
 ---  &  Q  &  10 45 55.2 & $-$59 59 51  &  23  &  $-$0.1$\pm$1.0  &    1.3$\pm$1.1  \\ 
 ---  &  R  &  10 46  05.4& $-$59 50  09  &  19  &  $-$1$\pm$1  &  1.7$\pm$0.9  \\ 
 ---  &  S  &  10 46 52.7 & $-$60 04 40  &  13  &  $-$2.8$\pm$0.7  &  $-$4.6$\pm$1.5  \\ 
 ---  &  T  &  10 47 12.5 & $-$60 05 58  &  33  &  $-$3.1$\pm$0.6  &  $-$3.7$\pm$0.8  \\ 
NGC 6357  &  A  &  17 24 43.7& $-$34 12  07  &  34  &  $-$1.1$\pm$0.5  &  3.1$\pm$0.5  \\ 
 ---  &  B  &  17 24 46.7& $-$34 15 23  &  23  &  $-$1.6$\pm$0.9  &  1.3$\pm$0.7  \\ 
 ---  &  C+D  &  17 25 34.3 & $-$34 23 10  &  27  &  1.0$\pm$0.6  &    2.4$\pm$1.2  \\ 
 ---  &  E  &  17 25 47.9& $-$34 27 12  &  11  &  $-$4.1$\pm$2.1  &  $-$2.7$\pm$0.8  \\ 
 ---  &  F  &  17 26  02.2 & $-$34 16 42  &  21  &  1.8$\pm$1.1  &  $-$1.0$\pm$0.8  \\ 
NGC 6530  &  A  &  18  03 23.8& $-$24 15 19  &  15  &  $-$3.4$\pm$1.2  &    1.9$\pm$0.8  \\ 
 ---  &  C  &  18  03 46.3 & $-$24 22  01  &  14  &  2.2$\pm$3.3  &  $-$1.9$\pm$0.9  \\ 
 ---  &  D  &  18  03 51.3 & $-$24 21  08  &  11  &  4.1$\pm$2.1  &  $-$1.1$\pm$0.8  \\ 
 ---  &  E  &  18  04  07.6 & $-$24 25 53  &  55  &  $-$0.9$\pm$1.1  &  $-$1.8$\pm$0.5  \\ 
 ---  &  F  &  18  04 13.3 & $-$24 18 27  &  173  &  1.5$\pm$0.4  &  1.1$\pm$0.3  \\ 
 ---  &  G  &  18  04 20.1 & $-$24 22 51  &  14  &  0.6$\pm$1.0  &  0.6$\pm$1.0  \\ 
 ---  &  H  &  18  04 23.3 & $-$24 21 13  &  65  &  0.03$\pm$0.4  &  0.8$\pm$0.2  \\ 
 ---  &  I  &  18  04 28.3 & $-$24 22 46  &  110  &  0.3$\pm$0.2  &  $-$0.5$\pm$0.5  \\ 
 ---  &  J  &  18  04 39.6 & $-$24 23 20  &  31  &  $-$0.7$\pm$0.5  &  0.04$\pm$0.3  \\ 
 ---  &  K  &  18  04 50.5 & $-$24 26 19  &  45  &  $-$2.3$\pm$0.9  &  $-$0.1$\pm$0.8  \\ 
NGC 6611  &  A+B  &  18 18 42.2 & $-$13 47  03  &  213  &  0.5$\pm$0.3  &  $-$0.1$\pm$0.3  \\ 
 ---  &  D  &  18 18 57.3 & $-$13 45 23  &  71  &  $-$1.7$\pm$0.4  &  0.9$\pm$0.7  \\ 
Cep OB3b  &  A  &  22 53 47.1& +62 35 47  &  195  &  2.2$\pm$0.4  &  $-$0.9$\pm$0.4  \\ 
 ---  &  B  &  22 54 58.4& +62 34  09  &  23  &  $-$0.1$\pm$0.5  &  $-$0.5$\pm$0.4  \\ 
 ---  &  C  &  22 56 40.3 &+62 42  06  &  401  &  $-$0.6$\pm$0.3  &  0.2$\pm$0.3  \\ 
\enddata
\tablecomments{Column~1: The names of the star-forming regions. Column~2: The names of the subclusters defined by \citet{2014ApJ...787..107K} and \citet{2018arXiv180405075G}. Columns~3--4: Coordinates of the subcluster centers. Column~5: The number of {\it Gaia} sources in each subcluster. Column~6--7: Subcluster velocity projected in the plane of the sky relative to the center-of-mass rest frame of the entire association.}
\end{deluxetable}

\section{Relationships Between Kinematics and Other Properties of Stellar Systems}\label{physical_properties.sec}

Physical properties of stellar systems, including their masses, sizes, and ages may be linked to their internal kinematics. These links may arise from stellar dynamics or through the star formation process. For example, the relationship between cloud size and cloud velocity dispersion \citep{1981MNRAS.194..809L} could yield a relationship between the mass of the resulting stellar system and its velocity dispersion.

We estimate characteristic masses, sizes, and ages for objects in our sample using the data and methods from the previous MYStIX/SFiNCS/NGC6231 studies, but updating the values with new distance estimates. Estimates of system mass, $M_{cl}$ (corrected for incompleteness), are calculated using the methods from \citet{2015ApJ...802...60K} -- the X-ray luminosity function for pre--main-sequence stars is used to extrapolate the completeness fraction and star counts are converted to masses using a mean mass of $\bar{m} = 0.61$~$M_\odot$ per star based on the \citet{2013MNRAS.429.1725M} initial mass function (IMF). \citet{2015ApJ...802...60K} estimated typical uncertainties of 0.25~dex on masses estimated with this method. Median ages for systems are calculated using the $Age_{JX}$ method from Getman et al. (2014). Half-mass radii, $r_{hm}$, are calculated by taking the median distance of stars in our samples from the center of the system. However, in the case of the ONC where our sample truncates the outer region of the cluster, we adopt the half-mass radius of $0.9$~pc from \citet{2014ApJ...795...55D}.

\begin{figure*}[t]
\centering
\includegraphics[width=1.0\textwidth]{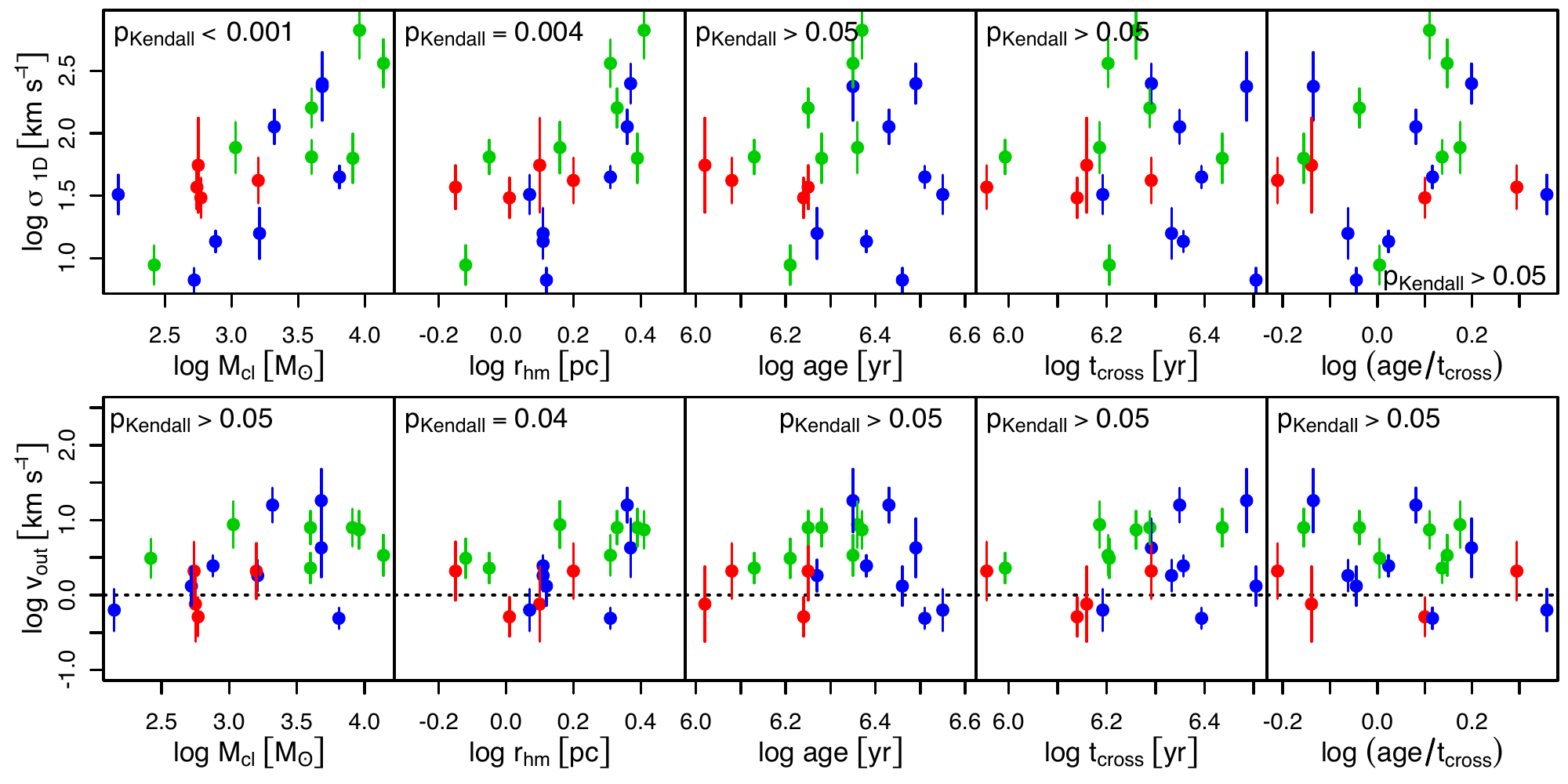} 
\caption{Scatter plots showing relationships between cluster kinematics and other physical cluster properties for the subset of stellar systems included in Table~\ref{velocities.tab}. Kinematics quantities include velocity dispersion  ($\sigma_{1D}$) and expansion velocity (median~$v_\mathrm{out}$), and other physical properties include system mass ($M_{cl}$), size ($r_{hm}$), age, crossing time ($t_\mathrm{cross}$), and the ratio of age to crossing time. Systems are color coded based on properties of the natal cloud: embedded (red), partially embedded (green), and revealed (blue). Statistical significance of correlation is assessed using the Kendall rank correlation test \citep{hollander1973nonparametric}, and the $p$-values are shown on the plot. Positive correlations exist between velocity dispersion, mass, and size, but no statistically significant correlations can be found for other properties. Data behind the figure are provided by the online version of this article.
 \label{explanatory.fig}}
\end{figure*}

These estimates may be subject to a variety of systematic uncertainties, whose effect is difficult to determine. For example, mass and age estimates can be affected by systematic errors in inference of stellar properties and choice of model isochrones \citep{2018MNRAS.477.5191R}. Furthermore, $M_{cl}$ and $r_{hm}$ could be underestimated due to difficulties determining the outer boundaries of clusters and the finite fields of view in the MYStIX/SFiNCS/NGC6231 studies.\footnote{See \citet{2017AJ....154..214K} for a detailed discussion of challenges involved in obtaining $M_{cl}$ and $r_{hm}$. } The derived physical properties of systems are shown in Figure~\ref{explanatory.fig} and the values used here are available from ``data behind the figure'' in the online version of this article.

The gravitational effects of the natal molecular clouds are also likely to affect young stellar systems. The systems in our sample are in various stages of gas dispersal. Systems are considered to be embedded when the stars are reddened and projected on the molecular cloud and revealed when there is little extinction from the cloud. Partially embedded clusters are an intermediate stage, and often represent systems at the edge of a cloud or within a bubble. The geometry of the system and projection effects may affect how systems are classified in ambiguous cases.

Figure~\ref{explanatory.fig} shows the relationships between kinematic properties ($\sigma_{1D}$ and median~$v_\mathrm{out}$) and $M_{cl}$, $r_{hm}$, and age. Points are color-coded by degree of ``embeddedness.'' Unsurprisingly, embedded systems tend to be younger while revealed systems tend to be older. There is no statistically significant relation between $\sigma_{1D}$ and embedded state. However, there is a statistical difference ($p_{KS} < 0.01$) between $\mathrm{median}~v_\mathrm{out}$ of embedded systems, which tend to have no expansion, and partially embedded/revealed systems, which tend to be expanding.

Statistically significant correlations are found between velocity dispersion and both $M_{cl}$ and $r_{hm}$ using Kendall's rank correlation test ($p < 0.01$). However, no statistical correlation is found between expansion velocity and the measured physical properties.  As mass and radius are themselves related to each other, multivariate regression analysis is needed to treat interdependencies.  We use the R package {\it relaimpo} \citep{gromping2006relative} to evaluate the importance of $\log M_{cl}$, $\log r_{hm}$ and $\log~\mathrm{age}$ for predicting $\log \sigma_{1D}$ and median~$v_\mathrm{out}$ in a linear regression.  This analysis identified $M_{cl}$ as the only important predictor of velocity dispersion and did not identify any of these variables as a statistically significant predictor of expansion. Although the relationship between system mass and velocity dispersion was expected, the result is interesting because it provides empirical evidence for the relationship based on independent estimates of system masses and velocity dispersions. 

The crossing time, defined as $t_\mathrm{cross}=2\,r_{hm}/\sigma_{1D}$, and the ratio of age to crossing time are two additional quantities that are important to stellar dynamics \citep{2008gady.book.....B}. 
Crossing times range from 0.4 to 3~Myr in our sample, while the ratios of system age to crossing time range from 0.6 to 3. This indicates that the systems are all dynamically very young, and would not have had time to relax through two-body interactions. Expansion can also drive the ratio of age to crossing time toward a small value, if a system expands at a rate proportional to its velocity dispersion. Surprisingly, there is no statistically significant correlation between either of these quantities and either velocity dispersion or expansion rate for the objects in our sample (Figure~\ref{explanatory.fig}). 

\subsection{Virial State}

The observed velocity dispersion of a system can be compared to the velocity dispersion needed for virial equilibrium $\sigma_{virial}$ to estimate whether a it is subvirial, virial, supervirial, or unbound. If $\sigma_{1D} > \sqrt{2}\sigma_{virial}$, the total energy of the system would be positive and the system would be unbound. Given a mass $M_{cl}$ and half-mass radius $r_{hm}$, the velocity dispersion of a virialized cluster is given by the equation 
\begin{equation}\label{virial_velocity.eqn}
\sigma_{virial} =  \left (\frac{G\,M_{cl}}{\eta\, r_{hm}} \right)^{1/2},
\end{equation}
where $G$ is the gravitational constant and $\eta$ is a constant that depends on the mass profile of a cluster. A Plummer model yields $\eta\approx 10$. Many young stellar clusters have $\eta < 10$ due to relatively broad den- sity profiles \citep[e.g.,][]{2010ARA&A..48..431P,2018MNRAS.475.3511G}.

\begin{figure*}[t]
\centering
\includegraphics[width=0.45\textwidth]{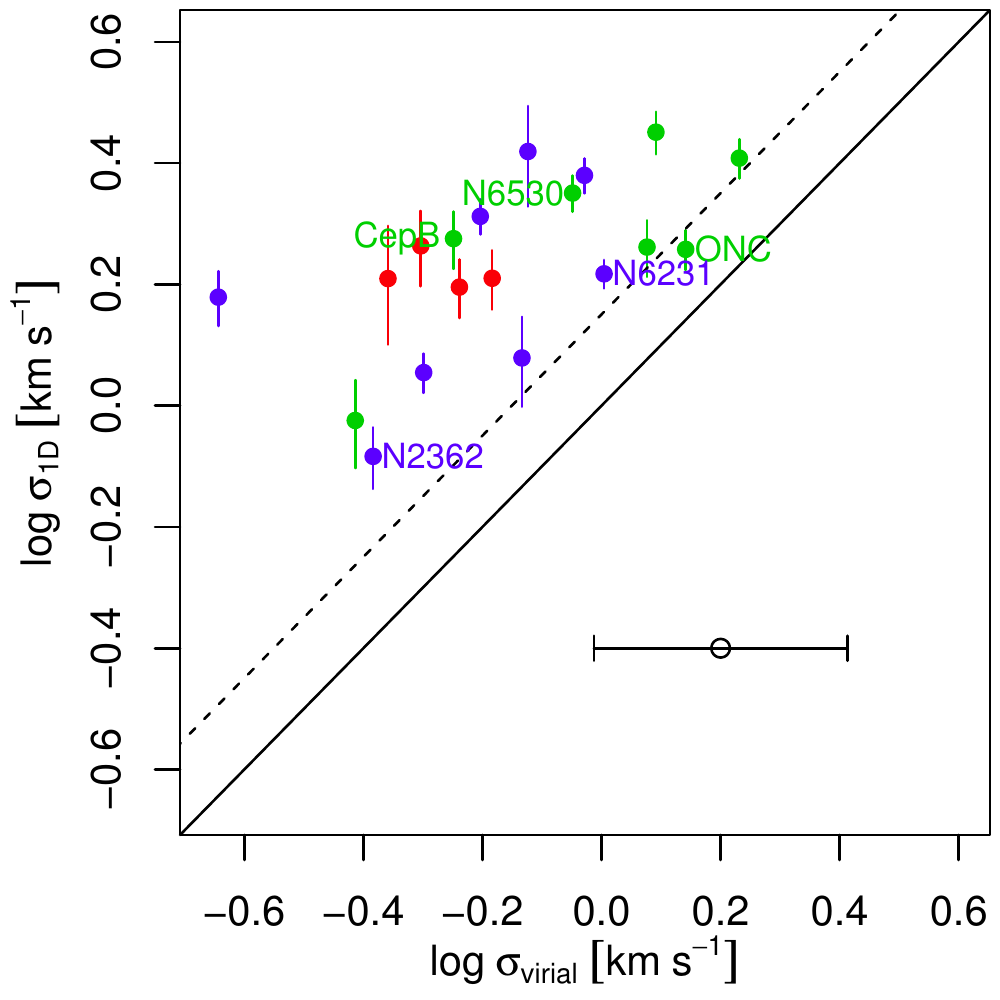} 
\caption{
Virial velocity dispersion calculated using observed $M_{cl}$ and $r_{hm}$ versus observed velocity dispersion. Two non-expanding systems (NGC~6231 and NGC~2362), one mildly expanding system (ONC), and two expanding systems (NGC~6530 and Cep~B) are labeled. Colors of points indicate the degree of ``embeddedness'' as in Figure~\ref{explanatory.fig}. The solid line shows the expected relationship for a cluster in virial equilibrium, while the dashed line shows the limit for a bound cluster. The values of $\sigma_{virial}$ are calculated assuming $\eta=10$ in Equation~\ref{virial_velocity.eqn}. The large error bar shows how a factor of 2 systematic uncertainty on mass or radius would affect $\sigma_{virial}$.
 \label{virial_vs_obs.fig}}
\end{figure*}

Figure~\ref{virial_vs_obs.fig} shows $\sigma_{1D}$ vs.\ $\sigma_{virial}$. There is a clear positive correlation between these two quantities ($p_{Kendall} < 0.001$). For the assumption that $\eta = 10$  (shown in the plot), all systems lie above the solid line showing the relationship for virial equilibrium, and most lie above the dashed line indicating zero total energy, which suggests that most of them are unbound. If we were to assume $\eta = 5$ instead, the ONC would be in approximate virial equilibrium, and several other systems, including Berkeley 59, NGC 2362, Tr~14, and NGC 6611, would have negative total energy, suggesting that they are likely bound. In addition, uncertainties on $M_{cl}$ and $r_{hm}$ could affect whether systems are in the ``bound'' or ``unbound'' regime of this plot. Although this plot is useful for demonstrating a statistical correlation between $\sigma_{1D}$ vs.\ $\sigma_{virial}$, in most cases whether or not a particular system is gravitationally bound remains ambiguous due to systematic uncertainties.

Two non-expanding systems (NGC 6231, and NGC 2362), one mildly expanding system (ONC), and two systems with clear expansion are labeled (NGC 6530 and Cep B) are labeled on the figure. The mildly/non-expanding systems lie along the bottom of the distribution, while the expanding systems lie toward the top, as would be expected if expansion is driven by systems being unbound. The ONC is within the region considered gravitationally bound, and NGC 6231 and NGC 2362 could be within this region given uncertainties in measurements or assumptions. At the other extreme, NGC 6530 and Cep B are sufficiently far from being in virial equilibrium that they can be classified as being unbound even in the presence of systematic uncertainties.

Given the large dynamical range in $M_{cl}$, which spans a factor of $\sim$100 in mass, it is notable that the relationship between $\sigma_{1D}$ and $\sigma_{virial}$ is relatively tight with only a factor of $\sim$1.5 scatter.

\subsection{Dynamical State of Expanding Systems}\label{dyn_expand.sec}

The two examples with the most statistically significant expansion are NGC~6530 and Cep~B, both of which have ``Hubble flow'' like expansion patterns (Figure~\ref{expand_rad.fig}) which hint at free expansion. They have one-dimensional velocity dispersions of 2.2$\pm$0.2~km~s$^{-1}$ and 1.9$\pm$0.2~km~s$^{-1}$ and approximate half-mass radii of $\sim$2~pc and $\sim$1.5~pc, respectively. For these values and the assumptions above, NGC~6530's virial mass would be $\sim$20,000~$M_\odot$ and Cep~B's virial mass would be $\sim$10,000~$M_\odot$. However, the estimated system masses are only $\sim$4000~$M_\odot$ and $\sim$1000~$M_\odot$ \citep{2015ApJ...802...60K}. Thus, both the gradient in expansion velocity and the inferred stellar populations indicate that these two systems are not gravitationally bound. This places these expanding associations near the top of the distribution on Figure~\ref{virial_vs_obs.fig}.

\begin{figure}[t]
\centering
\includegraphics[width=0.3\textwidth]{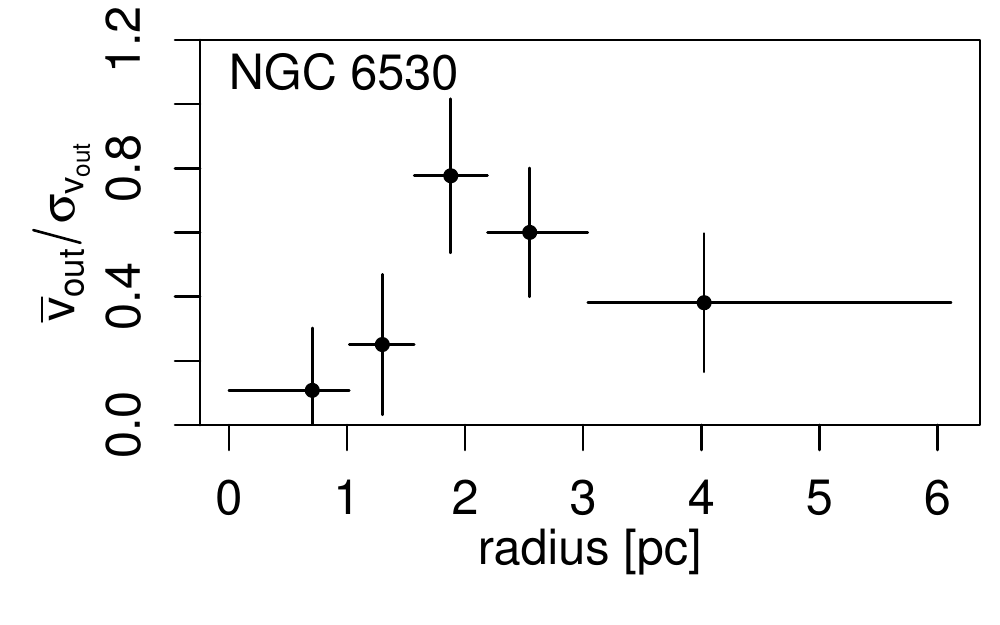} 
\caption{Ratio of expansion velocity to dispersion in $v_\mathrm{out}$ as a function of radius. Larger values indicate that velocity vectors are mostly directed outward, while smaller values indicate significant fractions of stars moving both outward and inward. 
 \label{sigmavout_rad.fig}}
\end{figure}

For a freely expanding association, stars will sort themselves by velocity as they move away from the center of the system, effectively decreasing the local velocity dispersion. This can be tested in  NGC~6530  by plotting the ratio of expansion velocity to the dispersion in $v_\mathrm{out}$ as a function of radius (Figure~\ref{sigmavout_rad.fig}). Values of this ratio $>$1 suggest that nearly all stars are moving outwards, while values $<$1 suggest some stars are moving inwards even as the system expands overall. In this case, it turns out that the expansion velocity is always less than or equal to the local velocity dispersion. 

Many of the expanding systems show internal substructure \citep{2014ApJ...787..107K}. We have not analyzed the kinematics within individual subclusters due to insufficient numbers of stars in our sample. In principal, it would be possible for subgroups of stars to be locally bound, even if the total energy of a region as a whole is positive. Future {\it Gaia} data releases are likely to provide more information about these groups due to larger samples of stars with good astrometry and higher overall precision.

\subsection{Dynamical State of Non-Expanding Systems}\label{bound.sec}

The three objects ONC, NGC~6231, and NGC~2362 may be the best candidates in our sample of 28 systems for being gravitationally bound clusters because the systems have little to no expansion and their surface density profiles are close to what would be expected for a system in virial equilibrium.

The core radius $r_c$ -- the radius where the apparent surface-density of stars drops by a factor of $\sim$2 -- can serve as a length scale with which to connect the spatial and kinematic cluster properties. 
For a dynamically relaxed cluster, with a surface density profile given by a King profile or an isothermal sphere, the density at the center of the clusters, $\rho_0$, is related to the cluster core radius, $r_c$, and the velocity dispersion, $\sigma_{1D}$, by the equation 
\begin{equation}
\rho_{0,{virial}} = \frac{9\sigma_{1D}^2}{4\pi G r_c^2},
\end{equation}
where $G$ is the gravitational constant \citep{2008gady.book.....B}.  For clusters in which the potential is dominated by the stars (versus needing to account for $M_\mathrm{stars}+M_\mathrm{gas}$), the mass density implied by the cluster dynamics can be compared to the observed cluster number density $n_0$, to infer the mean mass per star.  This, in turn, can be compared to the average mass per star predicted by a standard IMF. Adopting the \citet{2013MNRAS.429.1725M} form of the IMF over
the mass range 0.08--150~$M_\odot$ yields $\bar{m}=0.61$~$M_\odot$ for single stars or $\bar{m}=0.78$~$M_\odot$ assuming a population including unresolved binary systems.

The ONC is a smooth, centrally concentrated distribution of stars, which was modeled by \citet{1998ApJ...492..540H} using a King profile \citep{1966AJ.....71...64K,2008gady.book.....B} with a core radius of $\sim$0.15--0.20~pc.\footnote{The structure of the ONC core is more complex than accounted for by a single King profile. For example, there are small concentrations of stars with densities much greater than accounted for by the smooth King model \citep[e.g.,][]{1998AJ....116..322H,2013A&A...554A..48R,2014ApJ...787..107K}. We use the simpler model form from \citet{1998ApJ...492..540H} because it is easier to interpret dynamically.} Although the cluster is located within the Orion~A Cloud, the central region of the cluster is dominated by stars, not dense gas; an ionization front propagates back into the molecular cloud, located approximately 0.2 pc behind the cluster center, and a neutral gas ``lid'' is located $>$1 pc in the foreground \citep{2001ARA&A..39...99O}. 
\citet{1998ApJ...492..540H} used velocity measurements from \citet{1988AJ.....95.1755J} to compare the virial central density of the ONC to the observed number density of stars. We repeat this experiment, instead using the $\sigma_{1D}=1.8$~km~s$^{-1}$ from {\it Gaia} measurements and scaling their core radius to $r_c=0.14$~pc and their observed density of stars to $n_0=2.7\times 10^4$~stars~pc$^{-3}$ based on differences in distance assumptions. This gives $\rho_0=2.8\times 10^4$~$M_\odot$~pc$^{-3}=1.0~M_\odot\, n_0$. A value of 1.0~$M_\odot$ per star is slightly higher than the expected mean stellar mass, but some of the virial mass may be made up by accounting for some amount of remaining gas in the outer parts of the cluster. Thus, the ONC is close to virial equilibrium, although a slight discrepancy could account for the mild expansion. 

NGC~6231 and NGC~2362 are both larger, older, and less dense than the ONC. The large sizes suggest that they may have already expanded, but may have reached a turn-around radius where outward expansion has halted. 

For NGC~6231, values of $r_c=1.3$~pc and $n_0=180$~stars~pc$^{-3}$ were found by \citet{2017AJ....154..214K}, scaled for difference in assumed distances.\footnote{The estimation of the number of stars in the cluster (on which $n_0$ depends linearly) is based on extrapolation of the IMF to account for incompleteness in the observation and, thus, depends on assumptions of stellar age, pre--main-sequence isochrones, cluster distance, and the shape of the IMF. \citet{2017AJ....154..214K} estimated a total of 5700~stars down to 0.08~$M_\odot$ projected within the field of view, while \citet{2016A&A...596A..82D} estimated a similar cluster mass using different assumptions.} 
A velocity dispersion of $\sigma=1.6$~km~s$^{-1}$ values yield $\rho_{0,\mathrm{virial}}=260~M_\odot~\mathrm{pc}^{-3}=1.4~M_\odot\, n_0$, a slightly larger ratio than for the ONC. In this case, the cluster is completely devoid of molecular cloud, so gas mass cannot contribute significantly to the total mass. Assuming the average mass per star (or binary) is $\sim$0.6--0.8~$M_\odot$ this would place NGC~6231 near the threshold for being unbound, but whether the cluster is bound or unbound depends on the total mass of the cluster. 

For NGC~2362, values of $r_c=0.36$ and $n_0=600$~stars~pc$^{-3}$ were found by \citet{2015ApJ...812..131K}, scaled for difference in distance.\footnote{As for NGC~6231, the estimated density of stars at the center of NGC~2362 depends on model-based corrections for incompleteness in the observed sample.} A velocity dispersion of $\sigma=0.8$~km~s$^{-1}$ yields $\rho_{0,\mathrm{virial}}=870 ~M_\odot~\mathrm{pc}^{-3}=1.5~M_\odot\, n_0$. NGC~2362 is also a system from which gas has been expelled, so, again, the molecular cloud cannot provide the additional mass. However, uncertainty in mass estimation could make the difference between a bound cluster and unbound system. 

\section{Discussion}\label{discussion.sec}

The {\it Gaia} data reveal considerable diversity in kinematic properties in our sample of nearby young clusters, with examples of both expanding and non-expanding systems (Figures~\ref{arrow.fig}--\ref{vout_hist.fig}). The expected pattern of increased expansion velocity with distance from the cluster center is seen (Figure~\ref{expand_rad.fig}).  Velocity dispersions range from 1 to 3~km~s$^{-1}$, exhibit Gaussian distributions, and often exceed the expectations of virial equilibrium  (Figures \ref{phase.fig}--\ref{sigma_1d.fig}, \ref{virial_vs_obs.fig}). 

On larger scales, we examine the relative motions of subclusters in star forming complexes. Subcluster trajectories are typically divergent, reflecting their origins in turbulent clouds, rather than convergent as expected if clusters are currently assembling from smaller components (Figure~\ref{subcluster.fig}). 

\subsection{Expectations from Simulations}\label{simulations.sec}

Theoretical models of star cluster formation, informed by hydrodynamical and $N$-body codes, have led to predictions about the stellar dynamics of very young clusters -- even before sufficiently precise kinematic data were available to test these models. 

Cluster simulations typically show an initial collapse during the first crossing time where the global contraction is accompanied by increasing velocity dispersion. This brief phase is followed by a re-expansion with outward velocity accompanied by a slight decrease in velocity dispersion
\citep[e.g.,][their Figure 1]{proszkow2009}. Departure from spherical symmetry, such as elongated clusters, will produce observed kinematic structure dependent on  viewing angle. These projection effects can produce apparent velocity gradients and can influence the velocity dispersion by about a factor of two, mostly during the expansion phase \citep{proszkow2009}.

In gas-rich environments, subclusters are likely to merge into more massive young star clusters \citep{2009MNRAS.397..954F,2010MNRAS.404.1061M}. \cite{kuznetsova2015} find that the gas potential dominates on time scales $<$0.75--1 the free-fall time, encompassing the initial infall and star formation phases.  By the time the stellar potential begins to dominate, subclusters have already merged. \citet{kuznetsova2015} report stellar velocity dispersions of 3--4 km~s$^{-1}$ compared to gas velocity dispersion of 1--2 km~s$^{-1}$ for an isothermal (cold, sub-virial) model of an ONC-like cluster.  Other simulations \citep{2018ASSL..424..143B} imply that some observed massive young star clusters are too compact to be produced through hierarchical formation, thus suggesting the monolithic formation route. 

At the more advanced age at which young clusters are revealed in optical wavelengths, their initial kinematic state will have been altered by loss of residual gas, causing the system to expand and/or disperse \citep[e.g.,][]{2000ApJ...542..964A,2001MNRAS.321..699K}. Early studies focused on the role of star-formation efficiency in determining the final state of the system. These studies showed that for simple models of gas loss, there may be a threshold of 20\%--33\% star-formation efficiency that governed whether a cluster will survive \citep[e.g.,][]{1983ApJ...267L..97M,2006MNRAS.373..752G}, and that star-formation efficiency will affect the fraction of stars stripped from a surviving cluster  \citep[e.g.,][]{2001MNRAS.321..699K,2006MNRAS.369L...9B}. However, cluster survival is also strongly influenced by the initial dynamical state, with sub- virial initial states leading to higher survival probabilities and supervirial initial states leading to lower survival probabilities \citep{2009Ap&SS.324..259G}. The situation becomes even more complicated due to effects of cloud and star cluster structure. Simulations indicate that spatial decoupling between gas and stars \citep{2015MNRAS.451..987D} and highly substructured spatial distributions of stars \citep{2018MNRAS.476.5341F} will attenuate the effect of gas expulsion on a stellar system. Other factors such as dynamical ejection of massive stars from very dense clusters have also been identified as possible contributors to cluster mass loss leading to cluster expansion \citep{2013A&A...559A..38P}. Energy injected into clusters by the hardening or formation of binary systems can also contribute to cluster expansion \citep{2017A&A...597A..28B}.

The expansion of a cluster or association can follow different tracks depending on its mass and virial state. Even systems that have a positive total energy may leave behind a bound core of stars \citep[e.g.,][their Figure~4]{2006MNRAS.373..752G}. Escaping stars may lead to an excess number of stars at larger distances from a system center \citep[e.g.,][]{2006MNRAS.369L...9B}, and this could contribute to a positive radial expansion gradient. The timescale on which massive clus- ters may expand and/or evaporate in response to gas expulsion is on the order of 10 Myr \citep[][and references therein]{2014ApJ...794..147P}. Long term cluster survival (10~Myr to 1~Gyr) is positively correlated with cluster mass, but in clusters that arose from low star-formation efficiency environments the correlation may be weaker \citep{2018ApJ...863..171S}. The virial state of systems will also affect their structure as they evolve -- substructure will be erased in a few dynamical timescales from systems that are initially virial or subvirial, while supervirial systems may retain substructure as they expand \citep{parker2016}.

In the following discussion (Sections~\ref{disc_assembly.sec}-\ref{disc_expand.sec}), we use a simulation of cluster assembly from \citet{2018MNRAS.tmp..655S} as a benchmark to compare with the observed clusters in our study. The model used for comparison is the ``DR~21 Fiducial Model'' designed to understand the current gas and stellar distribution in DR~21, a massive filamentary star-forming region in Cygnus X. A full description of the model's initial conditions and assumptions is given by \citet{2018MNRAS.tmp..655S}. Briefly, this is a simulation of a 3000~$M_\odot$ star cluster in which stars originate in a chain of subclusters, spatially dispersed following the pattern of observed stars in DR~21. Each subcluster is modeled with an elongated Plummer sphere and a velocity dispersion based on the virial parameter. The subclusters have no initial bulk motion relative to each other in this simulation. Gas was modeled without star formation or stellar feedback, and moves gently out of the cluster center during the simulation. The model simulates the first 10~Myr of dynamical evolution, showing the progression from a clumpy assembly of subclusters to a single, virialized massive young star cluster. 

This simulation results in a bound system -- a different outcome than observed in many, but not all, of the systems in our sample -- and thus, the model is unlikely to be representative for the largely unbound systems in this study (Section~\ref{disc_expand.sec}). However, the simulation is useful for revealing what kinematic effects may be observed in bound clusters (Section~\ref{bound.sec}), and therefore allow us to evaluate which phenomena can be used as evidence for a system being unbound versus bound.

\subsection{Assembly of Clusters and Associations}\label{disc_assembly.sec}

The stellar systems included in our {\it Gaia}-based study are typically several million years old, and the clouds from which stars are forming have been partially or completely dispersed, so the epoch of cluster assembly will have mostly finished. However, kinematic properties such as  rotation and the motions of subclusters can provide constraints on how systems were assembled. 

In the simulation from \citet{2018MNRAS.tmp..655S}, starting at an earlier stage of evolution, subclusters rapidly converge to form a single cluster in $\sim$1~Myr. The evolution of the median $v_\mathrm{out}$ and velocity dispersion, calculated in the same way as above for the real clusters, is shown for the first 10~Myr in Figure~\ref{sim_v_evol.fig}. The infall of subclusters produces an initial negative median $v_\mathrm{out}$ for the central cluster for the first 2~Myr. The velocity distribution rebounds, yielding cluster expansion, before settling into a quasi-static state at $\sim$6~Myr. The infall velocity reaches $-0.75$~km~s$^{-1}$, while the rebound outward velocity peaks at $\sim$0.5~km~s$^{-1}$. The velocity dispersion increases during infall, peaking at the point of rebound, then settling down to just under 2~km~s$^{-1}$.

The only system in our study that may be caught in a phase of initial collapse is M~17, which appears to have a large (though uncertain) contraction velocity (median~$v_\mathrm{out}=-2.1\pm1.0$~km~s$^{-1}$). The spatial structure of stars in M~17 is also distinctive, being one of the few systems in the MYStIX survey with a ``clumpy'' structure suggesting that the merging of many subclusters is still underway \citep{2014ApJ...787..107K}. However, the other embedded stellar systems do not appear to be undergoing such rapid collapse. The systems NGC 1333 and NGC 2264 IRS1/2, much less massive than M17, are also gas rich and contain protostars, but neither of them have  statistically significant expansion or contraction.

For stellar systems in more evolved star-forming regions, where molecular gas has been partially or fully dispersed, the motions of subclusters are inconsistent with mergers. This supports the theoretical prediction that that hierarchical  cluster assembly, if it occurs, must happen promptly before gas is expelled by stellar feedback. 

Rotational properties of young stellar systems can also be used to test the hierarchical assembly scenario. Simulations of star formation by \cite{mapelli2017} indicate that bulk rotation should be common for clusters at ages of 1-2 Myr due to large-scale torques imparted on the gas as cluster assembly occurs. \cite{lee2016} also highlight the prominence of cluster rotation, which in their simulations results naturally from conservation of angular momentum during the global collapse.  Here rotation accounts for approximately 1/3 of the total kinetic energy, and together with turbulence acts to counteract gravity to keep the cluster globally virialized.  In both the Mapelli and Lee \& Hennebelle models, rotation signatures should be stronger than expansion signatures.  We do not find this result here, presenting an important constraint on the physical processes of cluster assembly. 

\subsection{Expanding Associations: Unbound or Bound}\label{disc_expand.sec}

The clear signs of expansion in many of the systems in our sample fit with the view that most star formation in massive star-forming complexes yields unbound systems, as described by \citet[][and references therein]{1538-3873-130-989-072001}. This view that many of these systems are not gravitationally bound is corroborated by the comparison of virial velocity dispersion to observed velocity dispersion (Figure~\ref{virial_vs_obs.fig}) that indicates that many of the systems are highly supervirial. In a few cases, specifically NGC 6530 and Cep B, the difference between the observed system mass and the mass needed to bind the system is sufficiently large (Section~\ref{dyn_expand.sec}) to definitively show that the systems are dispersing.

Expansion of stellar systems is correlated with gas loss in our sample of star-forming regions (Section~\ref{physical_properties.sec}). This correlation is consistent with a picture of system expansion as a reaction to changes in the gravitational potential due to the dispersal of the molecular cloud. However, it is also plausible that cloud dispersal and system expansion are both evolutionary processes that happen around the same time in star-forming regions but are independent of each other.

\begin{figure*}[t]
\centering
\includegraphics[width=0.40\textwidth]{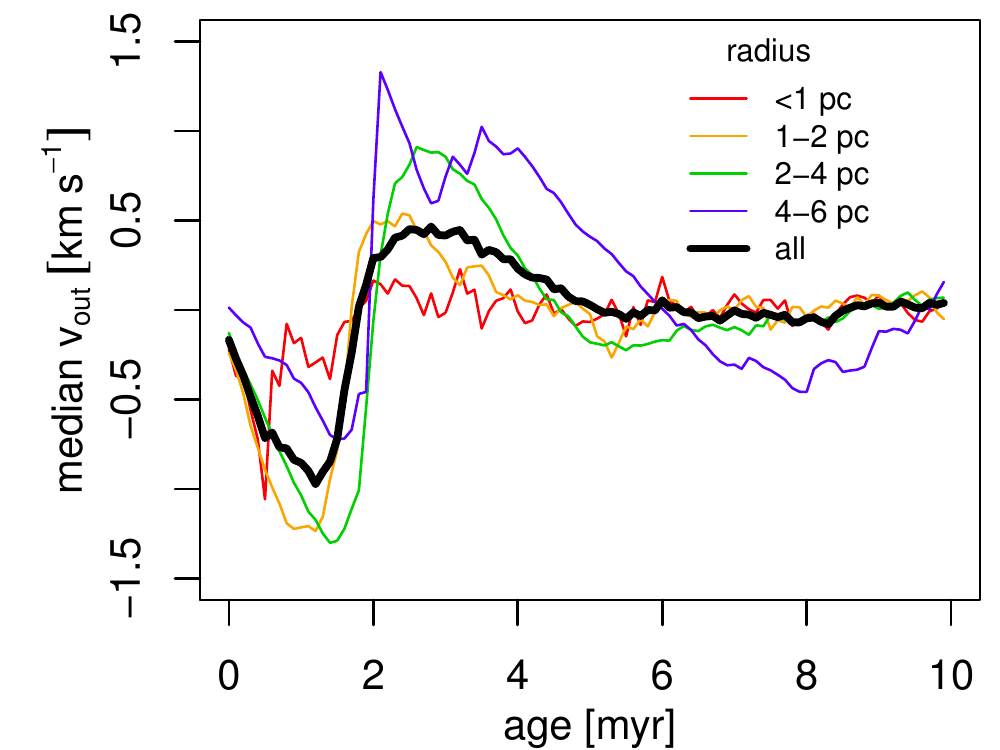} 
\includegraphics[width=0.40\textwidth]{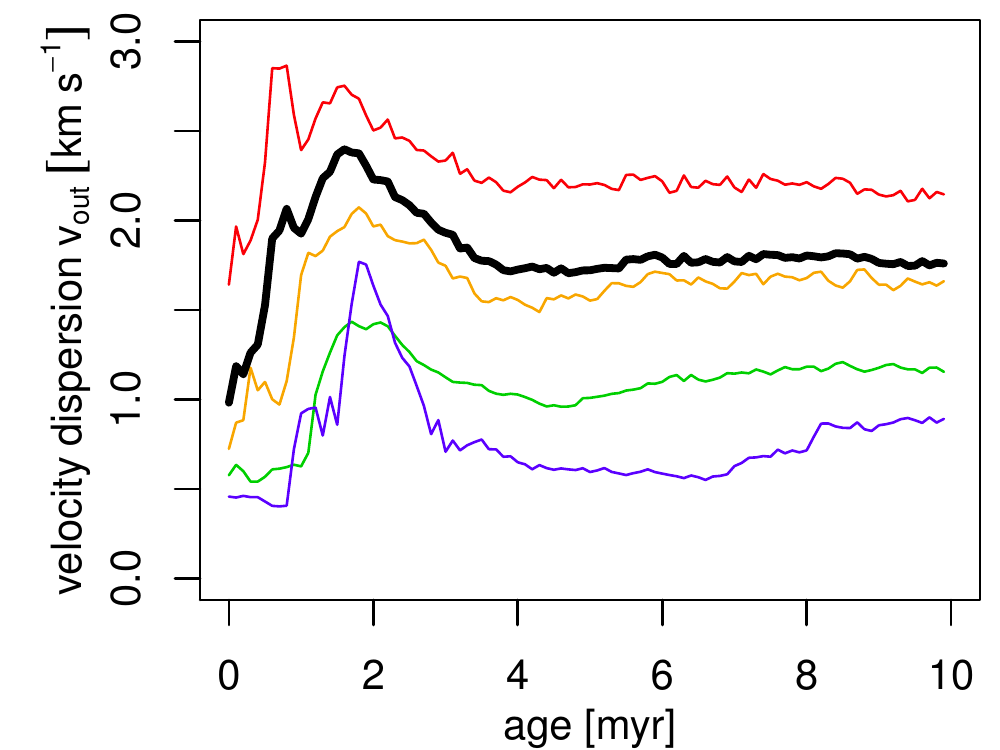} 
\caption{Characteristics of the velocity distribution for the stars in the \citet{2018MNRAS.tmp..655S} simulation as a function of time. The left panel shows median $v_\mathrm{out}$ while the right panel shows the standard deviation of $v_\mathrm{out}$. These statistics are calculated for stars in various radial bins as indicated by the legend. During the simulation, stars stream into the center of the cluster, yielding bulk inward velocities, before rebounding with a bulk outward velocity, and settling into a ``quasi-static'' state. 
 \label{sim_v_evol.fig}}
\end{figure*}

Observing expansion of a stellar system on the order of $\sim$0.5~km~s$^{-1}$ is insufficient evidence for demonstrating that a systems is unbound. In the simulated cluster of \citet{2018MNRAS.tmp..655S}, the median~$v_\mathrm{out}$ (Figure~\ref{sim_v_evol.fig}) increases after initial contraction, reaching an expansion speed of $\sim$0.5~km~s$^{-1}$, similar to the typical value observed for our expanding systems (Figures~\ref{vout_hist.fig} and \ref{sigma_1d.fig}). Velocity dispersion is also near maximum at this time, with a value of 2.5~km~s$^{-1}$. We also note that at the point of maximum expansion velocity in the simulation at $\sim$2.5~Myr, expansion velocity increases with radial distance from the system center (Figure~\ref{sim_v_evol.fig}, left), similar to the trends seen in Figure~\ref{expand_rad.fig} for NGC~6530 and Cep~B. However, a difference between the cluster expansion in the simulation and the pattern of expansion seen in NGC 6530 and Cep B is that, for the expanding associations, velocity dispersion increases with radius (Figure~\ref{sigma_rad.fig}) while velocity dispersion decreases with radius in the simulation of the bound cluster (Figure~\ref{sim_v_evol.fig}, right).

Apart from clear cases like NGC 6530 and Cep B, other expanding systems still lack sufficient information to definitively classify them as bound or unbound. To clearly distinguish between ``bound clusters'' and ``unbound associations'' we will require better constraints on the total mass of stars and gas in star-forming regions. In addition, some systems may have escaping stars, while a cluster core remains bound -- NGC 6611 in the Eagle Nebula is discussed below (Section~\ref{bound.sec}) as a possible example of this.

Considering slightly older systems, the conditions of formation of large OB associations like Sco-Cen have been uncertain, with debate about whether these were produced by expansion of an association or widely distributed star-formation events \citep{2016MNRAS.460.2593W,2018MNRAS.476..381W,2018MNRAS.475.5659W}. The systems in our sample are currently fairly compact, with sizes of several parsecs, but their kinematic properties mean that they will inevitably grow to sizes larger than 100 parsecs across. 

NGC~6530 may be a good analog of a precursors to a massive OB association. The bulk expansion velocity of NGC~6530 is 0.9~km~s$^{-1}$, but its velocity dispersion is 2.2~km~s$^{-1}$. The mass of the system compared to its virial mass is sufficiently low that self-gravity will have little effect on the evolution of stellar velocities, and the association will expand ballistically.  In 10 Myr, a star traveling at 2.2~km~s$^{-1}$ will travel 22~pc while a star with a velocity of 4.5~km~s$^{-1}$ (2$\sigma$ from the mean) would travel 45~pc, 
so that $\sim$95\% of the stars would be found within an region with a length of $\sim$90~pc. This size is quite similar to the size of Upper Scorpius in the Scorpius-Centaurus Association, which is $\sim$75--100~pc long along its longest axis \citep{2018MNRAS.477L..50G} and has an age of $\sim$10~Myr \citep{2016MNRAS.461..794P}. The estimated mass for NGC~6530 of $\sim$4000~$M_\odot$ is somewhat higher than 1400~$M_\odot$ for Upper Scorpius \citep{2008hsf2.book..235P}. 

Several of the expanding systems exhibit significant substructure in the spatial distribution of their stars \citep[see][their Figure~5]{2014ApJ...787..107K}. In NGC 6530, stars are not smoothly distributed but instead are clumped. In NGC 1893, the system is made up of a chain of subclusters, which likely traces the shape of the molecular filament from before gas was expelled from the system. If expansion of unbound stellar systems is truly ``Hubble-like'' this substructure would get expanded to larger sizes, as seen in large OB associations like Upper Scorpius. On the other hand, if spatial substructure is not reflected in kinematics, then spatial distributions of stars would tend to get smoother as stars drift further from their point of origin. The preservation of substructure during the dynamical evolution of a young stellar system is a result that has been predicted for supervirial systems (Section~\ref{simulations.sec}).

\subsection{Bound Clusters}\label{bound.sec}

Our sample includes a few systems that are likely to survive (temporarily) as bound open clusters (Section~\ref{bound.sec}). These systems are not currently expanding at a significant rate, and have total energies near the division between between bound and unbound systems. The ONC is still quite compact, while NGC 6231 and NGC 2362 have probably already expanded. Despite the young ages of these systems, which are all less than their virialization timescales for two-body interactions, they all show properties that would be expected for virialized systems. They are relatively well-fit by cluster models, and the cluster density, cluster core radius, and velocity distributions are consistent with theoretical expectations given the uncertainties. Velocity dispersions in these clusters are either constant with radius (expected for an isothermal structure) or decreasing.

In our sample, the ONC provides the best case for modeling velocity dispersions. For an isothermal sphere with equal-mass stars, the distribution of stellar velocities will be Maxwellian (i.e.\ a Gaussian distribution). Due to a number of assumptions that are violated in real young stellar systems, there is no reason to expect {\it a priori} that this distribution should accurately describe stellar velocities. Nevertheless, it turns out that the velocity distribution in the ONC is consistent with Gaussianity (Figure~\ref{qq.fig}), but the distribution is elongated north-south (Figure~\ref{phase.fig}). The elongation is parallel to the orientation of the Orion~A cloud, so it could result from perturbations of the gravitational potential from asymmetry in the star-forming complex, or it could be a residual effect of the cluster formation from a molecular filament  \citep[e.g.,][]{2009ApJS..185..486P,2015ApJ...815...27K}.

\begin{figure*}[t]
\centering
\includegraphics[width=0.40\textwidth]{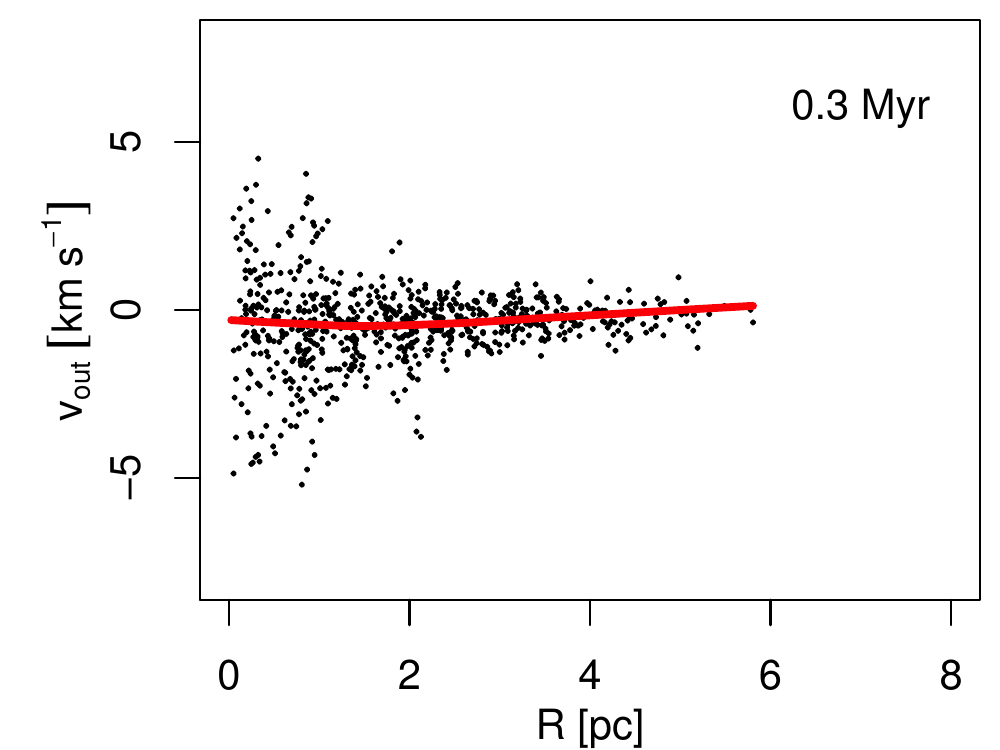} 
\includegraphics[width=0.40\textwidth]{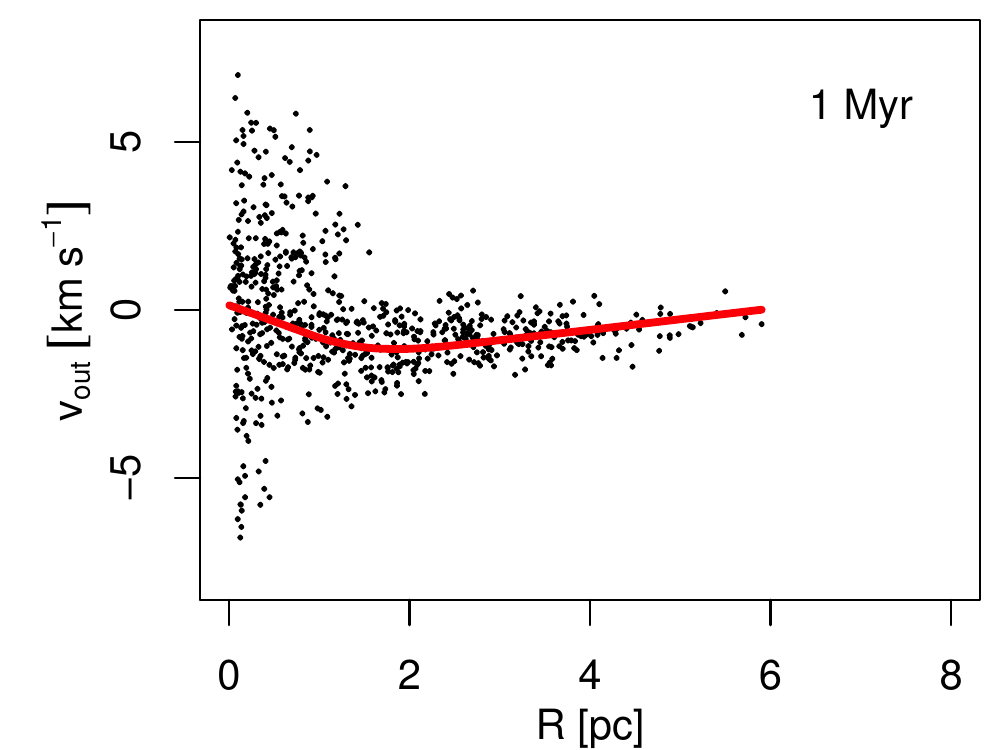} 
\includegraphics[width=0.40\textwidth]{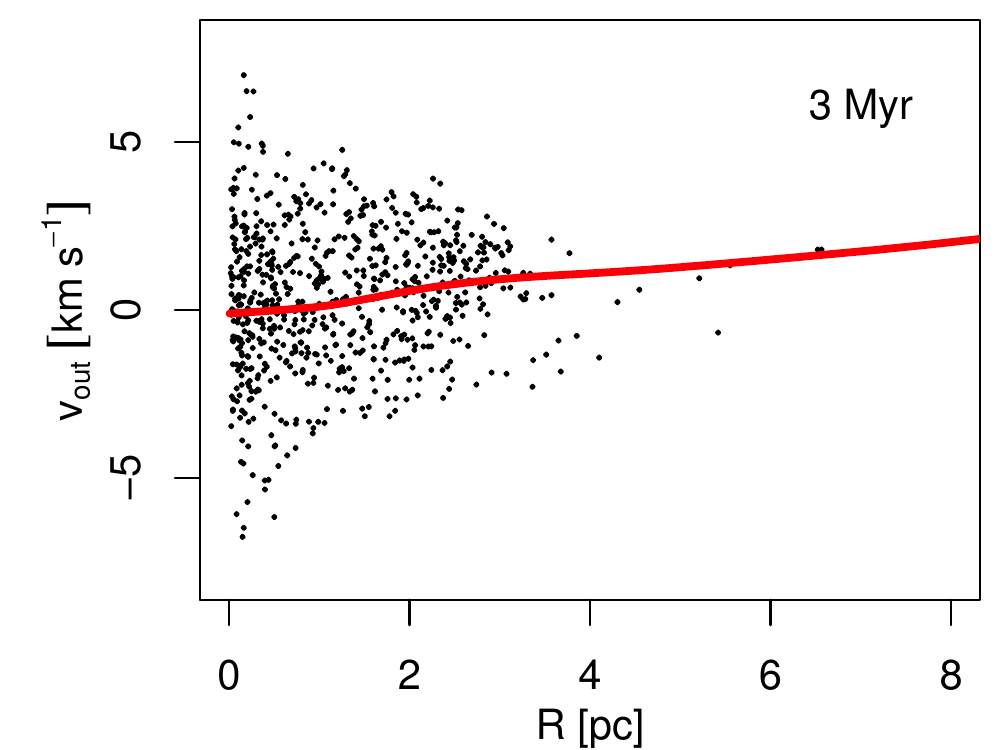} 
\includegraphics[width=0.40\textwidth]{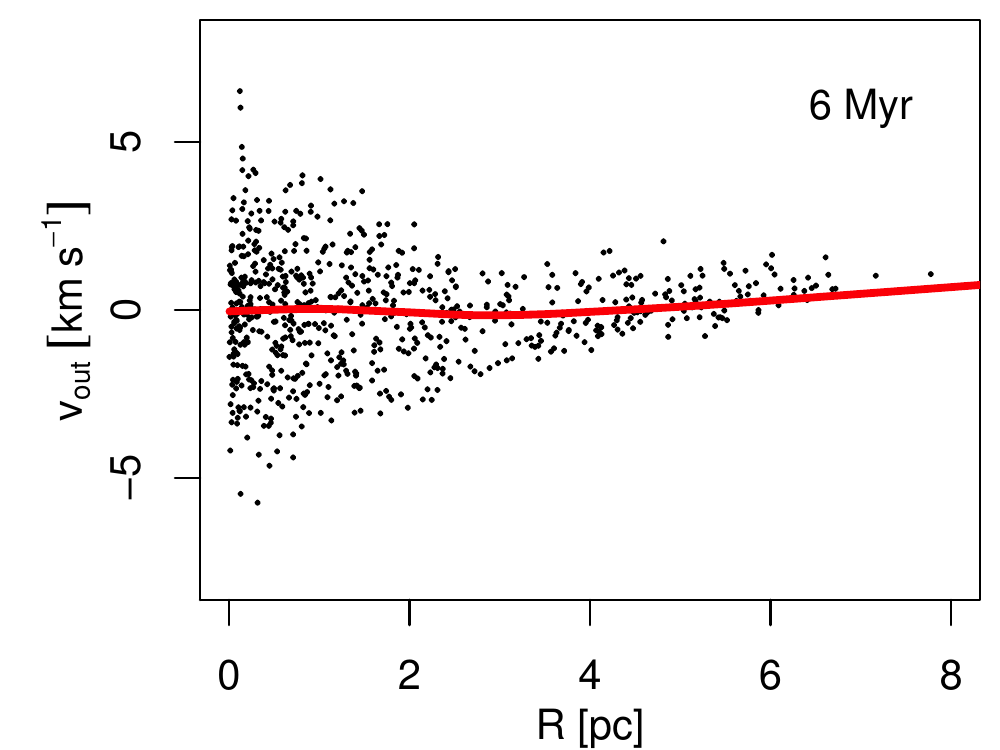} 
\caption{Snapshots of outward velocity $v_{out}$ as a function of distance $R$ from the cluster center in the \citet{2018MNRAS.tmp..655S} simulation at 0.3~Myr, 1.0~Myr, 3.0~Myr, and 6.0~Myr. Only stars with $M>1$~$M_\odot$ are shown for ease of plotting.  
 \label{sim_v.fig}}
\end{figure*}

The properties of probable bound clusters like the ONC, NGC 6231, and NGC 2362 can be compared to the simulated cluster from \citet{2018MNRAS.tmp..655S} to test the accuracy of its predictions. For several time points from the simulation, Figure~\ref{sim_v.fig} shows stellar velocity  $v_\mathrm{out}$ as a function of projected radius $R$. For each evolutionary stage shown, velocity dispersions are higher in the cluster center and lower in the cluster periphery. This is also clearly seen in Figure~\ref{sim_v_evol.fig}, where velocity dispersions are strongly affected by distance from the center of the cluster. This pattern is similar to radial dependence of velocity dispersion found in NGC 6231 (Figure~\ref{sigma_rad.fig}), but is different from the isothermal distributions in the ONC and NGC~2362. Future simulations can be used to test which initial conditions can produce the different types of observed velocity profile.

Bound systems seem to be in the minority in our sample based on the evidence from Figure~\ref{virial_vs_obs.fig}. Nevertheless, as discussed above, it is possible for a system with positive total energy to leave behind a bound core after most of the stars are lost. Given that many expanding systems show a radial gradient in expansion velocity, stars near the center might be less likely to be escaping from the system. A possible example could be NGC~6611, which is composed of a dense central cluster surrounded by lower-density groups of stars. Although the median expansion velocity of the system is $\sim$1~km~s$^{-1}$, the expansion velocity as measured within the inner 2~pc is lower at $\sim$0.5~km~s$^{-1}$ (Figure~\ref{expand_rad.fig}). Simulation of these systems may be helpful for determining whether a bound cluster is likely to remain in such cases.

%\vspace*{0.1in}
\section{Conclusions}\label{conclusions.sec}

The superb astrometric measurements of the {\it Gaia} spacecraft provided in DR2 have allowed us to examine the kinematics of nearby young star clusters and associations including many of the most massive star-forming complexes in our neighborhood of the Galaxy. The study also makes use of large samples of YSOs in these regions from the MYStIX and SFiNCs projects. Results from our sample of 28 systems likely represent the environments in which most star formation takes place in mature spiral galaxies.

The main scientific results of this study are: 
\begin{enumerate}
\item Bulk expansion is commonly seen for young stellar systems during the first few million years. At least 75\% of the systems in the sample have positive expansion velocities (likely 85--90\%), and expansion velocities range up to 2~km~s$^{-1}$, with a median value of $\sim$0.5~km~s$^{-1}$. Significant expansion around 1~km~s$^{-1}$ is measured for NGC~6530 in the Lagoon Nebula, Cep~B in the Cep~OB3b association, Tr~16 in the Carina Nebula, NGC~2244 in the Rosette Nebula, NGC~6611 in the Eagle Nebula, and NGC~1893.  

\item The most rapidly expanding systems have positive total energies, so their expansion can be explained as cluster dispersal. Velocity dispersions are sufficiently large for these associations to expand to the size of well-known OB associations like Sco-Cen in $\sim$10~Myr.

\item A positive radial gradient in expansion velocity is seen in some expanding systems like NGC~6530, Cep~B, NGC~2244, and Tr~16. Radial gradients in velocity can result from faster stars traveling larger distances, which yields positional sorting of stars by velocity. This phenomenon would be most prominent in systems where stellar trajectories are affected little by the gravitation of the cluster, so detection of this gradient provides evidence that these regions are unbound associations. In contrast, in a simulation of expansion in a gravitationally bound cluster, no radial dependence is seen.

\item Several systems appear likely to be gravitationally bound. Notable in this group is the still-compact ONC and the larger clusters NGC~6231 and NGC~2362 that may have expanded in the past. A mild expansion velocity is measured for the ONC, and there is no evidence for expansion of NGC~6231 or NGC~2362.  All three have velocity dispersions consistent with virial equilibrium; the ONC and NGC~2362 have isothermal velocity distributions while velocity dispersion decreases with radius in NGC~6231. The velocity dispersion in the ONC is approximately Gaussian. These findings are consistent with predictions of bound cluster models, and we expect that these clusters will emerge as gravitationally bound, main sequence, open clusters.

\item Stellar systems that are no longer embedded in their natal molecular clouds or only partially embedded are statistically more likely to be in a state of expansion than systems that are still embedded ($p < 0.001$). This result is consistent with expansion as a consequence of gas expulsion, but it is also possible that cloud dispersal and expansion of stellar systems occur simultaneously but independently.

\item Among the more embedded clusters, NGC 1333 and IC 348 in the Perseus cloud do not show signs of expansion or contraction. The embedded cluster M17 is unique in showing evidence (2$\sigma$ significance) of contraction with velocity $-2$~km~s$^{-1}$. This is consistent with its clumpy stellar structure and suggests it is still in the process of assembly. 

\item There is no evidence for rotation in all but one case, Tr~15. Theoretical simulations indicate
that rotation is expected to be present in clusters that inherit angular momentum from the merging of large subclusters, so its absence provides constraints on cluster formation. More sensitive measurements are required to further constrain rotation of young clusters.

\item In our sample of young stellar systems, one dimensional velocity dispersions, $\sigma_{1D}$, range from 1 to 3~km~s$^{-1}$. The velocity dispersion for the ONC (1.8$\pm$0.1~km~s$^{-1}$) is slightly lower than most previous estimates based on radial velocity studies. In the full cluster sample, velocity dispersions are typically greater than bulk expansion velocities by a factor of $\sim$2--3.

\item The relative motions of clusters within massive star forming regions generally show random motions, likely inherited from the parent molecular clouds.  They do not generally have convergent motions expected from hierarchical assembly, indicating that any cluster merging occurred during an embedded phase before the clusters were observed.
\end{enumerate}

While indirect evidence of cluster dispersal had been claimed previously \citep[e.g.,][]{2009A&A...498L..37P}, 
the {\it Gaia} observations provide direct evidence that stars produced in compact massive star-forming regions are more likely to immediately disperse after gas expulsion.  Only a few pre--main-sequence systems are likely to produce gravitationally bound, main sequence, open clusters. 
Future {\it Gaia} data releases are anticipated to dramatically improve astrometry for fainter sources, which will allow kinematics studies of stellar populations with higher absorption than the ones included in this study.  This will improve constraints on processes of cluster assembly and help distinguish which star-formation environments produce stellar systems of different dynamical fates. 

\appendix
\section{Distance to Orion}\label{orion.sec}

The Orion region is sufficiently close that distance measurements differ slightly depending on the location in the cloud. Based on our weighted median method described in the body of the paper, we find a median parallax of $\varpi=2.482\pm0.041$ for the X-ray selected probable cluster members in the ONC {\it Chandra} Field, corresponding to a distance of 403~pc. This field includes the dense central cluster, also known as the Trapezium Cluster, but excludes stars associated with Orion~A that lie more than 1.5~pc from the center. \citet{2018arXiv180504649K} report a {\it Gaia}-derived distance of 386$\pm$3~pc for the ONC, which is based on a much larger region spanning $\delta=-7^\circ$ to $-4^\circ$ in declination. Their analysis also differs in that they apply a correction to the {\it Gaia} parallaxes to account for systematic differences between astrometry from {\it Gaia} and the VLBA based on \citet{2017ApJ...834..142K}. This correction shifts objects at the distance of the ONC nearer by $\sim$10~pc. In our work, we have not applied this correction. 

\begin{figure}[t]
\centering
\includegraphics[width=0.75\textwidth]{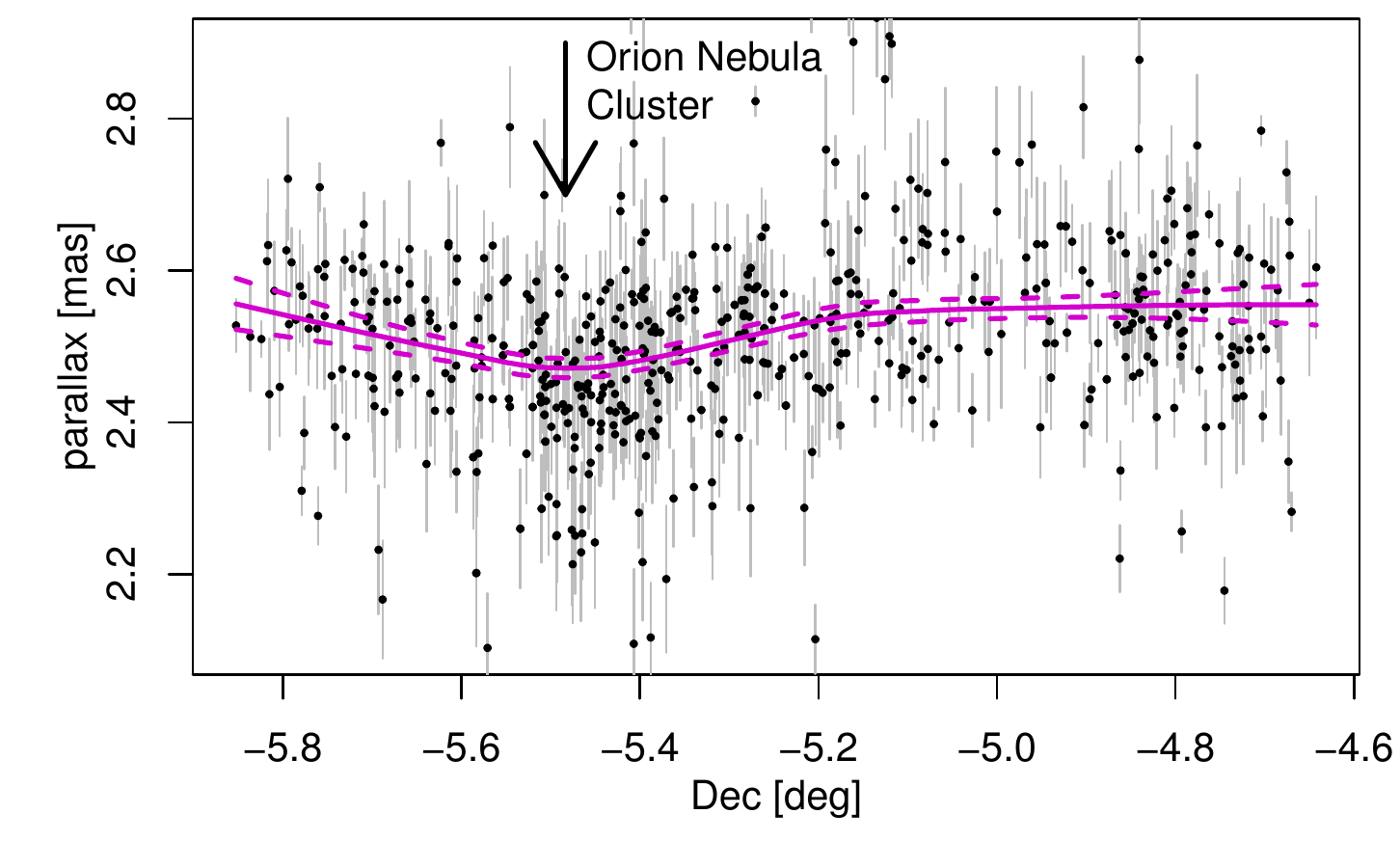} 
\caption{Parallax measurements for probable cluster members in the Orion A clouds, including the ONC. Only points with parallax uncertainties $<$0.1~mas are included. The magenta lines show a non-parametric regression line and 95\% confidence interval found using the loess regression in the R programming language with options span\,=\,0.6, degree\,=\,1, family\,=\,``symmetric'', iterations\,=\,4, and surface\,=\,``direct''. \label{onc_pardec.fig}}
\end{figure}

Figure~\ref{onc_pardec.fig} shows {\it Gaia} parallax measurements as a function of declination in the Orion~A cloud. We fit a non-parametric {\it loess} curve \citep{cleveland1979robust} and compute 95\% confidence intervals, which reveals some variation in mean parallax for stars in different parts of the Orion~A complex. The ONC itself is recessed (403~pc), while the stars to the north and south (including populations of stars both embedded in the filament and outside it) are nearer (395~pc). Given that this study is focused on the kinematics of the main cluster, we use the 403~pc distance. However, the discrepancy between our value and the distance from \citet{2018arXiv180504649K} can be explained by a combination of the different sizes of the region analyzed and whether additional correction factors are applied to {\it Gaia} astrometry.

\section{Proper Motion Corrections due to RVs}\label{radial_velocity.sec}

Calculating the contribution of perspective expansion to proper motions in Equations~\ref{perspective_expansion.eqn} and \ref{perspective_expansion2.eqn} requires measurements of the RVs of clusters. We have compiled a list of RVs for the clusters (Table~\ref{rv.tab}) based on both previously published RV measurements and RV measurements made by the {\it Gaia} spacecraft \citep{2018arXiv180409369C}. There are a few cases of discrepancies in RVs from the literature and/or median RVs from {\it Gaia}. In order to assign RV measurements to all clusters so that proper-motion corrections can be computed, we favor 1) the median {\it Gaia} RV when at least 5 stars are available, 2) the most recent literature RV, and finally 3) the {\it Gaia} RVs based on 1--4 stars. For Tr~15, which lacks its own independent RV measurement, we assign it an RV of $-20$~km~s$^{-1}$ based on the motion of the Carina Nebula as a whole. 

\begin{deluxetable}{lcccc}
\tablecaption{RVs of clusters\label{rv.tab}}
\tabletypesize{\small}\tablewidth{0pt}%\rotate
\tablehead{
 \colhead{Region} & \colhead{Sample} & \colhead{RV$_{Gaia}$} &  \colhead{RV$_\mathrm{lit}$}\\
 \colhead{} & \colhead{} & \colhead{km~s$^{-1}$} &  \colhead{km~s$^{-1}$}
}
\startdata
Berkeley~59  &   0 &  \ldots &             $-$14.5$^a$, $-$6.50$^b$\\
NGC~1333  & 0  & \ldots &             8.02$^c$ \\
IC~348  & 8  & 19.7$\pm$2.2&             14.00$^b$, 15.37$^d$\\
LkH$\alpha$ 101  & 2&   5.74$\pm$3 & 11--13$^{e\star}$             \\
NGC 1893    & 0 &  \ldots &             $-$2.25$^b$ \\
Orion A Cloud\\
--- ONC &  12 &  21.8$\pm$6.6&             26$^f$, 28.94$^b$\\
--- OMC 2--3 &  9 &  22.6$\pm$5.5\\             
--- Orion Flank S &  5 &  25.2$\pm$4.8\\          
--- Orion Flank N &  7&   27.4$\pm$3.3\\  
Mon R2&   1&   35.6$\pm$3&             28$^g$ \\
NGC~2244 &  1 &  19.1$\pm$0.96 &    26.16$^b$ \\    
NGC 2264&   7&   15.8$\pm$2.9&             22$^h$,  17.68$^b$\\
NGC~2362 &  1&   28.9$\pm$5.7  &       25.33$^b$     \\
Carina\\             
--- Tr~14 &   0 &  \ldots &             $-$6.7$^i$, $-$15.0$^b$  \\
--- Tr~15  &  0 &  \ldots &              	\ldots \\
--- Tr~16  &  0 &  \ldots &              	$-$25.00$^b$ \\
NGC~6231  & 1  & $-$19.1$\pm$1 &             $-$27.28$^b$\\
RCW 120 &  1 &  13$\pm$4.1 &         $-$16$^{j\star}$ \\   
NGC 6357   &   0 &  \ldots&            $-$12$^{k\star}$\\
M20   &  0 &  \ldots &             $-$2.13$^b$ \\
NGC~6530 &   0 &  \ldots&             $-$13.32$^b$\\
NGC~6611&   3&   16:&             18.00$^b$, 10$^l$\\
M17&    0 &  \ldots &             $-$17.00$^b$ \\
IC 5146  & 1  & 4.59$\pm$10  &           $-$5.5$^m$, $\sim$5$^n$\\
NGC 7160 &  1 &  $-$28$\pm$8.1  &            $-$23.8$^m$, $-$26.28$^b$ \\
Cep OB3b  & 4  & $-$28.7$\pm$4.4   &          $-$22$^o$, $-$15.12$^b$\\
\enddata
\tablecomments{Column~1: Region name. Column~2: Number of members with {\it Gaia} RV measurements. Column~3: Median RV (barycentric). Column~4: RV measurements from the literature (barycentric). Measurements for cloud gas are marked with an asterisk.\\
RV references:
($^a$) \citet{1989AJ.....98..626L},
($^b$) \citet{2005AaA...438.1163K},
($^c$) \citet{2015ApJ...799..136F},
($^d$) \citet{2015ApJ...807...27C},
($^e$) \citet{1982ApJ...255..564D},
($^f$) \citet{2008ApJ...676.1109F}, 
($^g$) \citet{2004PASJ...56..313K},
($^h$) \citet{2006ApJ...648.1090F},
($^i$) \citet{2002AaA...389..871D},
($^j$) \citet{1982ApJS...49..183B},
($^k$) \citet{1987AaA...171..261C},
($^l$) \citet{2005AaA...437..467E},
($^m$) \citet{1989AJ.....98..626L},
($^n$) \citet{1991AJ....102.1103L},
($^O$) \citet{2003MNRAS.341..805P}.
}
\end{deluxetable}

\acknowledgments We thank Alberto Krone-Martins and Lennart Lindegren for expert advice on {\it Gaia}, Floor van Leeuwen for advice about astrometry, and Fred Adams, Michael Grudi\'c, and Anushree Sengupta for useful feedback on the article. We would also like to thank the referee for helpful comments.  E.D.F.\ and K.V.G.\ are supported by NASA grant NNX15AF42G and {\it Chandra} grant AR7-18002X, GO7-18001X, G06-17015x, and the {\it Chandra} ACIS Team contract SV474018 (G.\ Garmire \& L.\ Townsley, Principal Investigators) issued by the {\it Chandra} X-ray Center under contract NAS8-03060. 

\facility{{\it Gaia}} 

\software{
          astro \citep{CRANastro},
          celestial \citep{2016ascl.soft02011R},
          MVN \citep{korkmaz2014mvn},
          R \citep{RCoreTeam2018}, 
          relaimpo \citep{gromping2006relative},
          spatstat \citep{baddeley2005spatstat},
          TOPCAT \citep{2005ASPC..347...29T}
          }

\clearpage\clearpage

\end{document}